\documentclass[aps,rmp,reprint,twocolumn,superscriptaddress]{revtex4-2}
\usepackage{enumerate,appendix,multirow,amsmath}
\usepackage{commath,braket}
\usepackage{color,calc,graphicx}
\usepackage[usenames,dvipsnames,svgnames,table,cmyk,hyperref]{xcolor}
\usepackage{epstopdf}
\usepackage[charter,cal=cmcal,sfscaled=false]{mathdesign}

\definecolor{oeawblue}{cmyk}{0.9,0.68,0,0}
\definecolor{iqoqiblue}{cmyk}{0.76,0.11,0,0}
\definecolor{iffsred}{cmyk}{0.12,0.94,0.87,0.34}
\usepackage{hyperref}
\newcommand{\bigzero}{\mbox{\normalfont\large\bfseries 0}}

\newcommand{\tilhv}{{\operatorname{TI-LHV}}}
\newcommand{\tins}{{\operatorname{TI-NS}}}
\newcommand{\tiq}{{\operatorname{TI-Q}}}
\newcommand{\ltins}[1]{{\operatorname{LTI_#1-NS}}}
\newcommand{\ltinpa}[2]{{\operatorname{LTI_#1-NPA_#2}}}

\def\be{\begin{equation}}
\def\ee{\end{equation}}

\hypersetup{
  pdftitle={Contextuality in infinite 1D TI local Hamiltonians},
  pdfstartview=Fit,
  pdfpagelayout=SinglePage,
  colorlinks,
  citecolor=iffsred,
  linkcolor=iffsred,
  urlcolor=iffsred}
\usepackage[figure]{hypcap}

\def\C{\mathbb{C}}
\def\Z{\mathbb{Z}}
\def\tr{\mbox{tr}}
\def\id{\mathbb{I}}

\begin{document}

\title{Contextuality in infinite one-dimensional translation-invariant local Hamiltonians: strengths and limits}

\author{Kaiyan Yang}
\affiliation{Institute of Fundamental and Frontier Sciences, \\University of Electronic Science and Technology of China, Chengdu 610054, China}
\affiliation{School of Mathematical Sciences, \\University of Electronic Science and Technology of China, Chengdu 611731, China}
\author{Xiao Zeng}
\affiliation{Institute of Fundamental and Frontier Sciences, \\University of Electronic Science and Technology of China, Chengdu 610054, China}
\affiliation{School of Computer Science and Engineering, \\University of Electronic Science and Technology of China, Chengdu 611731, China}
\author{Yujing Luo}
\affiliation{Institute of Fundamental and Frontier Sciences, \\University of Electronic Science and Technology of China, Chengdu 610054, China}
\author{Guowu Yang}
\affiliation{School of Computer Science and Engineering, \\University of Electronic Science and Technology of China, Chengdu 611731, China}
\author{Lan Shu}
\affiliation{School of Mathematical Sciences, \\University of Electronic Science and Technology of China, Chengdu 611731, China}
\author{Miguel Navascu\'es}
\affiliation{Institute for Quantum Optics and Quantum Information - IQOQI Vienna,\\ Austrian Academy of Sciences, Boltzmanngasse 3, 1090 Vienna, Austria}
\author{Zizhu Wang}
\affiliation{Institute of Fundamental and Frontier Sciences, \\University of Electronic Science and Technology of China, Chengdu 610054, China}

\begin{abstract}
In recent years there has been a growing interest in treating many-body systems as Bell scenarios, where lattice sites play the role of distant parties and only near-neighbor statistics are accessible. We investigate contextuality arising from three Bell scenarios in infinite, translation-invariant 1D models: nearest-neighbor with two dichotomic observables per site; nearest- and next-to-nearest neighbor with two dichotomic observables per site and nearest-neighbor with three dichotomic observables per site. For the first scenario, we give strong evidence that it cannot exhibit contextuality, not even in non-signaling physical theories beyond quantum mechanics. For the second one, we identify several low-dimensional models that reach the ultimate quantum limits, paving the way for self-testing ground states of quantum many-body systems. For the last scenario, which generalizes the Heisenberg model, we give strong evidence that, in order to exhibit contextuality, the dimension of the local quantum system must be at least 3.
\end{abstract}

\maketitle

\section*{Introduction}
In a many-body quantum system, correlations appear as one of the most common manifestations of the quantum nature of the system. Different types of correlations, such as entanglement, EPR steering and nonlocality, were identified over the years and found applications in various quantum information processing tasks. Out of these types of correlations, nonlocality is the strongest and most difficult to test~\cite{RevModPhys.86.419,ScaraniBook}. While experiments exploiting entanglement to teleport photons date back to the 90s~\cite{Bouwmeester1997Teleportation}, nonlocality passed the most stringent experimental tests only in 2015~\cite{loopholefree_Delft,PhysRevLett.115.250401,PhysRevLett.115.250402}. One of the key assumptions in any nonlocality experiment is keeping the different parties space-like separated. In a many-body quantum system, however, this assumption may be too formidable to be overcome.

In recent years, the exploration of contextuality in quantum many-body systems has been a fruitful endeavor. Contextuality witnesses adapted from Bell inequalities have been tested in Bose-Einstein condensates~\cite{JordiBEC,Tura13062014}. Translation and permutation symmetry allowed the full characterization of contextuality witnesses in many-body systems through bipartite correlators only~\cite{Marginal1D,Tura2015370,jordi_ti,PhysRevX.7.021005} (see~\cite{Sanpera2018Review} for a review). However, very little is known about the strengths and limits of the quantum models which violate these witnesses. 

In this work, building on earlier characterizations of classical local behavior in many-body systems~\cite{Marginal1D}, we investigate under which conditions the $n$-nearest neighbor statistics of a translation-invariant 1D quantum system evidence that the latter is contextual, by exploiting the connection between contextuality and Bell nonlocality. We focus on three different Bell scenarios, which differ on the number of measurement settings available to each party and the size of the near-neighbor marginals considered.

Our results show that some \emph{a priori} promising Bell scenarios are unlikely to show any form of contextuality, even if we allow greater-than-quantum correlations. In other scenarios, we give evidence that some Bell inequalities require quantum systems of high enough local dimension to be violated. More interestingly, for several Bell inequalities, we find the maximum violation compatible with the laws of quantum mechanics, and identify the Hamiltonians achieving it. The ground states of contextual Hamiltonians are entangled, even though locally they may appear separable. As shown in~\cite{Marginal1D}, it is possible to estimate the size at which the reduced density matrix of a quantum state becomes entangled, just by computing the difference between its average energy and the classical bound. Since the former can be easily estimated in essentially any Noisy Intermediate-Scale Quantum (NISQ) device, one can think of our contextuality witnesses as robust entanglement benchmarks for future quantum simulators. From a more fundamental perspective, identifying the maximum quantum violation of a $k$-local Bell functional opens the possibility to falsify quantum theory in the many-body regime. Indeed, given access to a NISQ device, we can represent the local measurements and the state preparation of the corresponding Bell test through a vector of lab controls $\theta$. By estimating the gradient of the Bell functional with respect to the variables $\theta$, we can sequentially update the latter so as to minimize the observed value of the Bell functional in the device (effectively mimicking the working principle of the variational quantum eigensolver \cite{VQE}). If it so happened that the final value of the Bell functional were below the quantum limit, then we would have disproven the universal validity of quantum theory, despite the lack of an alternative theoretical model.


One-dimensional quantum systems are the simplest many-body ensembles one can control in the lab. They can be found in natural condensed matter systems, as well as implemented via optical lattices or ion traps. For some such systems the only experimentally available data are near-neighbor correlators averaged over the whole chain (the so-called \emph{structure factors}). As shown in \cite{Marginal1D}, in the regime of large system size, structure factors correspond to the near-neighbor correlators of an infinite, translation-invariant chain. Comprehending Bell nonlocality in large 1D systems hence requires us to characterize near-neighbor correlations in classical, quantum and supra-quantum translation-invariant systems.

The correlations $P(a_1,\ldots,a_n|x_1,\ldots,x_n)$ generated by $n$ space-like separated classical systems with (classical) inputs $x_i, i\in{1,\ldots,n}$ and outputs $a_i, i\in{1,\ldots,n}$ admit a decomposition of the form:
\begin{align}
P(a_1,\ldots,a_n&|x_1,\ldots,x_n)\nonumber\\
&=\int P(a_1|x_1,\lambda)\ldots P(a_n|x_n,\lambda)P(\lambda) d\lambda,\label{LHV}
\end{align}
where $\lambda$ is a set of hidden variables with probability distribution $P(\lambda)$. The distributions $P(\lambda)$ and $\{P(a_i|x_i,\lambda)\}_{i=1}^n$ are hence a \emph{local hidden variable model} for the observed correlations $P(a_1,\ldots,a_n|x_1,\ldots,x_n)$. In Bell tests, different parties are required to be space-like separated, which can be seen as the physical realization of the independence of probabilities in~(\ref{LHV}). However, in a many-body quantum system, such a requirement is too formidable to overcome. As a result, when we assume that Eq.~(\ref{LHV}) holds in a quantum many-body system, what we are actually testing is contextuality~\cite{kochen1967problem,Budroni2021ContextualityReview}. The connection between contextuality and Bell nonlocality had been known since the 70s. Every Bell inequality can be regarded as a contextuality witness; the other direction is less systematic~\cite{Cabello2021KSBell}. The role of contextuality, especially of the Kochen-Specker type, in quantum computation has been actively investigated in recent years (for a review see~\cite{Budroni2021ContextualityReview}). It can be shown to be the source of the quantum advantage in several scenarios in quantum computation~\cite{NoisyShallowCircuits,ShallowCircuits,ContextualityMagic,PhysRevLett.119.120505}. Most of these scenarios are constructed from the stabilizer formalism with magic states. While a hidden variable model for this formalism has been found recently~\cite{PhysRevLett.125.260404}, the model itself is contextual~\cite{Budroni2021ContextualityReview}.

Our starting point is thus a Bell scenario with infinitely many parties in a chain, labeled by the integer numbers. At site $i\in\Z$, the corresponding party can conduct a measurement $x_i\in\{1,...,X\}$, obtaining the result $a_i\in A$. Since we assume translation invariance, the measurement statistics observed by any $m$ consecutive parties equal those of parties, $1,...,m$, that is, $P_{1,..,m}(a_1,...,a_m|x_1,...,x_m)$. Any Bell scenario in the translation-invariant chain can therefore be fully specified by the three natural numbers $m, X, |A|$. Consequently, in this paper, a Bell scenario where only nearest-neighbor correlations are available and each party can conduct two dichotomic observables will be called the \emph{222-scenario}. Extending the interaction distance to next-to-nearest neighbors gives us the \emph{322-scenario}. Heisenberg-like Bell scenarios, with nearest-neighbor interactions but three dichotomic observables per site correspond to the \emph{232-scenario}.

We say that an $m$-partite distribution $P_{1,..,m}(a_1,...,a_m|x_1,...,x_m)$ is no-signaling~\cite{popescu1994quantum} if, for all $i\in\{1,...,m\}$,

\begin{align}
&\sum_{a_i}P_{1,..,m}(a_1,...,a_m|x_1,...,x_i,...,x_m)=\nonumber\\
&\sum_{a_i}P_{1,..,m}(a_1,...,a_m|x_1,...,\tilde{x}_i,...,x_m),
\end{align}
\noindent for all pairs of measurement settings $x_i,\tilde{x}_i\in X$. Intuitively, this condition signifies that the statistics of the remaining $m-1$ parties are not affected by the choice of measurement setting of party $i$. Hence, party $i$ cannot instantaneously transmit information to others.

Some no-signalling distributions $P_{1,..,m}(a_1,...,a_m|x_1,...,x_m)$ can be shown not to arise out of an infinite no-signalling TI system. This idea is formalized in the following definition: we say that $P_{1,..,m}(a_1,...,a_m|x_1,...,x_m)$ admits a TI no-signalling extension if there exists a mapping $Q$ from finite sets $B\subset \Z$ to no-signalling $|B|$-partite measurement statistics $Q_{B}(a_B|x_B)$ with the following properties:
\begin{enumerate}
    \item 
    \begin{align}
    &\sum_{a_{B\setminus C}} Q_{B}(a_{B\setminus C},a_{C}|x_{B\setminus C},x_C)=\nonumber\\
    &\sum_{a_{D\setminus C}} Q_{D}(a_{D\setminus C},a_{C}|x_{D\setminus C},x_C),\nonumber
    \end{align}
    for all finite sets $B,C,D\subset\Z$, with $C\subset B,D$ (compatibility).
    \item $Q_B=Q_{B+z}$, for all $z\in\Z$ (translation invariance). 
    \item $Q_{1,...,n}=P_{1,...,n}$ (consistency with observed statistics).
\end{enumerate}
\noindent We call $\tins$ the set of all distributions $P_{1,..,m}(a_1,...,a_m|x_1,...,x_m)$ admitting a no-signalling, translation-invariant extension.

The existence of a no-signalling extension is just a pre-requisite for the existence of an overall infinite translation-invariant state. Whether such an entity exists at all depends also on the physics generating the observed correlations. We say that $P_{1,..,m}(a_1,...,a_m|x_1,...,x_m)$ admits a TI classical extension if it admits a NS extension $Q$ and there exist distributions $P(\lambda), \{P_i(a|x,\lambda):i\in\mathbb{Z}\}$ such that, for all $N$,

\be
Q_{-N,..,N}(a_{-N},...,a_N|x_{-N},...,x_{N})=\sum_\lambda P(\lambda)\prod_{i=-N}^NP_i(a_i|x_i,\lambda).
\ee
\noindent We call $\tilhv$ the set of all distributions $P_{1,..,m}(a_1,...,a_m|x_1,...,x_m)$ admitting a TI classical extension.

Analogously, $P_{1,..,m}(a_1,...,a_m|x_1,...,x_m)$ admits a TI quantum extension if it admits a NS extension $Q$ and there exist a Hilbert space ${\cal H}$, measurement operators $E_{a|x}:{\cal H}\to {\cal H}$, with $\sum_a E_{a|x}=\id$, and a translation-invariant quantum state $\rho$ on the infinite chain with local Hilbert space ${\cal H}$ such that, for all $N$,

\be
Q_{-N,..,N}(a_{-N},...,a_N|x_{-N},...,x_{N})=\tr\left\{\bigotimes_{j=-N}^N E_{a_j|x_j}\rho_{-N,....,N}\right\}.
\ee
\noindent We call $\tiq$ the set of all distributions $P_{1,..,m}(a_1,...,a_m|x_1,...,x_m)$ admitting a TI quantum extension.

In \cite{Marginal1D}, two of us provided a full characterization of the set of $m$-nearest neighbor correlations admitting a TI classical extension. This set happens to be a polytope, i.e., a convex set defined by a finite number of linear inequalities or \emph{facets}. When all local measurements are dichotomic ($|A|=2$), one can regard any measurement $x$ by party $i$ as an observable $\sigma^i_x$ with possible values $\pm 1$, and specify any no-signaling $m$-nearest neighbor distribution $P(a_1,...,a_m|x_1,...,x_m)$ through the averages of the different products of the observables $\sigma^{1}_{x_1},...,\sigma^{m}_{x_m}$. For $m=2$, in this `observable representation' a facet would take the form

\be
\sum_{x=1,...,X}J_{x}\langle\sigma^1_x\rangle+\sum_{x,y=1,...,X}J_{xy}\langle\sigma^1_x\sigma^2_y\rangle\geq L_J.
\label{facet}
\ee
\noindent Should the observed one-particle averages $\{\langle\sigma^1_x\rangle:x \in X\}$ and nearest-neighbor two-point correlators $\{\langle\sigma^1_x\sigma^2_y\rangle:x,y\in X\}$ of a TI system violate a facet of the classical (also called `local') polytope, the corresponding many-body system would be shown not to admit a description compatible with classical physics.

The left hand side of Eq.~(\ref{facet}) can be interpreted as a \emph{Bell functional} that acts linearly on the distribution $P_{1,...,m}(a_1,...,a_m|x_1,...,x_m)$. Minimizing it over all distributions admitting a TI quantum extension, we obtain the \emph{quantum limit} $Q_J$ of the Bell functional $J$.

In the following, we describe a method that, for any $d\in \mathbb{N}$, carries such a minimization variationally over TI quantum systems of local dimension $d$, thus obtaining an upper bound $\mathcal{Q}_d$ on $Q_J$. The method also returns a concrete TI quantum system, with $\mbox{dim}(H)=d$, achieving the Bell value $\mathcal{Q}_d$ with measurement operators $\{E_{a|x}:a,x\}$. Most statistical models studied in the literature use projective measurements, i.e. $\{E_{a|x}:a,x\}$ are projectors. For most of our results we only consider projective measurements, with one notable exception. When we need to verify that no 232-type Hamiltonian can violate the classical bound when $d=2$, only considering projective measurements is too restrictive. Therefore for these Hamiltonians we allow fully general complex positive operator-valued measurements (POVMs) as their local observables, using a modified version of the algorithm presented in the following section to perform the optimization.

\section*{Results}\label{sec:methods}

\subsection*{Upper bounding the ground state energy density}\label{sec:methods:upperbound}
To minimize the left-hand side of expressions of the form (\ref{facet}), we start from the following observation: let $\{\sigma_x:\C^d\to \C^d:x\in X\}$ be a set of $d$-dimensional Hermitian operators with spectrum contained in $\{1,-1\}$. Then, the minimum value of (\ref{facet}) over all TI quantum states corresponds to the minimum energy-per site of the TI Hamiltonian

\be
\mathcal{H}_{222}(\sigma_1,...,\sigma_m)=\sum_{i\in\Z}\sum_{x=1,...,X}J_{x}\sigma^i_x+\sum_{x,y=1,...,X}J_{xy}\sigma^i_x\sigma^{i+1}_y.
\label{hamilton}
\ee
\noindent Tools from condensed matter physics such as uniform matrix product states (uMPS)~\cite{MPSReview} allow us to compute the desired energy density efficiently. In order to minimize (\ref{facet}) for a given local dimension $d$, all we have to do is suitably explore the manifold of the set of local observables, e.g.: via gradient descent.


Our first step consists of finding a parametrization of all the local observables. Consider observables $\{\sigma_a|a\in X\}$, each of which can be diagonalized by an unitary matrix $U_a$ as
\begin{equation}
	\sigma_a = U_a \Lambda_a U_a^{\dagger},
	\label{unitary diagonalized}
\end{equation}
where $\Lambda_a$ is a diagonal matrix with entries $\pm 1$. To make this more explicit, we use the vector $[n_x, n_y]$ to describe number of $-1$ in the eigenvalues $\sigma_{x}$ and $\sigma_{y}$.

We can then use the space of skew-Hermitian matrices to effectively parameterize each $U_a$ as
\begin{equation}
	U_a = e^{S_a}, 
	\label{unitary Lie algebra}
\end{equation}
where $S_a$ is skew-Hermitian. Let $\{B_1, B_2, \dots, B_n\}$ be a basis of the vector space of skew-Hermitian matrices. Here $n=d^2-d$ denotes the dimension of the space. Expanding $S_a$ in this basis gives
\begin{equation}
	S_a(W_a) = \sum_{k=1}^{n}w_{ak} B_k,
	\label{skew-matrix linear combination}
\end{equation}
where $W_a \equiv \{w_{a1}, w_{a2}, \dots, w_{an}\}$ are scalars. Our optimization parameters are therefore $\{w_{ak}|a\in X; k=1,\ldots,n\}$. 

Using the method above, observables $\sigma_a (a=x,y)$ in ${\cal{H}}_{222}$ can be parameterized as
\begin{equation}
	\sigma_a(W_a) = (e^{\sum_{k=1}^nw_{ak} B_k}) \Lambda_a (e^{\sum_{k=1}^nw_{ak} B_k})^{\dagger}.
\label{general form of paramterized observables}
\end{equation}
Consequently, ${\cal{H}}_{222}$ is parameterized as ${\cal{H}}_{222}(W_x, W_y)$.

Using Jordan's lemma~\cite{ScaraniBook}, the number of real parameters can be reduced when $|X|=2$. For example, applying Jordan's lemma to ${\cal{H}}_{222}$ when $d=4$ yields a basis in which both $\sigma_x$ and $\sigma_y$ are block-diagonal: 
\begin{equation}
	\sigma_x = 
	\begin{bmatrix}
		\sigma_{x,1} &  \bigzero \\
		\bigzero & \sigma_{x,2} \\
	\end{bmatrix}, \quad
	\sigma_y = 
	\begin{bmatrix}
		\sigma_{y,1} &  \bigzero \\
		\bigzero & \sigma_{y,2} \\
	\end{bmatrix}
\end{equation}
where $\sigma_{x,1}, \sigma_{x,2}, \sigma_{y,1}, \sigma_{y,2}$ are ${2\times 2}$ Hermitian matrices. 

%

We are now ready to present our MPS based gradient descent method. The method is iterative. For $a=x,y$, let $W_a(k)$ denote the parametrization of observable $\sigma_a$ at the $k$-th iteration. We will refer to the parametrization $\{W_x(k), W_y(k)\}$ of both observables as $W(k)$. At each iteration $k$, the parameters $W(k)$ are updated to $W(k+1)$ through the following procedure.

First, we minimize the energy-per-site of the Hamiltonian ${\cal{H}}_{222}(k) \equiv {\cal{H}}_{222}(\sigma_a(k), \sigma_b(k))$ over the manifold of uMPS. The result $e(k)$ can be computed using, e.g., the Time-dependent Variational Principle (TDVP) algorithm~\cite{Haegeman2011TDVP,TDVP2,MPSReview} or the Variational Uniform Matrix Product State (VUMPS) algorithm~\cite{Valentine2018VUMPS,MPSReview}. We mainly use the TDVP algorithm for its good numerical stability and reasonable speed of convergence.

Following the TDVP algorithm, $e(k)$ can be expressed as
\begin{equation}\label{determined energy density equation}
	\begin{split}
		e(k) = \sum_{s,t,u,v} h(k)_{st}^{uv}
		{\rm Tr} [{A^t(k)}^{\dagger}
		{A^s(k)}^{\dagger}
		l(k) {A^u(k)} {A^v(k)} r(k)],
	\end{split}
\end{equation}
\begin{figure}[htbp!]
	\centering
	\includegraphics{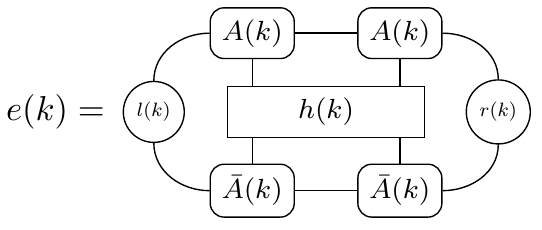}
	\caption{Energy density}
\end{figure}
\\ where $\sum_{s,t,u,v}h(k)^{uv}_{st}\ket{s}\bra{u}\otimes \ket{t}\bra{v}$ is the local term of ${\cal{H}}_{222}(k)$; $\{A^s(k)\}_s \subset \mathbb{C}^{D \times D}$ is the tensor defining the optimal uMPS, and $l(k), r(k)$ are the left and right leading eigenvectors of the transfer matrix $T(k) = \sum_{s=1}^{d}\bar{A}^{s}(k) \otimes A^{s}(k)$.

Next, we seek to find observables leading to a Hamiltonian with a smaller energy-per-site, when evaluated over the uMPS with tensor $\{A^s(k)\}$ just identified. Hence, with $A(k)$ fixed, we replace the local term $h(k)$ by $h(\sigma_x(W),\sigma_y(W))$ in Eq.(\ref{determined energy density equation}). This leads to a function $e(W;k)$ of the parameters $W$ defining the observables. To update the parameters $W(k)$, we move away from $W(k)$ in the direction of maximum function decrease at point $W(k)$. That is, we move against the gradient of $e({W};k)$:
\begin{equation}
	{W}(k+1) = {W}(k) - {\gamma}(k) \cdot \nabla_W e({W};k).
	\label{eq:update parameters}
\end{equation}
Here ${\gamma}(k)$ is a scaling parameter, which we take to be of the form ${\gamma}(k)= {\rm max}(\gamma_0\alpha^{q(k)}, \gamma_{\rm{min}})$, where $\alpha \in (0,1)$ and $q(k)$ is linear with respect to the iteration number $k$. 

Starting from an initial seed $W(0)$, we iterate the two steps above, hence generating a sequence of parameter values $(W(0),W(1),...)$. At every iteration $k$, we check the condition ${\Vert \nabla e(W;k) \Vert}_2 <\epsilon^*$, for some desired convergence threshold $\epsilon^*$. If the condition holds, we stop the algorithm and return the optimal parameters $W^* \equiv W(k)$.

In our experience, the quantity $e(W^*)$ is typically a very good estimate of the lowest quantum value of the considered contextuality functional over TI quantum systems of local dimension $d$. If $e^*$ happens to be smaller than the classical bound of the corresponding facet inequality, then we can state that the found quantum system characterized by the TI Hamiltonian ${\cal{H}}_{222}({W}^*)$ exhibits contextuality.

\begin{figure}[htbp!]\label{fig:322_iteration}
	\scalebox{0.49}{\includegraphics{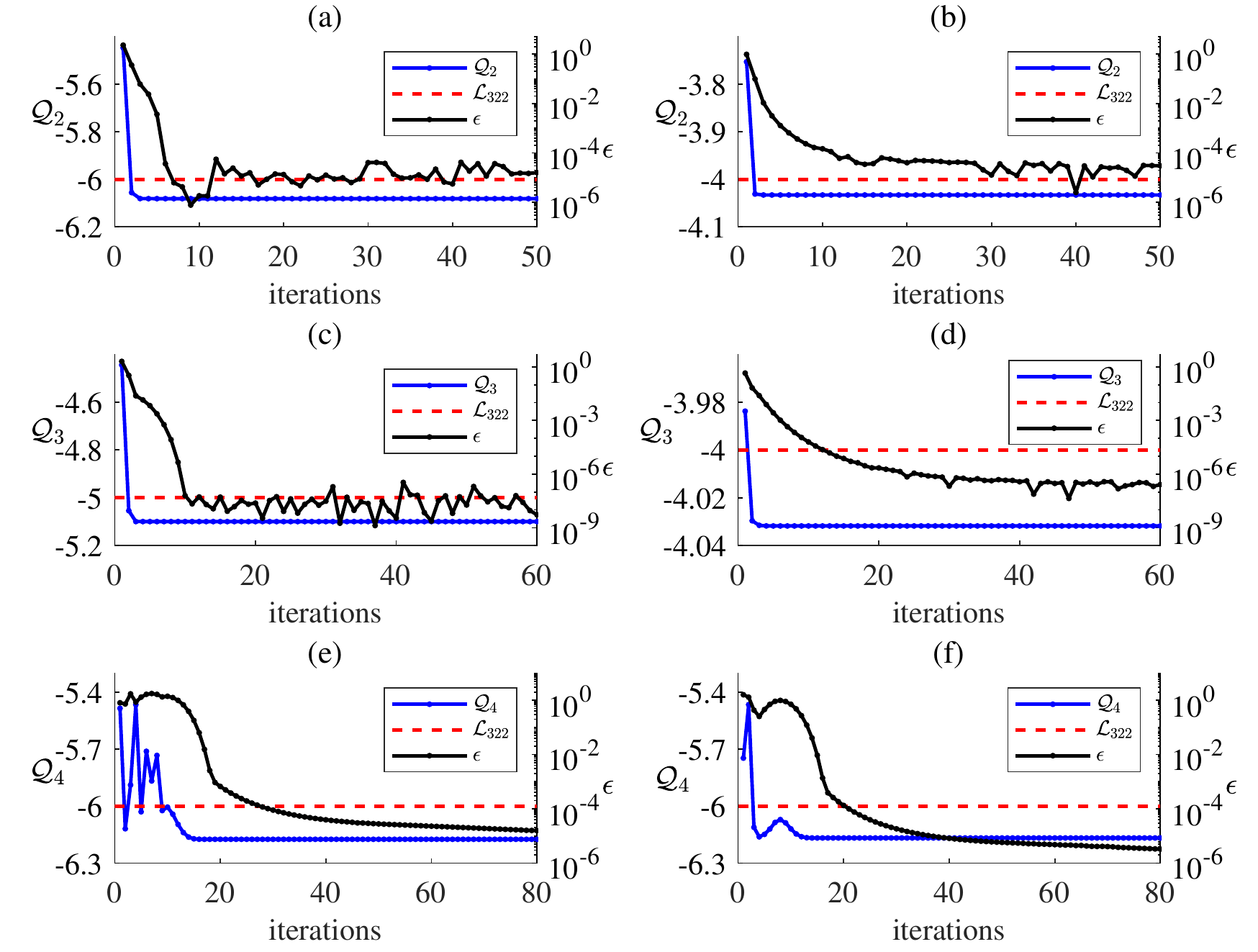}}
	\caption{Convergences of ground state energy density ${\cal{Q}}_d$ (blue line) and two-norm of gradient $\epsilon$ (black line) for 322-type TI quantum systems with the local observable dimension $d=2,3,4$. The red dashed line represents the classical bound, and the regions below show contextuality. (a) ${\cal{Q}}_2$ of No.1 in Table~\ref{all_322_parameters}. (b) ${\cal{Q}}_2$ of No.38 in Table~\ref{all_322_parameters}. (c) ${\cal{Q}}_3$ of No.7 in Table~\ref{all_322_parameters}. (d) ${\cal{Q}}_3$ of No.30 in Table~\ref{all_322_parameters}. (e) ${\cal{Q}}_4$ of No.37 in Table~\ref{all_322_parameters}. (f) ${\cal{Q}}_4$ of No.54 in Table~\ref{all_322_parameters}. }
\end{figure}

To test the algorithm, we apply it to compute the minimum ground state energy densities ${\cal{Q}}_d$ $(d=2,3,4)$ of six 322-type TI quantum systems introduced in Sec.~\ref{sec:322-type Hamiltonian}. All the results are plotted in Fig.~\ref{fig:322_iteration}. We find that the initial ground state energy densities determined by random parameters typically do not violate the classical bound ${\cal{L}}_{322}$ (red line). As the iteration number increases, ${\cal{Q}}_2$ and ${\cal{Q}}_3$ decrease approximately linearly and begin to show contextuality. In Fig.~\ref{fig:322_iteration}(e) and Fig.~\ref{fig:322_iteration}(f), ${\cal{Q}}_4$ oscillates during the first several iterations. As the optimization process continues, ${\cal{Q}}_4$ also begins to cross ${\cal{L}}_{322}$ after. The ground state energy densities of all six models converge to values below their classical bounds within 20 iterations.
 
\subsection*{Lower bounding the ground state energy density}
\begin{figure}[htbp!]
\centering
\includegraphics[width=\columnwidth]{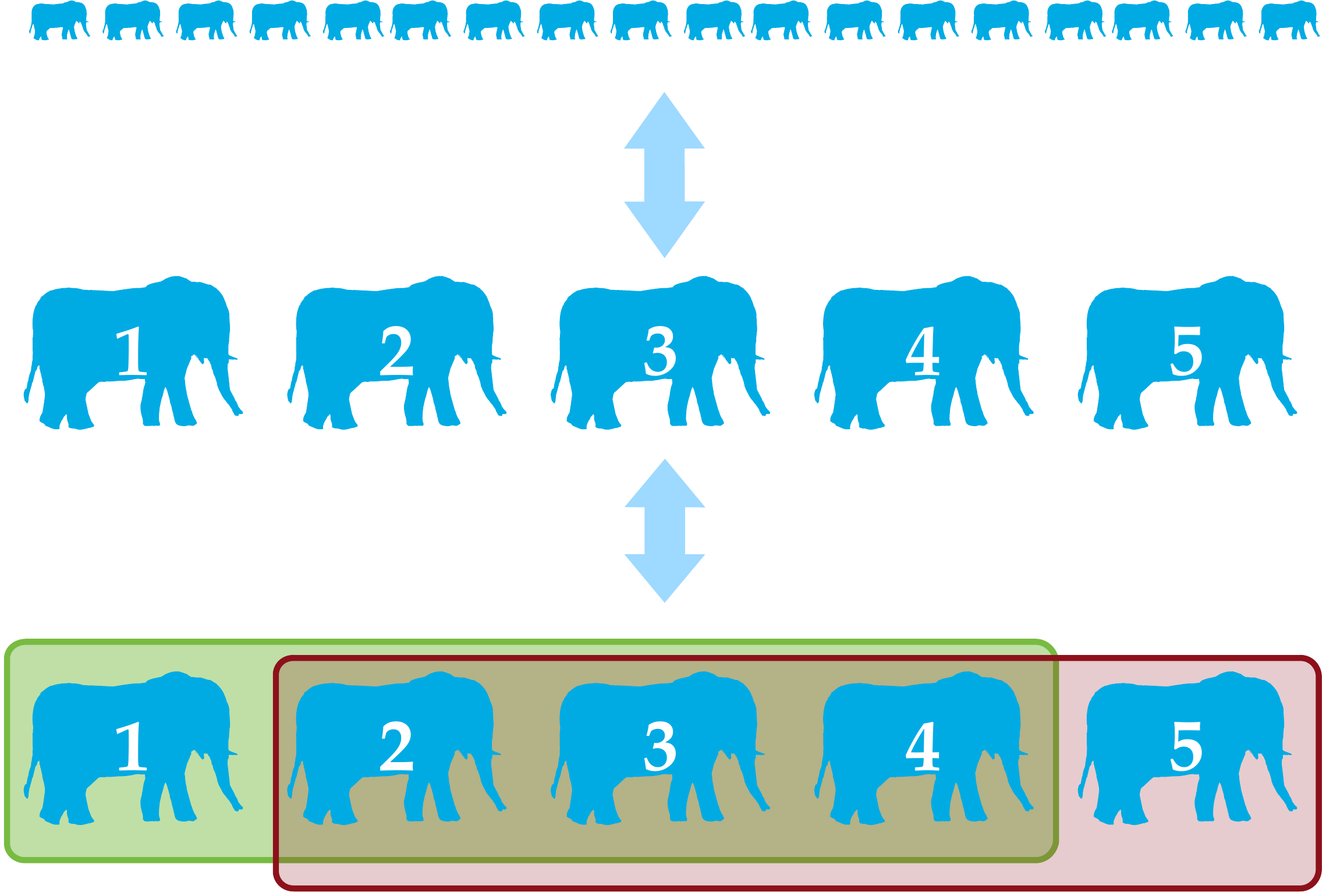}
\caption{Local translation invariance. A necessary condition from the requirement that the entire system is translation-invariant.}
\label{fig:lti}
\end{figure}

Consider the scenario shown in Fig.~\ref{fig:lti}: an infinite chain of elephants, each of which represents a physical system, be it quantum, classical or else. Call $\varepsilon$ the overall state of the chain. Depending on the context, $\varepsilon$ will be a classical probability distribution, a quantum state or a no-signaling box. Because  $\varepsilon$ is TI, the marginal distribution or the reduced state of each of the $5$ marked elephants, taken from an arbitrary contiguous subset of the chain, should be equal: $\varepsilon_1=\ldots=\varepsilon_5$. Moreover, the reduced state of any contiguous subset of elephants should also be equal: $\varepsilon_{1,\ldots,1+k}=\varepsilon_{2,\ldots,2+k}, \forall 1\leq k\leq 3$. When $k=3$, the marginals/reduced states are shown in Fig.~\ref{fig:lti} as green and red rectangles. For any contiguous subset of $\varepsilon$ of length $l$, the marginals/reduced states are said to be \emph{locally translation-invariant} (LTI) if
\begin{align}
\varepsilon_{1,\ldots,l-1}=\varepsilon_{2,\ldots,l}.\label{eq:lti}
\end{align}

Clearly, LTI is a necessary condition for $\varepsilon$ to be TI. For classical probability distributions in 1D, LTI is also sufficient: any LTI marginal can be extended to an infinite TI distribution~\cite{Goldstein2017}. In fact, this property is the key to the characterization the set $\tilhv$ presented in~\cite{Marginal1D}. 

Unfortunately, LTI is not enough to characterize the near-neighbor density matrices of TI quantum states or even the near-neighbor marginals of TI no-signaling systems. In those scenarios LTI can be used use to relax the set of such marginals, rather than to fully characterize it. Define thus $\ltins{n}$ as the set of boxes admitting an extension to an $n$-partite no-signaling box with local translation invariance. As shown in \cite{Marginal1D}, the distance between any element of the set $\ltins{n}$ and its subset $\tins$ is upper bounded by $O\left(\frac{1}{n}\right)$.

A straightforward extension to bound $\tiq$ is impossible, as the approximate characterization of general multipartite quantum correlations is an undecidable problem \cite{ji2020mipre}. One can, however, relax the existence of quantum states and observables reproducing the observed correlations to that of positive semidefinite \emph{moment matrices}. Those are matrices $\Gamma$ whose rows and columns are labeled by monomials of measurement operators with at most $s$ (the order of the relaxation) measurement operators per party, and where each entry $\Gamma_{\alpha \beta}$ is supposed to represent the quantity $\langle \alpha^\dagger \beta\rangle$, see \cite{NPA2,NPA1} for details. In order to bound $\tiq$, we demand the existence of a moment matrix for an $n$-partite Bell scenario and then impose LTI over the said moment matrix. Call $\ltinpa{n}{s}$ the corresponding relaxation. 

For any Bell functional, we can thus find a lower bound on its minimal value in $\tins$ and $\tiq$ by respectively optimizing over $\ltins{n}$ (with linear programming techniques~\cite{LPBook}) and $\ltinpa{n}{s}$ (with semidefinite programming techniques~\cite{SDPBook}). Moreover, one can improve those lower bounds by increasing the values of $n,s$.


\subsection*{Contextuality in 222-type Hamiltonians}\label{sec:222-type Hamiltonian}
The LTI-LHV polytope for 2 dichotomic observables has 36 facets. Computing their $\ltins{4}$ lower bounds reveals that most of them coincide with the corresponding classical bounds. In fact, there is only one inequality, up to local relabeling, which can potentially show contextuality. In its 1D TI quantum Hamiltonian form, it reads
\begin{equation}
	\begin{split}
		{\cal H}_{222} = & \sum_{i=1}^{\infty}(2\sigma_x^i +  \sigma_x^i\sigma_x^{i+1} + \sigma_x^i\sigma_y^{i+1} -\sigma_y^i\sigma_x^{i+1} - \sigma_y^i \sigma_y^{i+1}),
	\end{split}
	\label{Hamiltonin222_example}
\end{equation}
with classical bound $-2$.

Lower bounding Eq.~\eqref{Hamiltonin222_example} with $\ltins{n}$ with increasing $n$, we observed some curious phenomena. Because exact optimal solutions of linear programs are rational numbers, we obtain the solutions in Table~\ref{tab:lti-ns}.

\begin{figure}[htbp!]
\centering
\includegraphics[width=\columnwidth]{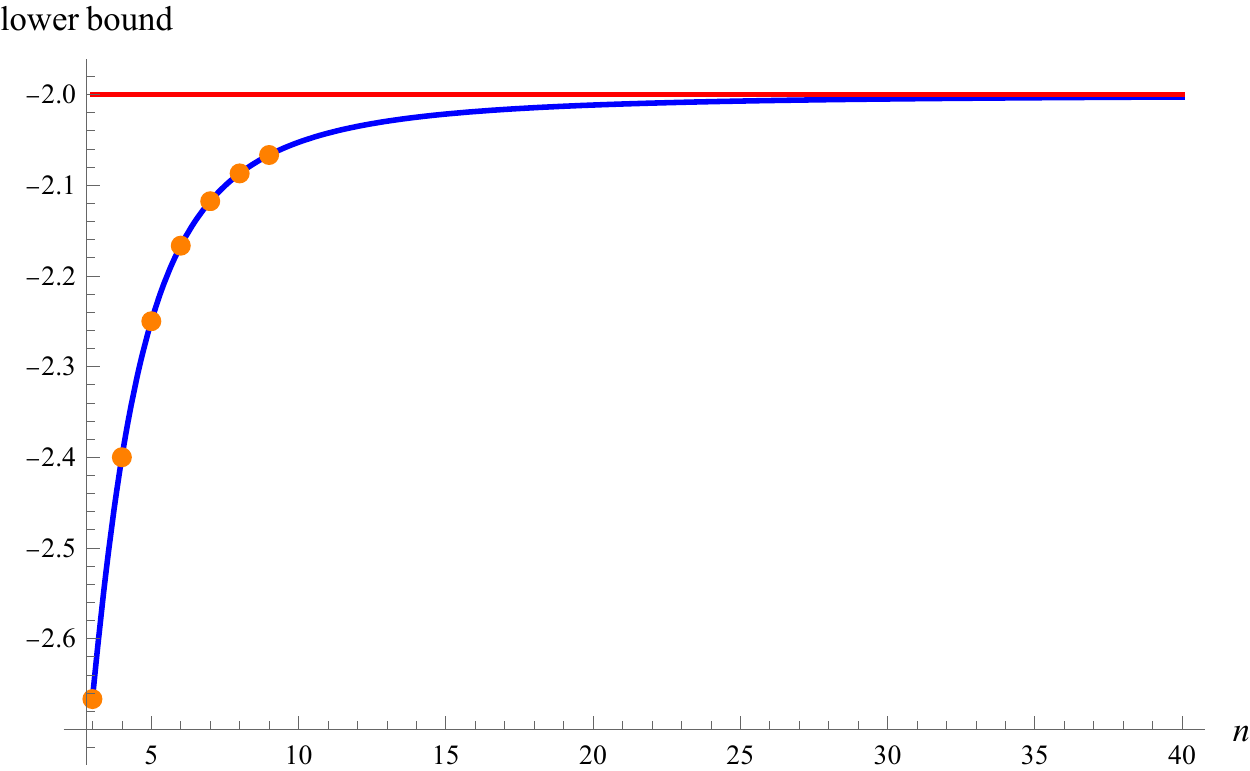}
\caption{Exact solutions for the $\ltins{n}$ approximation of the lower bound of Eq.~\ref{Hamiltonin222_example} (orange dots), the fitted function (\ref{ltins-fitted-function}) (blue line) when $3\leq n\leq 40$, and the classical bound (red line).}
\label{fig:lti-ns-exact}
\end{figure}

\begin{table}[htbp!]
\renewcommand{\arraystretch}{1.2}
\begin{tabular}{|c|c|c|c|c|c|c|c|}
\hline
$n$ & 3 & 4 & 5 & 6 & 7 & 8 & 9 \\
\hline
$\ltins{n}$ lower bound & $-\frac{8}{3}$ &$-\frac{12}{5}$  & $-\frac{9}{4}$ & $-\frac{13}{6}$ & $-\frac{36}{17}$ & $-\frac{48}{23}$ & $-\frac{62}{30}$ \\
\hline
\end{tabular}
\caption{Exact solutions of $\ltins{n}$ approximations of the lower bound of \eqref{Hamiltonin222_example} as a function of $n$.}
\label{tab:lti-ns}
\end{table}
The numerators and denominators in the table form two integer sequences: A027691~\cite{oeisNum} and A152948~\cite{oeisDenom} in \textit{The On-Line Encyclopedia of Integer Sequences}. Moreover, the displaced inverse of a quadratic function 
\begin{align}
-2-\frac{4}{n^2-3n+6},\, n\in\mathbb{N},\, n\geq3\label{ltins-fitted-function}
\end{align}
\noindent perfectly fits the sequence of lower bounds in Table~\ref{tab:lti-ns}, see Fig.~\ref{fig:lti-ns-exact}.

In the limit $n\to\infty$, this function converges to the classical bound $-2$. In other words, if the solution of the optimization over $\ltins{n}$ satisfies (\ref{ltins-fitted-function}) for all $n\geq 3$, then no Hamiltonian of the form (\ref{Hamiltonin222_example}), quantum or otherwise, can possibly violate the classical bound.

Proving that a series of rational numbers, the solutions of linear programs of exponentially increasing size, converges to a certain value is very hard. However, we do have additional numerical evidence to support our claim that the lowest possible ground state energy density of 1D TI quantum Hamiltonians of the form (\ref{Hamiltonin222_example}) is $-2$. We used our algorithm, described in Section~\ref{sec:methods}, to search for the quantum Hamiltonian with the lowest ground state energy density, for local observables of dimension $2\leq d\leq 6$. For each $d$, $\sigma_{x}$ and $\sigma_{y}$ are parameterized by the method described below, and we find the lowest quantum value ${\cal{Q}}_d$ among all possible systems is $-2$. Moreover, the corresponding two-body reduced density matrix of the quantum system for the ground state is a rank $1$ projector, which shows that the ground state is in fact a product state.  We present these ground states and the parameters for observables in Table~\ref{table:222 pure state and parameters}.
\begin{table}[htbp!]
	\caption{\label{table:222 pure state and parameters}Ground state and parameters of ${\cal{H}}_{222}$ under the different observable dimension settings when the ground state energy density equals to the classical bound $-2$.}
		\renewcommand\arraystretch{1.3}
		\begin{ruledtabular}
			\begin{tabular}{ccccccc}
				$d$ & $w_1$  & $w_2$   &  $w_3$  & $[n_x, n_y]$ & Ground state 	\\ \hline
				2 &  0.00001 &   --    &   --    &  $[1, 1]$    & $\ket{1}^{\otimes\infty}$ \\ \hline
				3 &  0.65129 &   --    &   --    &  $[2, 2]$    & $\ket{2}^{\otimes\infty}$ \\ \hline
				4 &  0.00004 & 0.59946 &   --    &  $[2, 2]$    & $\ket{3}^{\otimes\infty}$ \\ \hline
				5 &  0.98662 & 0.00001 &   --    &  $[2, 2]$    & $\ket{4}^{\otimes\infty}$ \\ \hline
				6 &  0.90451 & 0.91913 & 0.00005 &  $[3, 3]$           & $\ket{5}^{\otimes\infty}$ \\ 
		\end{tabular}
	\end{ruledtabular}
\end{table}


To make sense of Table \ref{table:222 pure state and parameters}, we next explicitly write the parametrization of the $d$-dimensional observables achieving the classical bound. Two $2\times 2$ matrices having eigenvalues one $1$ and one $-1$ will repeatedly appear below: $\Lambda$ is the diagonal matrix with diagonal entries $\pm1$, $B(w)$ is a matrix governed by one parameter $\{w\}$:
\begin{equation}\label{2-dimensional block matrices}
	\Lambda = \begin{bmatrix}
		1 & 0 \\ 
		0 & -1\\
	\end{bmatrix}, \quad
	B(w) = 
	\begin{bmatrix}
		\cos(2w) & -\sin(2w) \\ 
		-\sin(2w) & -\cos(2w)\\
	\end{bmatrix}.
\end{equation}

When $d=2$ is assigned to local observables in ${\cal{H}}_{222}$, $\sigma_{x}$ is a diagonal matrix with diagonal entries $1$ and $-1$, i.e., $\sigma_{x} = \Lambda$, and $\sigma_y$ determined by one parameter $\{w_1\}$ has the same parameterized form as $B(w)$, i.e., $\sigma_y(w_1)=B(w_1)$. In this case, $[n_x, n_y] = [1, 1]$ and $\sigma_{x}, \sigma_{y}$ are of the form
\begin{equation}\label{222_d2_sigma_x_sigma_y}
	\sigma_{x} = \Lambda, \quad
	\sigma_{y}(w_1) = 
	B(w_1).
\end{equation}

When $d=3$ is assigned to local observables in ${\cal{H}}_{222}$, $\sigma_{x}$ is a diagonal matrix with diagonal entries $(1,-1,-1)$.  $\sigma_{y}$ is a block diagonal matrix, where the main-diagonal blocks are one matrix $B(w_1)$ and one numerical value $-1$. Then, $[n_x, n_y] = [2, 2]$ and $\sigma_{x}, \sigma_{y}$ are given by 
\begin{equation}\label{222_d3_sigma_x_sigma_y}
	\sigma_{x} = 
		\begin{bmatrix}
			\Lambda  & \bigzero \\ 
			\bigzero & -1 \\	
		\end{bmatrix}, \quad
    \sigma_{y}(w_1) = 
		\begin{bmatrix}
			B(w_1)  & \bigzero \\ 
			\bigzero & -1 \\			
		\end{bmatrix}.
\end{equation}

When $d=4$ is assigned to local observables in ${\cal{H}}_{222}$, $\sigma_{x}$ is a diagonal matrix with diagonal entries two $1$ and two $-1$, and $\sigma_{y}$ is a block diagonal matrix with main-diagonal blocks being two $2\times2$ matrices $B(w_1)$ and $B(w_2)$. In this case, $[n_x, n_y] = [2, 2]$ and $\sigma_{x}, \sigma_{y}$ are of the forms 
\begin{equation}\label{222_d4_sigma_x_sigma_y}
	\sigma_{x} = 
		\begin{bmatrix}
				\Lambda  & \bigzero \\ 
			\bigzero & \Lambda \\			
		\end{bmatrix}, \quad
	\sigma_{y}(w_1, w_2) =
		\begin{bmatrix}
			B(w_1)  & \bigzero \\ 
			\bigzero & B(w_2) \\			
		\end{bmatrix}.
\end{equation}
	
When $d=5$ is assigned to local observables in ${\cal{H}}_{222}$, $\sigma_{x}$ is a diagonal matrix with diagonal entries two $-1$ and three $1$. $\sigma_{y}$ is a block diagonal matrix, where the main-diagonal blocks are two $2\times2$ matrices $B(w_1)$ and $B(w_2)$ and one numerical number $1$. Then, $[n_x, n_y] = [2, 2]$ and $\sigma_{x}, \sigma_{y}$ are given by 
\begin{equation}\label{222_d5_sigma_x_sigma_y}	
	\sigma_{x} = \begin{bmatrix} 
		\Lambda    &      & \multicolumn{2}{c}{\multirow{2}*{{\Large\bfseries0}}}\\
		&  \Lambda  &  \\
		\multicolumn{2}{c}{\multirow{1}*{{\Large\bfseries0}}} &  & 1\\
	\end{bmatrix}, \quad
	\sigma_{y}(w_1, w_2) = \begin{bmatrix}
		B(w_1)    &      & \multicolumn{1}{c}{\multirow{2}*{{\Large\bfseries0}}}\\
		&  B(w_2)  &  \\
		\multicolumn{2}{c}{\multirow{1}*{{\Large\bfseries0}}} &  & 1\\
	\end{bmatrix}.
\end{equation}

When $d=6$ is assigned to local observables in ${\cal{H}}_{222}$, $\sigma_{x}$ is a diagonal matrix, where three $-1$ and three $1$ are alternately arranged in the diagonal. $\sigma_{y}$ is a block diagonal matrix with main-diagonal blocks being three $2\times2$ matrices $B(w_1)$, $B(w_2)$ and $B(w_3)$ . Then, $[n_x, n_y] = [3, 3]$ and $\sigma_{x}, \sigma_{y}$ have the forms
\begin{equation}\label{222_d6_sigma_x_sigma_y}
	\begin{split}
		\sigma_{x} & = \begin{bmatrix}
			\Lambda    &      & \multicolumn{2}{c}{\multirow{2}*{{\Large\bfseries0}}}\\
			&  \Lambda  &  \\
			\multicolumn{2}{c}{\multirow{1}*{{\Large\bfseries0}}} &  & \Lambda\\
		\end{bmatrix}, \\
		\sigma_{y}(w_1, w_2,w_3) & =  
		\begin{bmatrix}
			B(w_1)   & & \multicolumn{2}{c}{\multirow{2}*{{\Large\bfseries0}}}\\
			&B(w_2)\\
			\multicolumn{2}{c}{\multirow{1}*{{\Large\bfseries0}}}     & & B(w_3)\\
		\end{bmatrix}.
	\end{split}
\end{equation}

\subsection*{Contextuality in 322-type Hamiltonians} \label{sec:322-type Hamiltonian}
The $\tilhv$ polytope for the 322-type Hamiltonians has been characterized in~\cite{Marginal1D}: it has 32372 facets which can be sorted into 2102 equivalence classes. The general form of the 322-type Hamiltonian is given by
\begin{equation}
	\begin{split}
		{\cal H}_{322} & = \sum_{i=1}^{\infty} J_x{\sigma_x^i} + J_y{\sigma_y^i} + J_{xx}^{AB}{\sigma_x^i \sigma_x^{i+1}} + J_{xy}^{AB}{\sigma_x^i \sigma_y^{i+1}} \\ 
		& 
		+  J_{yx}^{AB}{\sigma_y^i \sigma_x^{i+1}} + J_{yy}^{AB}{\sigma_y^i \sigma_y^{i+1}} + J_{xx}^{AC}{\sigma_x^i \sigma_x^{i+2}} \\
		&
		+ J_{xy}^{AC}{\sigma_x^i \sigma_y^{i+2}} + J_{yx}^{AC}{\sigma_y^i \sigma_x^{i+2}} +
		J_{yy}^{AC}{\sigma_y^i \sigma_y^{i+2}},
	\end{split}
	\label{Hamiltonian322}
\end{equation}
where $\{J_x, J_y, J_{xx}^{AB}, J_{xy}^{AB}, J_{yx}^{AB}, J_{yy}^{AB}, J_{xx}^{AC},  J_{xy}^{AC}, J_{yx}^{AC}, J_{yy}^{AC}\}$ are the couplings given by the facet inequalities and $\sigma_x, \sigma_y$ are local observables.

Using our uMPS based gradient descent algorithm, a total of 63 Hamiltonians exhibit contextuality. The explicit parameterization of observables $\sigma_{x}$ and $\sigma_{y}$ is explained at the end of this section. All the contextual Hamiltonians and ground state energy densities are listed in Table~\ref{all_322_couplings} and Table~\ref{all_322_parameters} respectively. Among them, we identify some quantum models whose ground state energy density matches the $\ltinpa{5}{1}$ lower bounds.  For all these contextuality witnesses, we have thus identified translation-invariant quantum models exhibiting the strongest quantum violation. All the matched models are summarized in Table~\ref{table:322 quantum values match NPA}. As the reader can appreciate, the first five inequalities seem to require local dimension $d=3$ to be saturated; inequality $6$, dimension $4$; and the last four inequalities, dimension $5$.

\begin{table}[htbp]
	\scriptsize
	\caption{\label{table:322 quantum values match NPA} The ground state energy density ${\cal{Q}}_d$ of ten 322-type TI quantum systems matches the $\ltinpa{5}{1}$ lower bound. The second column gives the number of each model in this table in Table~\ref{all_322_parameters}. }
	\renewcommand\arraystretch{1.2}
	\setlength{\tabcolsep}{0.3mm}{
\begin{ruledtabular}
	\begin{tabular}{cccccccc}
		No. & Table~\ref{all_322_parameters} & ${\cal{L}}$ & ${\cal{Q}}_2$ & ${\cal{Q}}_3$ & ${\cal{Q}}_4$ & ${\cal{Q}}_5$ & $\ltinpa{5}{1}$ 	\\ \hline
		1 & No.1 & -6 & -6.08108  & -6.32747  &  &  & -6.32747 \\
		2 & No.2 & -6 & -6.10943  & -6.33712  &  &  & -6.33712 \\
		3 & No.3 & -3 & -3.04150  & -3.20711  &  &  & -3.20711 \\
		4 & No.4 & -4 & -- & -4.14623  &  &  & -4.14623 \\
		5 & No.5 & -8 & -- & -8.12123  &  &  & -8.12123 \\
		6 & No.6 & -4 & -- & -4.02415  & -4.10310  &  & -4.10310 \\
		7 & No.7 & -5 & -- & -5.09951  & -5.09951  & -5.29852  & -5.29852 \\
		8 & No.8 & -4 & -- & -4.18655  & -4.18655  & -4.33137  & -4.33137 \\
		9 & No.9 & -4 & -- & -4.11581  & -4.11581  & -4.41421  & -4.41421 \\
		10 & No.10 & -5 & -- & -5.07058  & -5.07058  & -5.26969  & -5.26969 \\	
	\end{tabular}
\end{ruledtabular}}
\end{table}

\begin{figure}[!ht]
	\centering
		\begin{minipage}[b]{0.48\textwidth}
			\centering
			(a) No.3 in Table~\ref{table:322 quantum values match NPA}, ${\cal{L}}=-3$ \\
			\includegraphics[width=\textwidth]{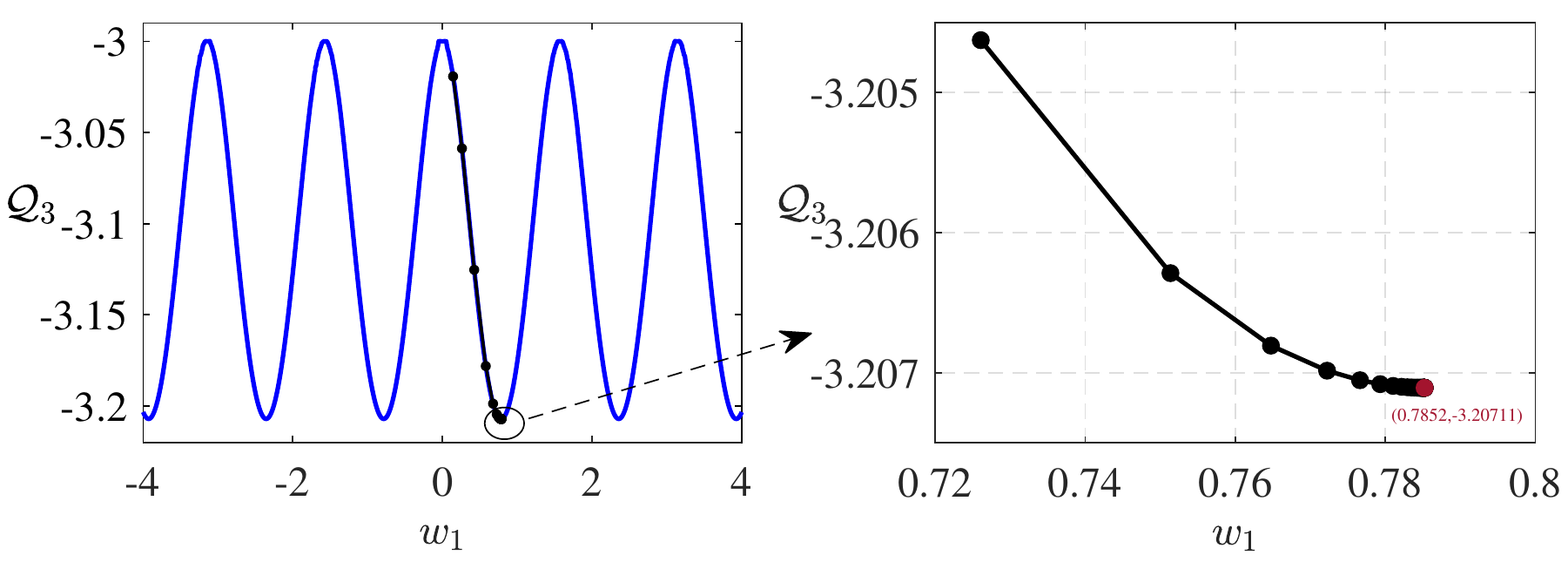}
			\label{orbit:d3_No3}
		\end{minipage}
	 \hfill
	\begin{minipage}[b]{0.48\textwidth}
		\centering
		(b) No.6 in Table~\ref{table:322 quantum values match NPA}, ${\cal{L}}=-4$ \\
		\includegraphics[width=\textwidth]{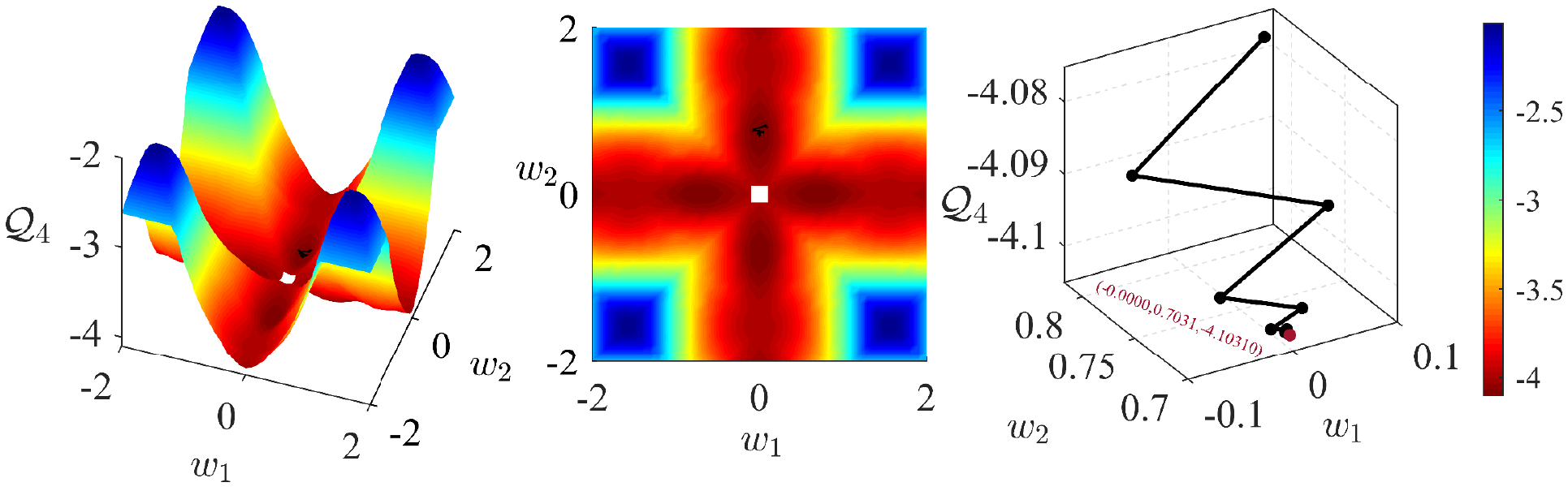}
		\label{orbit:d4_N06}
	\end{minipage}
	\caption{The trajectory of the ground state energy density (black dotted line) of two models in Table~\ref{table:322 quantum values match NPA}. (a) has two subplots, and the right one is an enlarged view for the trajectory in the left one when ground state energy density converges to the minimum. (b) has three subplots, the the middle one is the top view of the leftmost one, and the rightmost one only depicts the trajectory.}
	\label{fig:extreme points orbits}
\end{figure}

In Fig.~\ref{fig:extreme points orbits}, the reader can see the trajectories in parameter space followed by two quantum systems, of dimensions $d=3$ and $d=4$, undergoing our gradient descent method. This is possible because the number of free continuous parameters in one and another case are $1$ and $2$.

In Fig.~\ref{fig:extreme points orbits}(a), ${\cal{Q}}_3$ surfaces (blue lines) show that the Hamiltonian exhibits contextuality no matter which values the parameter takes. The trajectory of ground state energy density (black dotted line) in the left subplot decreases along the ${\cal{Q}}_3$ surface to the bottom. Besides, the right enlarged subplot of the trajectory demonstrates that ground state energy density eventually converges to the respective $\ltinpa{n}{s}$ lower bound. In Fig.~\ref{fig:extreme points orbits}(b), the leftmost subplot shows the trajectory of the ground state energy density on the 3D ${\cal{Q}}_4$ surface, the middle 2D-subplot is the top view of the leftmost one, and the rightmost one only depicts the trajectory of the ground state energy density. Iteratively, our methods guide the initial random ground state energy density converging to the lowest possible one.

\begin{figure}[!ht]
	\centering
	\begin{minipage}[b]{0.48\textwidth}
		\centering
		(a) No.1 in Table~\ref{table:322 quantum values match NPA}, ${\cal{L}}=-6$ \\
		\includegraphics[width=\textwidth]{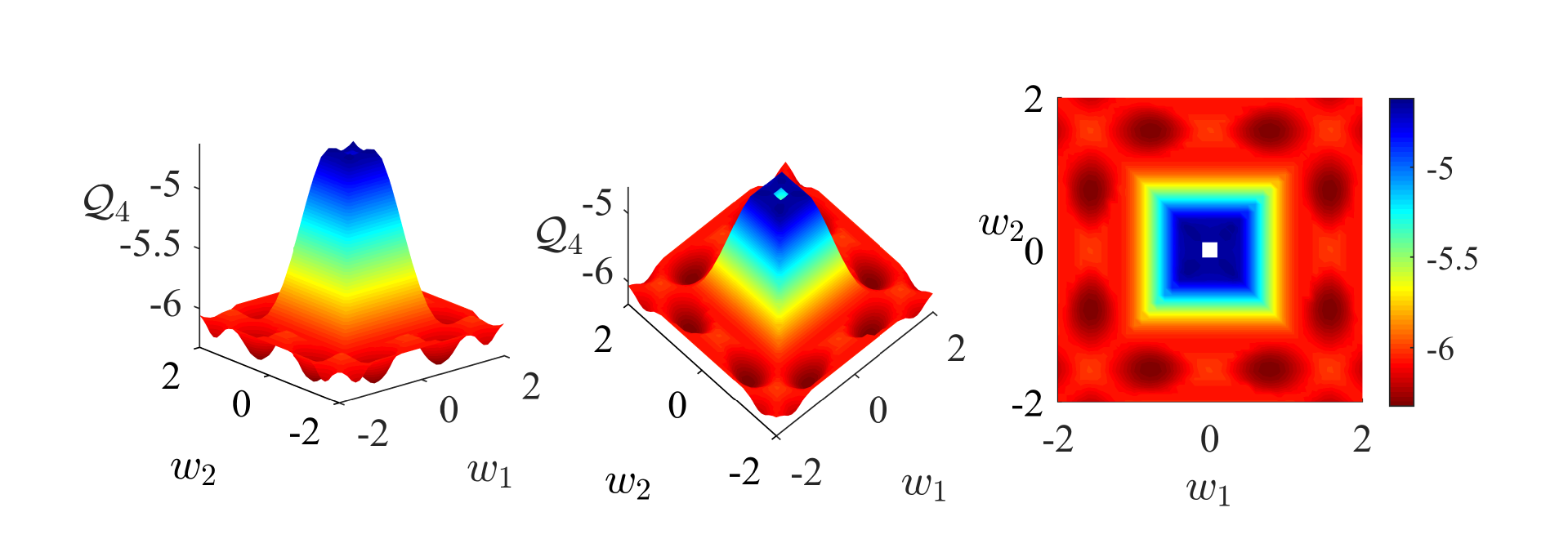}
		\label{surface:d4_No1}
	\end{minipage}
	\hfill
	\begin{minipage}[b]{0.48\textwidth}
		\centering
		(b) No.5 in Table~\ref{table:322 quantum values match NPA}, ${\cal{L}}=-8$ \\
		\includegraphics[width=\textwidth]{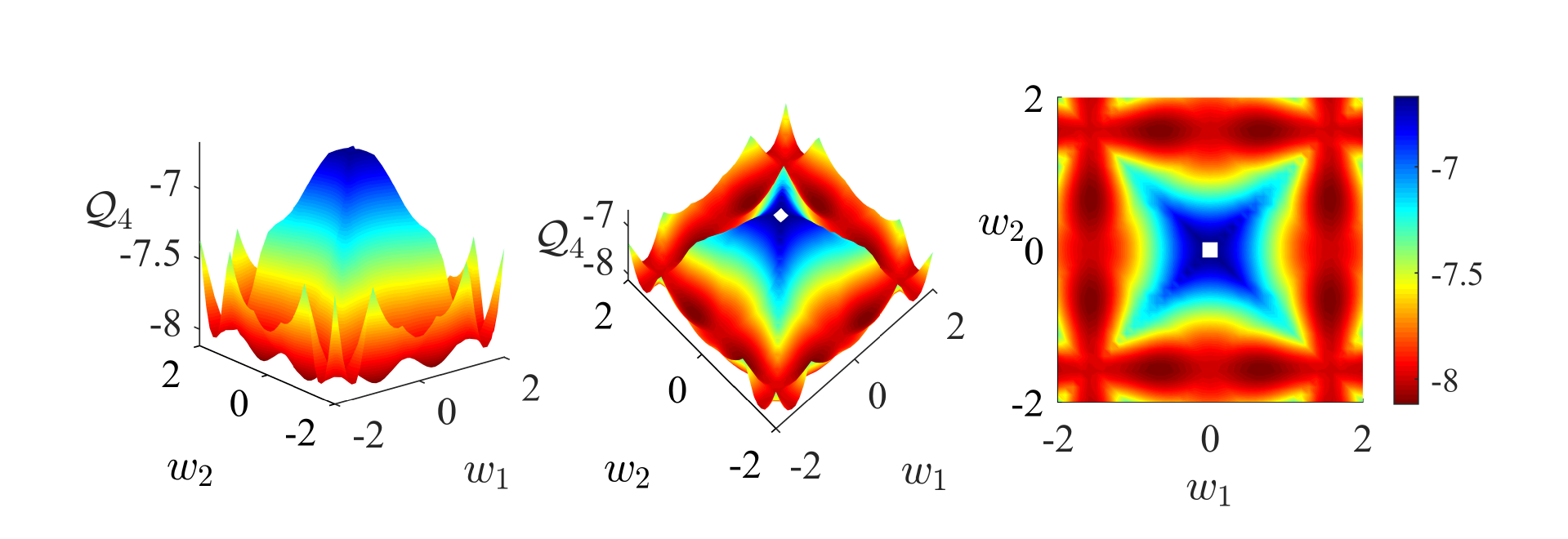}
		\label{surface:d4_No5}
	\end{minipage}
	\hfill
	\begin{minipage}[b]{0.48\textwidth}
		\centering
		(c) No.8 in Table~\ref{all_322_parameters}, ${\cal{L}}=-4$ \\
		\includegraphics[width=\textwidth]{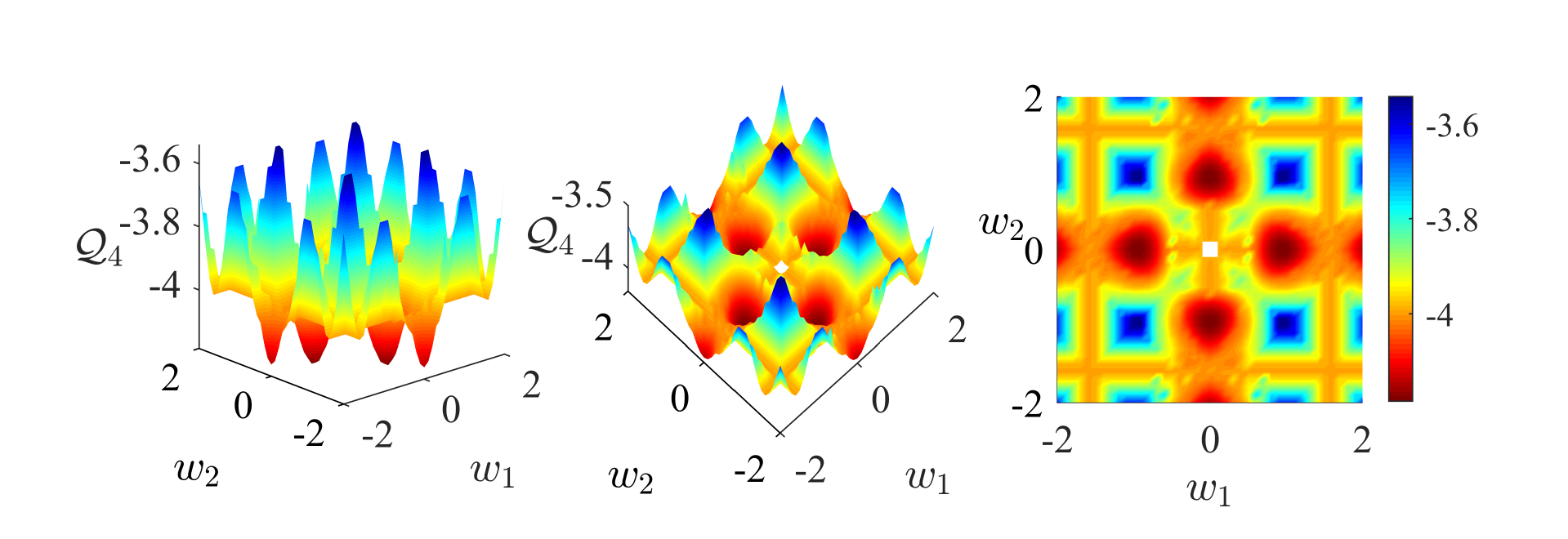}
		\label{surface:d4_No8}
	\end{minipage}
	\hfill
	\begin{minipage}[b]{0.48\textwidth}
		\centering
		(d) No.19 in Table~\ref{all_322_parameters}, ${\cal{L}}=-11$ \\
		\includegraphics[width=\textwidth]{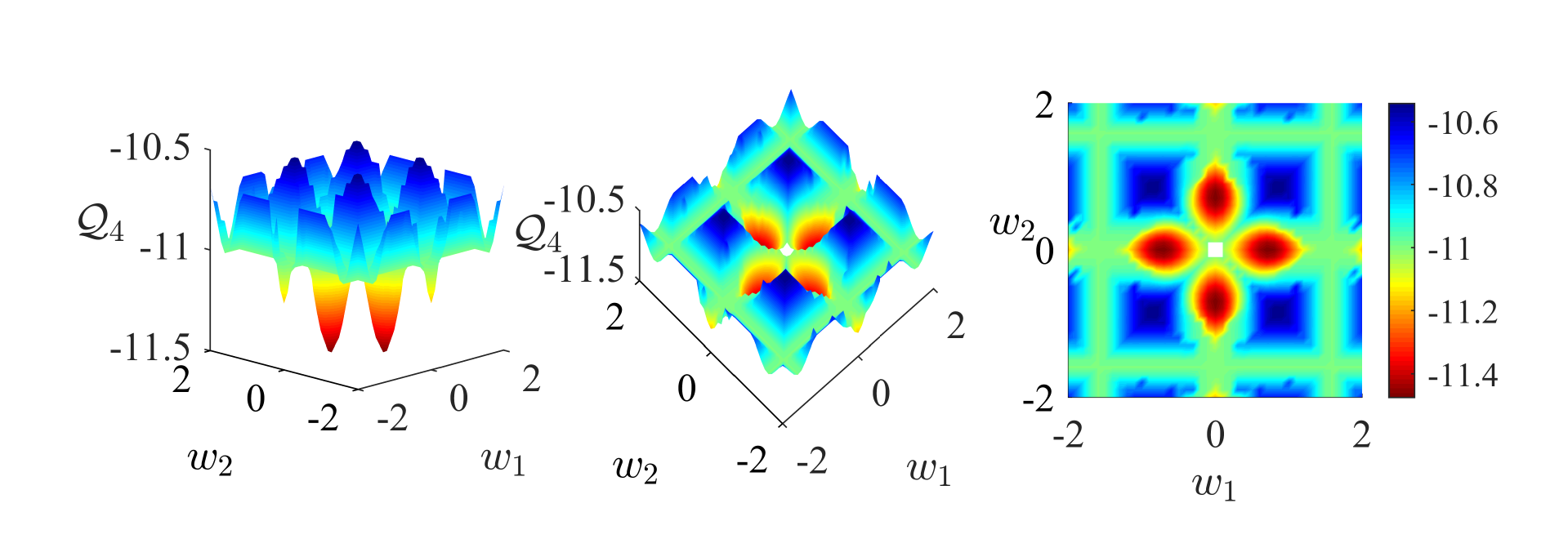}
		\label{surface:d4_No19}
	\end{minipage}
	\hfill
	\begin{minipage}[b]{0.48\textwidth}
		\centering
		(e) No.23 in Table~\ref{all_322_parameters}, ${\cal{L}}=-8$ \\
		\includegraphics[width=\textwidth]{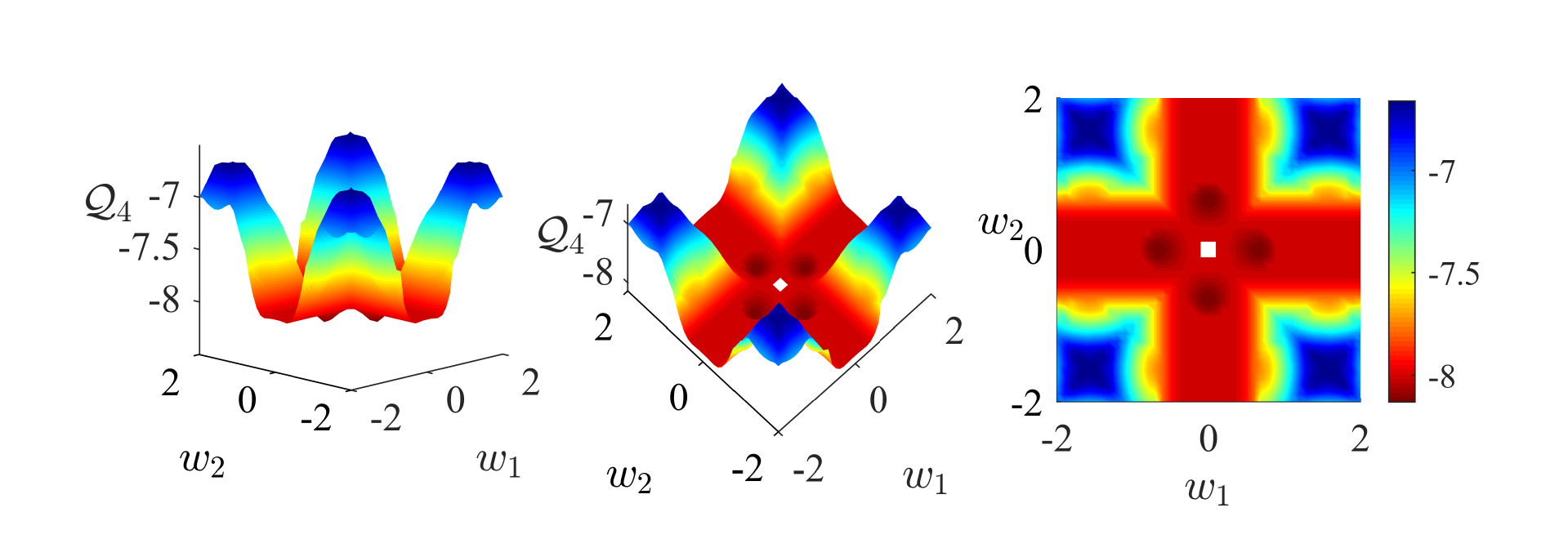}
		\label{surface:d4_No23}
	\end{minipage}
	\caption{${\cal{Q}}_4$ surface of five models in Table~\ref{table:322 quantum values match NPA} and Table~\ref{all_322_parameters}, where $w_1, w_2$ both take discrete values on $[-2, 2]$ at the interval $0.1$. Each model has three subplots. The leftmost is a 3D-surface, the middle figure is from an another perspective of the left side one to view the internal structure, and the third one is the top view of the leftmost image.} 
	\label{fig:322_d4_surface}
\end{figure}

\begin{figure}[!ht]
	\centering
	\begin{minipage}[b]{0.48\textwidth}
		\centering
		(a) No.7 in Table~\ref{table:322 quantum values match NPA}, ${\cal{L}}=-5$ \\
		\includegraphics[width=\textwidth]{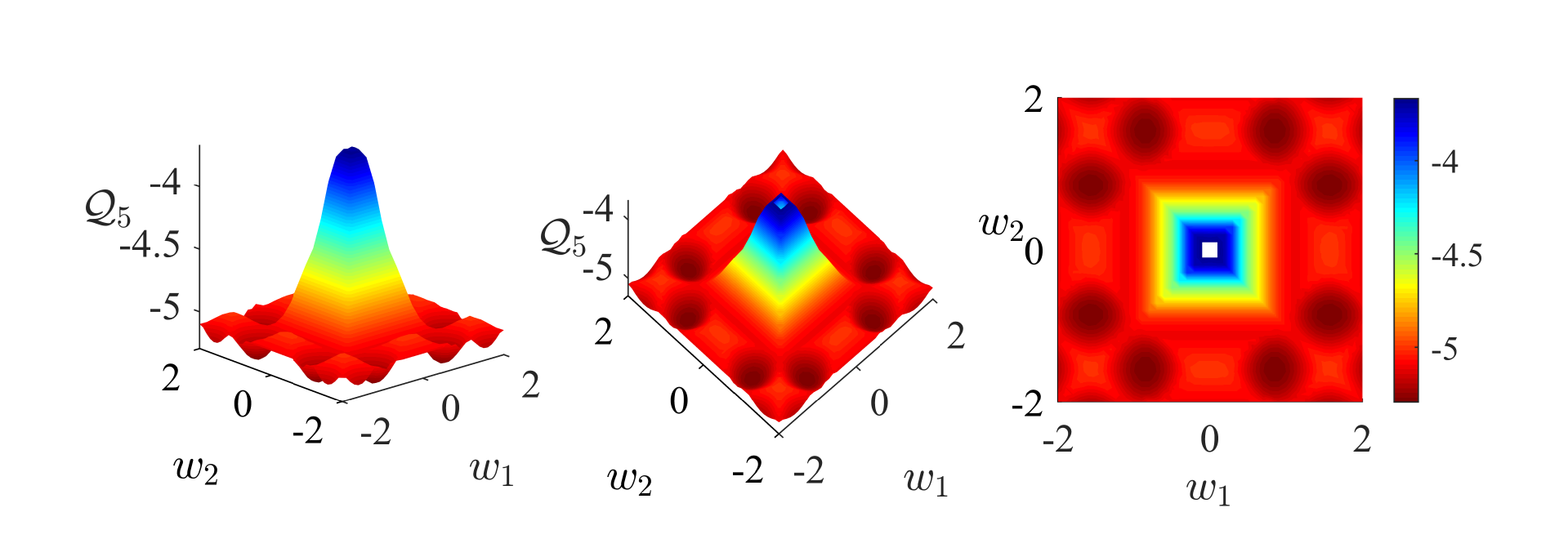}
	\end{minipage}
	\hfill
	\begin{minipage}[b]{0.48\textwidth}
		\centering
		(b) No.8 in Table~\ref{table:322 quantum values match NPA}, ${\cal{L}}=-4$ \\
		\includegraphics[width=\textwidth]{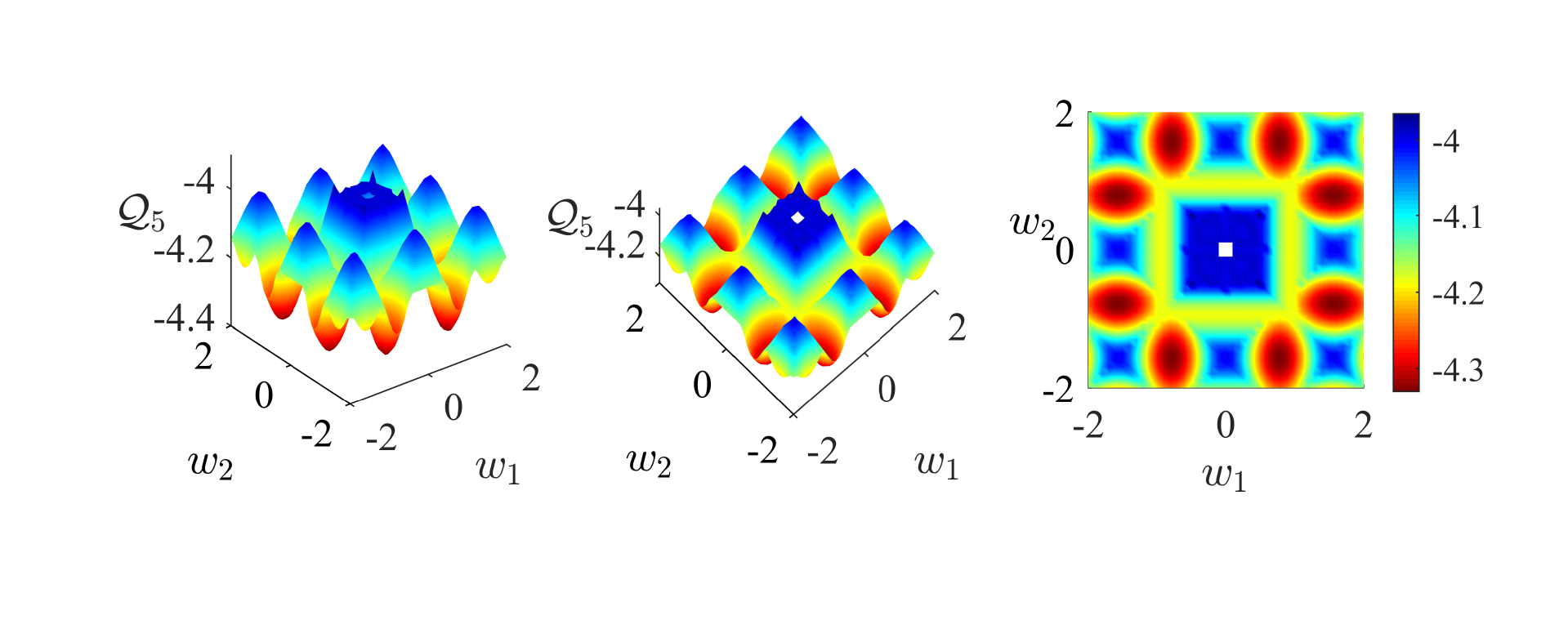}
	\end{minipage}
	\hfill
	\begin{minipage}[b]{0.48\textwidth}
		\centering
		(c) No.9 in Table~\ref{table:322 quantum values match NPA}, ${\cal{L}}=-4$ \\
		\includegraphics[width=\textwidth]{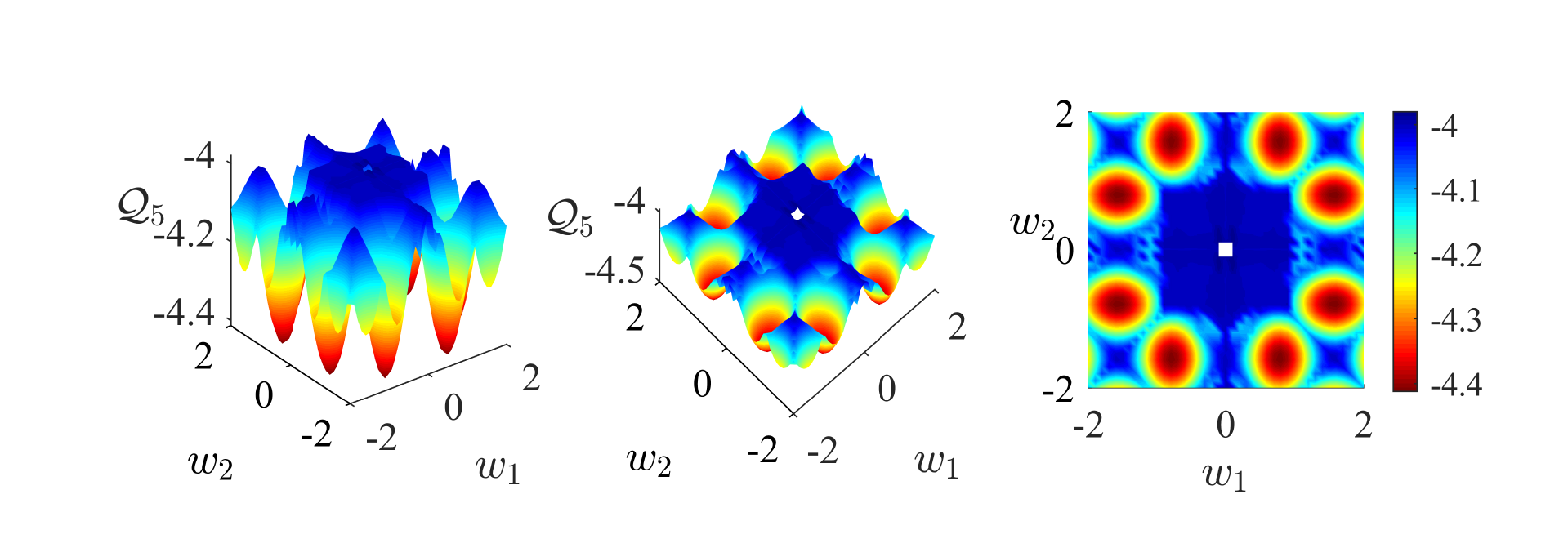}
	\end{minipage}
		\hfill
	\begin{minipage}[b]{0.48\textwidth}
		\centering
		(d) No.10 in Table~\ref{table:322 quantum values match NPA}, ${\cal{L}}=-5$ \\
		\includegraphics[width=\textwidth]{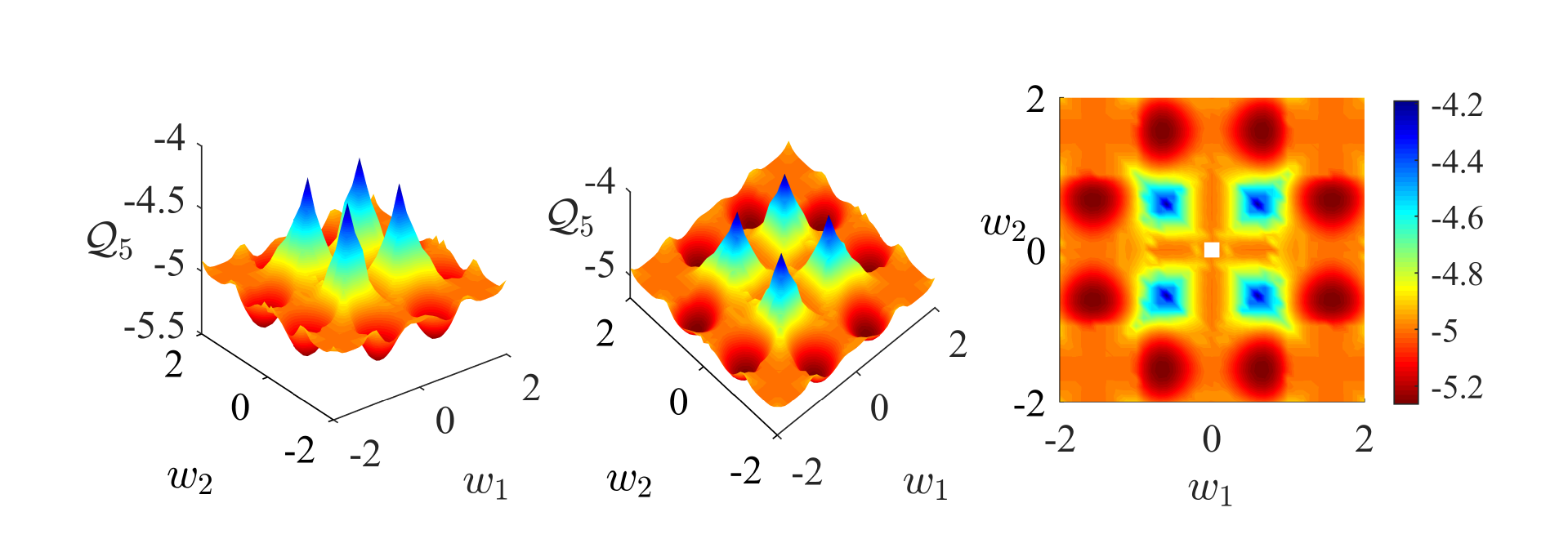}
	\end{minipage}
		\hfill
	\begin{minipage}[b]{0.48\textwidth}
		\centering
		(e) No.11 in Table~\ref{all_322_parameters}, ${\cal{L}}=-3$ \\
		\includegraphics[width=\textwidth]{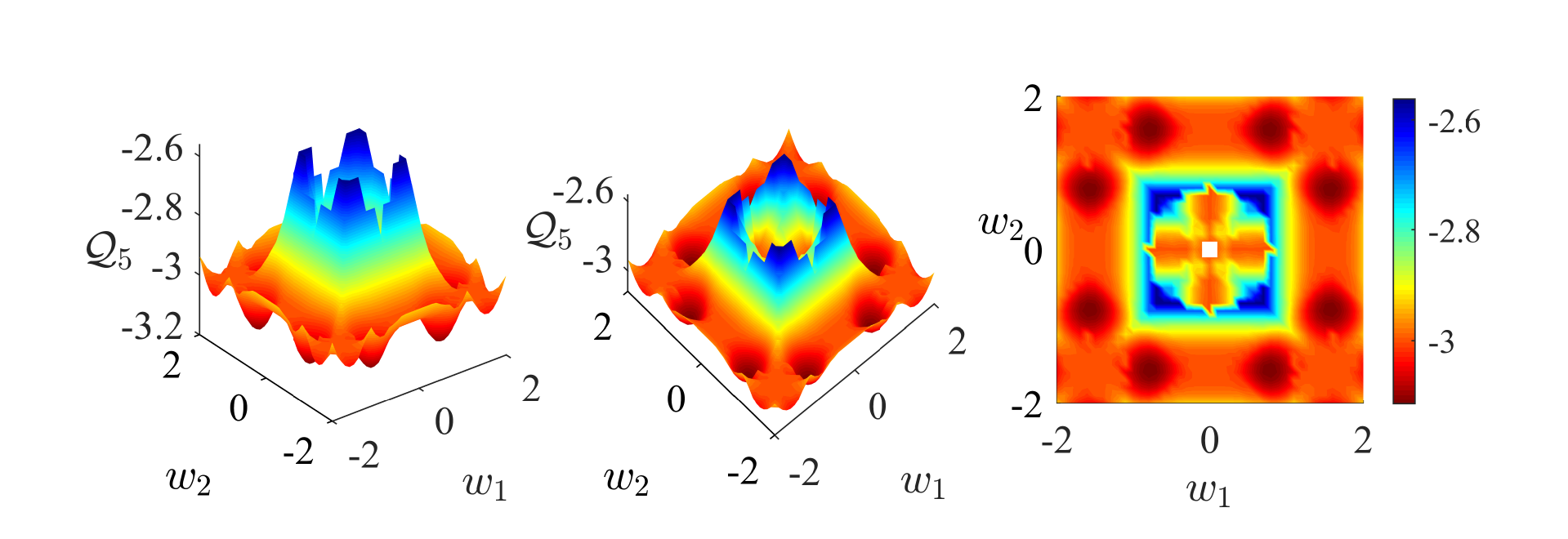}
	\end{minipage}
	\caption{${\cal{Q}}_5$ surface of five models in Table~\ref{table:322 quantum values match NPA} and Table~\ref{all_322_parameters}, where $w_1, w_2$ both take discrete values on $[-2, 2]$ at the interval $0.1$. Each model has three subplots. The leftmost is a 3D-surface, the middle figure is from an another perspective of the left side one to view the internal structure, and the third one is the top view of the leftmost image.} 
	\label{fig:322_d5_surface}
\end{figure}

We plot the ground state energy density as functions of the parameters defining the local observables for some of the Hamiltonians in Table~\ref{table:322 quantum values match NPA} and Table~\ref{all_322_parameters} to gauge the robustness of the contextuality violations. Five models for $d=4$ and another five models for $d=5$ are shown in Fig.~\ref{fig:322_d4_surface} and Fig.~\ref{fig:322_d5_surface} respectively. Note that the first two models in Fig.~\ref{fig:322_d4_surface}(a)-(b) and the first four models in Fig.~\ref{fig:322_d5_surface}(a)-(d) exhibit the strongest contextuality.

It can be seen that some Hamiltonians are much more susceptible to small changes in parameters that define the local observables than others. For the Hamiltonians in Fig.~\ref{fig:322_d4_surface}(b), Fig.~\ref{fig:322_d5_surface}(b) and Fig.~\ref{fig:322_d5_surface}(c), keeping the ground state energy density above the classical bound is unstable, small perturbations in the parameters will make them violate it. In contrast, the remaining Hamiltonians need carefully engineered parameters to violate the classical bound. Especially for the Hamiltonian in Fig.~\ref{fig:322_d5_surface}(e), square-like parameter regions exist in which the corresponding ground state energy density could not violate the classical bound no matter how many times the perturbations are given. These plots help us find suitable Hamiltonians for simulation in trapped-ion or optical lattice systems, where witnessing contextuality (or the strongest contextuality) simply involves cooling the corresponding Hamiltonian to the ground state.

We next describe the parameterization of the 322-type Hamiltonians achieving the minimum quantum values in Table \ref{table:322 quantum values match NPA}. For $d=2$ and $d=4$, $\sigma_{x}$, $\sigma_{y}$ have the exact same parameterized forms as the observables in the 222-type Hamiltonians in Eq.~\eqref{222_d2_sigma_x_sigma_y} and Eq.~\eqref{222_d4_sigma_x_sigma_y} respectively. 

For $d=3$ and $d=5$, depending on the number of $1$'s and $-1$'s of each matrix $\Lambda_a (a=x,y)$ in \eqref{unitary diagonalized}, two different classes of pairs of local observables $\sigma_{x}$ and $\sigma_y$ are considered. Here, we continue using the notations $\Lambda$ and $B(w)$ introduced in \eqref{2-dimensional block matrices}.

For local dimension $d=3$, the first class of pairs of local observables is of the form $[n_x,n_y]=[1,2]$. More specifically:
\begin{equation}\label{322_d3_sigma_x_sigma_y_type1}
	\sigma_{x} = 
	\begin{bmatrix}
		\Lambda  & \bigzero \\ 
		\bigzero & 1 \\	
	\end{bmatrix}, \quad
	\sigma_{y}(w_1) = 
	\begin{bmatrix}
		B(w_1)  & \bigzero \\ 
		\bigzero & -1 \\			
	\end{bmatrix}.
\end{equation} 
The second class is of the form $[n_x,n_y]=[1,1]$, with
\begin{equation}\label{322_d3_sigma_x_sigma_y_type2}
	\sigma_{x} = 
	\begin{bmatrix}
		1  & \bigzero \\ 
		\bigzero & \Lambda \\	
	\end{bmatrix}, \quad
	\sigma_{y}(w_1) = 
	\begin{bmatrix}
		1  & \bigzero \\ 
		\bigzero & B(w_1) \\			
	\end{bmatrix}.
\end{equation} 

For local dimension $d=5$, the first class of observable pairs is of the form $[n_x,n_y]=[2,2]$, with
\begin{equation}\label{322_d5_sigma_x_sigma_y_type1}
	\sigma_{x} = \begin{bmatrix}
		\Lambda    &      & \multicolumn{2}{c}{\multirow{2}*{{\Large\bfseries0}}}\\
		&  \Lambda  &  \\
		\multicolumn{2}{c}{\multirow{1}*{{\Large\bfseries0}}} &  & 1\\
	\end{bmatrix}, \quad
	\sigma_{y}(w_1, w_2) = \begin{bmatrix}
		B(w_1)    &      & \multicolumn{1}{c}{\multirow{2}*{{\Large\bfseries0}}}\\
		&  B(w_2)  &  \\
		\multicolumn{2}{c}{\multirow{1}*{{\Large\bfseries0}}} &  & 1\\
	\end{bmatrix}. 
\end{equation}
The second class of observable pairs satisfies $[n_x,n_y]=[3,3]$, and $\sigma_{x}$ and $\sigma_{y}$ are given by
\begin{equation}\label{322_d5_sigma_x_sigma_y_type2}
	\sigma_{x} = \begin{bmatrix}
		\Lambda    &      & \multicolumn{2}{c}{\multirow{2}*{{\Large\bfseries0}}}\\
		&  \Lambda  &  \\
		\multicolumn{2}{c}{\multirow{1}*{{\Large\bfseries0}}} &  & -1\\
	\end{bmatrix}, \quad
	\sigma_{y}(w_1, w_2) = \begin{bmatrix}
		B(w_1)    &      & \multicolumn{2}{c}{\multirow{2}*{{\Large\bfseries0}}}\\
		&  B(w_2)  &  \\
		\multicolumn{2}{c}{\multirow{1}*{{\Large\bfseries0}}} &  & -1\\
	\end{bmatrix}. 
\end{equation}

\subsection*{Contextuality in 232-type Hamiltonians} \label{sec:232-type Hamiltonian}
The LTI-LHV polytope for 3 dichotomic observables has 92694 facets, which can be classified into 652 equivalent classes~\cite{DenisThesis}. The general form of this type of Hamiltonian is given by
\begin{align}
{\cal H}_{232} & = \sum_{i=1}^{\infty} J_x{\sigma_x^i} + J_y{\sigma_y^i} + J_z{\sigma_z^i} + J_{xx}{\sigma_{x}^{i}\sigma_{x}^{i+1}} + J_{xy}{\sigma_{x}^{i}\sigma_{y}^{i+1}} \nonumber\\
& + J_{xz}{\sigma_{x}^{i}\sigma_{z}^{i+1}} + J_{yx}{\sigma_{y}^{i}\sigma_{x}^{i+1}} +J_{yy}{\sigma_{y}^{i}\sigma_{y}^{i+1}} + J_{yz}{\sigma_{y}^{i}\sigma_{z}^{i+1}} \nonumber\\ 
&+ J_{zx}{\sigma_{z}^{i}\sigma_{x}^{i+1}} + J_{zy}{\sigma_{z}^{i}\sigma_{y}^{i+1}} +
J_{zz}{\sigma_{z}^{i}\sigma_{z}^{i+1}}.
\label{Hamiltonian232}
\end{align}

We consider ${\cal{H}}_{232}$ when $d=2,3,4$ and perform the optimizations on one representative facet from each of the 652 classes. For $d=3,4$, we only consider real parameters. For $d=2$, we allow the most general measurements in quantum theory: complex POVMs. In the Method section, we present a projected gradient descent algorithm to optimize over the set of complex POVMs. All 652 Hamiltonians can only reach the classical bound up to numerical precision of $10^{-5}$ when $d=2$, while for $d\geq 3$ there are many Hamiltonians which can violate the classical bound. The ground state energy densities of some contextual Hamiltonians are shown in Table~\ref{table:232violations}, see Table~\ref{232couplings} for the couplings defining the contextuality witnesses. In addition, the parameters specifying the optimal local observables are listed in Table~\ref{232_d2_d3} for $d=3$ and in Table~\ref{232_d4} for $d=4$. 
\begin{table}[htbp!]
	\caption{Ground state energy density for 5 232-type TI Hamiltonians under the dimension of local observables $d=2,3,4$.}
	\label{table:232violations}
	\scriptsize
	\begin{ruledtabular}
		\renewcommand\arraystretch{1.2}
		\begin{tabular}{cccccc}
		No. & ${\cal{L}}$ & ${\cal{Q}}_2$ &  ${\cal{Q}}_3$ & ${\cal{Q}}_4$ & $\ltinpa{4}{1}$
			\\ \hline
			1 & -9 & -9.00000  & -9.01875  & -9.01875  & -9.27833   \\
			2 & -4 & -4.00000  & -4.03928  & -4.03928  & -4.20626   \\
			3 & -5 & -5.00000  & -5.04162  & -5.04162  & -5.14754   \\
			4 & -4 & -4.00000  & -4.01336  & -4.11562  & -4.23786  \\
			5 & -2 & -2.00000  & -2.08094  & -2.08749  & -2.28767  \\					
		\end{tabular}
	\end{ruledtabular}
\end{table}

The local observables $\sigma_{x}, \sigma_{y}$ and $\sigma_{z}$ are parametrized using the method presented in Section~\ref{sec:methods:upperbound}: for given $\Lambda_{a}, S_a$ the local observable $\sigma_{a}$ can be written as 
\begin{equation}
	\sigma_{a} = e^{S_a}\Lambda_{a}(e^{S_a})^\dagger.
\label{232_parameterized_observables}
\end{equation}
While this step is straightforward, different combinations of $\pm1$ in $\Lambda_{a}$ may lead to different ground state energy density.

Since the number of $1$ and $-1$ on the diagonal of $\Lambda_{a}$ in three local observables is not necessarily the same, there is more than one combination of three parameterized local observables. We consider every possible combination of $1$ and $-1$ in $\Lambda_a$ for each $a\in\{x,y,z\}$. We only show combinations of parameterized local observables used in Table~\ref{table:232violations} below. Here, we denote the $2\times2$ identity matrix by $I$, and continue using the notation $\Lambda$ introduced in \eqref{2-dimensional block matrices}.

When $d=2$, the classical bound can be achieved via three local observables determined by two parameters, where $[n_x, n_y, n_z] = [2,1,1]$. The first local observable $\sigma_{x}$ is minus the identity matrix:
\begin{equation}\label{232_d2_sigma_x}	
	\sigma_{x} = 
	\begin{bmatrix}
		-1  &  0 \\ 
		0  & -1 \\	
	\end{bmatrix}.
\end{equation} 

\noindent For the second and third local observables $\sigma_{a}(a=y,z)$, $\Lambda_{a}$ has entries one $1$ and one $-1$ on the main diagonal, and $S_a$ is determined by one parameter $\{w\}$. Hence, $\sigma_{y}$ and $\sigma_{z}$ are specified by
\begin{equation}\label{232_d2_sigma_x_sigma_y_sigma_z}
	\begin{split}	
	    & \Lambda_y = 
	    \begin{bmatrix}
	    	1  & 0 \\ 
	    	0  & -1\\			
	    \end{bmatrix}, \quad    
		S_y(w_1) = 
		\begin{bmatrix}
			0  & w_1 \\ 
			- w_1 & 0\\			
		\end{bmatrix}; \\
		& \Lambda_z = 
		\begin{bmatrix}
			1  & 0 \\ 
			0  & -1\\			
		\end{bmatrix}, \quad    
		S_z(w_2) = 
		\begin{bmatrix}
			0  & w_2 \\ 
			- w_2 & 0\\			
		\end{bmatrix}.	
	\end{split}
\end{equation} 

For $d=3$, two different combinations of three parameterized local observables are used, where the difference arises from $\Lambda_{a}$, but $S_a$ of each local observable share the same parameterized form as
\begin{equation}
	\begin{split}
		& S_x(w_1,w_2,w_3) = 
		\begin{bmatrix}
			0  & w_1  & w_2 \\
			-w_1 &  0   & w_3 \\
			-w_2 & -w_3 & 0 \\		
		\end{bmatrix}, \\
		& S_y(w_4,w_5,w_6) = 
		\begin{bmatrix}
			0  & w_4  & w_5 \\
			-w_4 &  0   & w_6 \\
			-w_5 & -w_6 & 0 \\		
		\end{bmatrix}, \\
		& S_z(w_7,w_8,w_9) = 
		\begin{bmatrix}
			0  & w_7  & w_8 \\
			-w_7 &  0  & w_9 \\
			-w_8 & -w_9 & 0 \\		
		\end{bmatrix}.
	\end{split}
\label{232_d3_S_x_S_y_S_z}
\end{equation} 

In the first combination, $[n_x,n_y,n_z]=[2,1,1]$, where $\Lambda_x$ has one $1$ and two $-1$ on the main diagonal, and $\Lambda_y$ and $\Lambda_z$ both have two $1$ and one $-1$ being main diagonal entries. Then, $\Lambda_{x}$, $\Lambda_{y}$, and $\Lambda_{z}$ are given by
\begin{equation}
	\Lambda_{x} = 
		\begin{bmatrix}
			\Lambda  & \bigzero \\ 
			\bigzero & -1 \\	
		\end{bmatrix}, \quad
	\Lambda_{y} = 
		\begin{bmatrix}
			1  & \bigzero \\ 
			\bigzero & \Lambda \\			
		\end{bmatrix}, \quad
	\Lambda_{z} = 
		\begin{bmatrix}
			1  & \bigzero \\ 
			\bigzero & \Lambda \\			
		\end{bmatrix}.
\label{232_d3_Lambda_x_Lambda_y_Lambda_z_type1}
\end{equation} 

In the second combination, $[n_x,n_y,n_z]=[2,1,2]$, where $\Lambda_x$ and $\Lambda_z$ both have one $1$ and two $-1$ on the main diagonal, and $\Lambda_y$ has two $1$ and one $-1$ being main diagonal entries. Then, $\Lambda_{x}$, $\Lambda_{y}$, and $\Lambda_{z}$ are given by
\begin{equation}
	\Lambda_{x} = 
	\begin{bmatrix}
	\Lambda  & \bigzero \\ 
		\bigzero & -1 \\	
	\end{bmatrix}, \quad
	\Lambda_{y} = 
	\begin{bmatrix}
		1  & \bigzero \\ 
		\bigzero & \Lambda \\			
	\end{bmatrix} \quad
	\Lambda_{z} = 
	\begin{bmatrix}
		\Lambda  & \bigzero \\ 
		\bigzero & -1 \\			
	\end{bmatrix}.
	\label{232_d3_Lambda_x_Lambda_y_Lambda_z_type2}
\end{equation} 

For $d=4$, four different classes of triples of local observables are used. These classes differ from each other on the structure of the matrices $\Lambda_a$ in (\ref{232_parameterized_observables}). The matrices $S_a$ have, nonetheless, the same form in the three classes, namely:
\begin{equation}
	\begin{split}
		S_x(w_1,w_2,w_3, w_4,w_5,w_6) & = 
		\begin{bmatrix}
			  0  & w_1  & w_2  & w_3\\
			-w_1 &  0   & w_4  & w_5 \\
			-w_2 & -w_4 & 0    & w_6 \\	
			-w_3 & -w_5 & -w_6 & 0 \\
		\end{bmatrix}, \\
		S_y(w_7, w_8, w_9, w_{10}, w_{11}, w_{12}) & = 
		\begin{bmatrix}
			0  & w_7  & w_8  & w_9\\
			-w_7 &  0   & w_{10}  & w_{11} \\
			-w_8 & -w_{10} & 0    & w_{12} \\	
			-w_9 & -w_{11} & -w_{12} & 0 \\
		\end{bmatrix}, \\
		S_z(w_{13}, w_{14}, w_{15}, w_{16}, w_{17}, w_{18}) & = \begin{bmatrix}
			0  & w_{13}  & w_{14}  & w_{15}\\
			-w_{13} &  0   & w_{16}  & w_{17} \\
			-w_{14} & -w_{16} & 0    & w_{18} \\	
			-w_{15} & -w_{17} & -w_{18} & 0 \\
		\end{bmatrix}.
	\end{split}
	\label{232_d4_S_x_S_y_S_z}
\end{equation} 

The first class is of the form $[n_x,n_y,n_z]=[3,1,1]$, where $\Lambda_x$ has one $1$ and three $-1$ on the main diagonal, and $\Lambda_y$ and $\Lambda_z$ both have three $1$ and one $-1$ being main diagonal entries. Then, $\Lambda_{x}$, $\Lambda_{y}$, and $\Lambda_{z}$ are given by
\begin{equation}
	\Lambda_{x} = 
	\begin{bmatrix}
		\Lambda  & \bigzero \\ 
		\bigzero & -I \\	
	\end{bmatrix}, \quad
	\Lambda_{y} = 
	\begin{bmatrix}
		I  & \bigzero \\ 
		\bigzero & \Lambda \\			
	\end{bmatrix}, \quad
	\Lambda_{z} = 
	\begin{bmatrix}
		I  & \bigzero \\ 
		\bigzero & \Lambda \\			
	\end{bmatrix}.
	\label{232_d4_Lambda_x_Lambda_y_Lambda_z_type1}
\end{equation} 

The second class is of the form $[n_x,n_y,n_z]=[3,2,3]$, where $\Lambda_x$ and $\Lambda_z$ both have one $1$ and three $-1$ on the main diagonal, and $\Lambda_y$ has two $1$ and two $-1$ being main diagonal entries. Then, $\Lambda_{x}$, $\Lambda_{y}$, and $\Lambda_{z}$ are given by
\begin{equation}
	\Lambda_{x} = 
	\begin{bmatrix}
		\Lambda  & \bigzero \\ 
		\bigzero & -I \\	
	\end{bmatrix}, \quad
	\Lambda_{y} = 
	\begin{bmatrix}
		\Lambda  & \bigzero \\ 
		\bigzero & \Lambda \\			
	\end{bmatrix} \quad
	\Lambda_{z} = 
	\begin{bmatrix}
		\Lambda  & \bigzero \\ 
		\bigzero & -I \\			
	\end{bmatrix}.
	\label{232_d4_Lambda_x_Lambda_y_Lambda_z_type2}
\end{equation} 

The third class is of the form $[n_x,n_y,n_z]=[3,2,1]$, where $\Lambda_x$ has one $1$ and three $-1$ on the main diagonal, $\Lambda_y$ has two $1$ and two $-1$ being main diagonal entries, and $\Lambda_z$ takes three $1$ and one $-1$ on the main diagonal. Then, $\Lambda_{x}$, $\Lambda_{y}$, and $\Lambda_{z}$ are given by
\begin{equation}
	\Lambda_{x} = 
	\begin{bmatrix}
		\Lambda  & \bigzero \\ 
		\bigzero & -I \\	
	\end{bmatrix}, \quad
	\Lambda_{y} = 
	\begin{bmatrix}
		\Lambda  & \bigzero \\ 
		\bigzero & \Lambda \\			
	\end{bmatrix} \quad
	\Lambda_{z} = 
	\begin{bmatrix}
		I  & \bigzero \\ 
		\bigzero & \Lambda \\			
	\end{bmatrix}.
	\label{232_d4_Lambda_x_Lambda_y_Lambda_z_type3}
\end{equation} 

The fourth class is of the form $[n_x,n_y,n_z]=[2,1,2]$, where $\Lambda_x$ and $\Lambda_z$ has two $1$ and two $-1$ on the main diagonal, $\Lambda_y$ has three $1$ and one $-1$ being main diagonal entries. Then, $\Lambda_{x}$, $\Lambda_{y}$, and $\Lambda_{z}$ are given by
\begin{equation}
	\Lambda_{x} = 
	\begin{bmatrix}
		\Lambda  & \bigzero \\ 
		\bigzero & \Lambda \\	
	\end{bmatrix}, \quad
	\Lambda_{y} = 
	\begin{bmatrix}
		I  & \bigzero \\ 
		\bigzero & \Lambda \\			
	\end{bmatrix} \quad
	\Lambda_{z} = 
	\begin{bmatrix}
		\Lambda  & \bigzero \\ 
		\bigzero & \Lambda \\			
	\end{bmatrix}.
	\label{232_d4_Lambda_x_Lambda_y_Lambda_z_type4}
\end{equation} 

\section*{Discussion}\label{sec:conclusions}
In this paper, we investigate the contextuality of several types of infinite one-dimensional translation-invariant local quantum Hamiltonians. We found that it is very likely that all quantum Hamiltonians with nearest-neighbor only interactions and two dichotomic observables per site admit local hidden variable models. Violation of contextuality witnesses are only possible when we either increase the interaction distance to include next-to-nearest neighbor terms or have three dichotomic observables per site. In the former case, we identified several Hamiltonians with the lowest possible ground state energy density in quantum theory. In the latter case, we give strong evidence that contextuality is only present if the dimension of local observables is greater than 2, which excludes the usual Heisenberg-type models where local observables are Pauli matrices. 

States and measurements which exhibit the strongest violations of Bell inequalities are essential ingredients in device-independent certifications and self-testing~\cite{MayersYao2004SelfTesting,Supic2020SelfTestingReview}. So far the possibility of self-testing in quantum many-body systems has not been thoroughly established, due to a lack of tools to certify the strongest violation of Bell inequalities or contextuality witnesses, without having to solve the corresponding quantum model analytically. Our results pave the way for self-testing quantum many-body systems in the thermodynamic limit.

The ground states of our models are computed using uMPS, and they are global approximations of the true ground state of the corresponding quantum models. However, in applications such as quantum simulation, we will only have access to local approximations of the ground state. Moreover, the qualities we are interested in, such as the ground state energy density and the expectation values of local observables all depended on the accuracy of the local description. Finding locally accurate approximations of properties of one-dimensional local quantum Hamiltonians has yielded many interesting results~\cite{Landau2015RRG,Kuzmin2021FiniteSimulation,HuangConstantBondDimension,HuangEnergyDensity,Dalzell2019locallyaccuratemps}. However, most of these results assume the models to have nearest-neighbor interactions. As we can see from our results, the models with next-to-nearest neighbor interactions are surprisingly the most interesting in terms of contextuality.

In two dimensions, very little is known about the contextuality of translation-invariant local Hamiltonians. We know that when the number of inputs and outputs is unrestricted, the set of local hidden variable models becomes non-semi-algebraic and eventually characterization of the set is impossible~\cite{ti2d}. Properties of 2D classical and quantum models differ so markedly from their 1D counterparts that most intuitions and tools we gained in 1D break down. However, in 2D a powerful mathematical tool, tiling, has been repeatedly employed to solve question about computability and complexity of classical and quantum models~\cite{Cubitt2015SpectralGap,GottesmanIrani2013Tiling,ti2d,Huang2DQMA}. The number of tiles in an aperiodic tiling would correspond to the number of states in a local hidden variable model, so it would be interesting to explore the connection between tiling and contextuality.

\section*{Methods}

We describe an extension of the algorithm used to minimize the ground state energy density to include the most general quantum measurements: positive operator-valued measure (POVM) measurements. The extended algorithm is used to minimize 232-type Hamiltonians when the dimension of local observables is 2. These local observables $\{\sigma_a:\mathbb{C}^2 \to \mathbb{C}^2, a \in X\}$ are constructed from POVM elements $M_{a0},\;M_{a1}$. In a gradient descent algorithm, at iteration $k$ the current gradient is subtracted from the parameters, which may take the local observables out of the space of POVMs. To correct this issue we project the local observables after the gradient has been subtracted onto the closest POVM found via semidefnite programming:
\begin{equation}
	\begin{split}
		min \quad & \Vert \tilde{\sigma}_a - \sigma_a \Vert_2 \\ 
		s.t. \quad & \sigma_a = M_{a0} - M_{a1} \\
		 		   & M_{a0} + M_{a1} = I \\
			       & M_{a0} \succeq 0,\; M_{a1} \succeq 0.
	\end{split}
\end{equation}
Here, $\tilde{\sigma}_a=\tilde{\sigma}_a(k+1)$ and $\sigma_a =\sigma_a(k+1)$. The parameters $W(k)$ are complex decision variables for the SDP, whose value at each iteration will be given by the solver.

Even though the extended algorithm based on projected gradient descent works in principle, we have encountered a number of numerical issues which require additional tweaks. The main issue affecting convergence is that it takes many iterations to traverse a nearly flat region in the parameter space. It is one of the most common problems affecting the performance of gradient descent algorithms, and it is very common to encounter such regions in our Hamiltonians. We use a well-known remedy, using momentum to speed up the traversal of nearly flat regions. At each iteration $k$, the parameters  $W(k)$ defining the local observables $\{\sigma_a(k)|a\in X\}$ are updated by
\begin{align}
	V(k+1)&=\eta \cdot V(k)- {\gamma}(k) \cdot \nabla_W e(W;k),\\
	\tilde{W}(k+1) &= W(k) + V(k+1),
	\label{POVM:update W}
\end{align}
where $V(k)$ is the momentum and $\eta$ is the decay factor.

Beginning with random initial $W(0)$ and $V(0)=0$, iterating the steps above, we obtain the a sequence of parameter values $(W(0), W(1), \dots)$ defining a sequence of local observables $(\sigma_a(0), \sigma_a(1), \dots)$, each of which is constructed from POVM elements. If the convergence criterion $|e(k+1)-e(k)|\leq \epsilon^* $ is met at a iteration $k$, then the algorithm stops and returns the optimal parameters $W^* \equiv W(k+1)$. For 100 out of the 652 Hamiltonianswe tested, even though the random initial parameters are allowed to be complex, the converged values are all real. Having nonzero imaginary parts in the local observables meant the ground state energy of the Hamiltonian will stall at a value higher than the classical bound, often resulting in 10000 iterations only decreasing the ground state energy marginally. When this happens, a new set of random initial complex parameters are generated and the algorithm restarts. When the algorithm converges, the imaginary parts of all the parameters are smaller than $10^{-10}$.

\vspace{10pt}
\textit{Acknowledgments}---This work is supported by the National Key R\&D Program of China (No.2018YFA0306703, No.2021YFE0113100). Z.W. is supported by the Sichuan Innovative Research Team Support Fund (No. 2021JDTD0028). G.Y. is supported by the National Natural Science Foundation of China (No.62172075).


\begin{thebibliography}{47}%
\makeatletter
\providecommand \@ifxundefined [1]{%
 \@ifx{#1\undefined}
}%
\providecommand \@ifnum [1]{%
 \ifnum #1\expandafter \@firstoftwo
 \else \expandafter \@secondoftwo
 \fi
}%
\providecommand \@ifx [1]{%
 \ifx #1\expandafter \@firstoftwo
 \else \expandafter \@secondoftwo
 \fi
}%
\providecommand \natexlab [1]{#1}%
\providecommand \enquote  [1]{``#1''}%
\providecommand \bibnamefont  [1]{#1}%
\providecommand \bibfnamefont [1]{#1}%
\providecommand \citenamefont [1]{#1}%
\providecommand \href@noop [0]{\@secondoftwo}%
\providecommand \href [0]{\begingroup \@sanitize@url \@href}%
\providecommand \@href[1]{\@@startlink{#1}\@@href}%
\providecommand \@@href[1]{\endgroup#1\@@endlink}%
\providecommand \@sanitize@url [0]{\catcode `\\12\catcode `\$12\catcode
  `\&12\catcode `\#12\catcode `\^12\catcode `\_12\catcode `\%12\relax}%
\providecommand \@@startlink[1]{}%
\providecommand \@@endlink[0]{}%
\providecommand \url  [0]{\begingroup\@sanitize@url \@url }%
\providecommand \@url [1]{\endgroup\@href {#1}{\urlprefix }}%
\providecommand \urlprefix  [0]{URL }%
\providecommand \Eprint [0]{\href }%
\providecommand \doibase [0]{https://doi.org/}%
\providecommand \selectlanguage [0]{\@gobble}%
\providecommand \bibinfo  [0]{\@secondoftwo}%
\providecommand \bibfield  [0]{\@secondoftwo}%
\providecommand \translation [1]{[#1]}%
\providecommand \BibitemOpen [0]{}%
\providecommand \bibitemStop [0]{}%
\providecommand \bibitemNoStop [0]{.\EOS\space}%
\providecommand \EOS [0]{\spacefactor3000\relax}%
\providecommand \BibitemShut  [1]{\csname bibitem#1\endcsname}%
\let\auto@bib@innerbib\@empty
\bibitem [{\citenamefont {Bermejo-Vega}\ \emph {et~al.}(2017)\citenamefont
  {Bermejo-Vega}, \citenamefont {Delfosse}, \citenamefont {Browne},
  \citenamefont {Okay},\ and\ \citenamefont
  {Raussendorf}}]{PhysRevLett.119.120505}%
  \BibitemOpen
  \bibfield  {author} {\bibinfo {author} {\bibnamefont {Bermejo-Vega},
  \bibfnamefont {Juan}}, \bibinfo {author} {\bibfnamefont {Nicolas}\
  \bibnamefont {Delfosse}}, \bibinfo {author} {\bibfnamefont {Dan~E.}\
  \bibnamefont {Browne}}, \bibinfo {author} {\bibfnamefont {Cihan}\
  \bibnamefont {Okay}}, and\ \bibinfo {author} {\bibfnamefont {Robert}\
  \bibnamefont {Raussendorf}}} (\bibinfo {year} {2017}),\ \bibfield  {title}
  {\enquote {\bibinfo {title} {{Contextuality as a Resource for Models of
  Quantum Computation with Qubits}},}\ }\href
  {https://doi.org/10.1103/PhysRevLett.119.120505} {\bibfield  {journal}
  {\bibinfo  {journal} {Phys. Rev. Lett.}\ }\textbf {\bibinfo {volume} {119}},\
  \bibinfo {pages} {120505}}\BibitemShut {NoStop}%
\bibitem [{\citenamefont {Bouwmeester}\ \emph {et~al.}(1997)\citenamefont
  {Bouwmeester}, \citenamefont {Pan}, \citenamefont {Mattle}, \citenamefont
  {Eibl}, \citenamefont {Weinfurter},\ and\ \citenamefont
  {Zeilinger}}]{Bouwmeester1997Teleportation}%
  \BibitemOpen
  \bibfield  {author} {\bibinfo {author} {\bibnamefont {Bouwmeester},
  \bibfnamefont {Dik}}, \bibinfo {author} {\bibfnamefont {Jian-Wei}\
  \bibnamefont {Pan}}, \bibinfo {author} {\bibfnamefont {Klaus}\ \bibnamefont
  {Mattle}}, \bibinfo {author} {\bibfnamefont {Manfred}\ \bibnamefont {Eibl}},
  \bibinfo {author} {\bibfnamefont {Harald}\ \bibnamefont {Weinfurter}}, and\
  \bibinfo {author} {\bibfnamefont {Anton}\ \bibnamefont {Zeilinger}}}
  (\bibinfo {year} {1997}),\ \bibfield  {title} {\enquote {\bibinfo {title}
  {{Experimental quantum teleportation}},}\ }\href
  {https://doi.org/10.1038/37539} {\bibfield  {journal} {\bibinfo  {journal}
  {Nature}\ }\textbf {\bibinfo {volume} {390}}~(\bibinfo {number} {6660}),\
  \bibinfo {pages} {575--579}}\BibitemShut {NoStop}%
\bibitem [{\citenamefont {Bravyi}\ \emph {et~al.}(2018)\citenamefont {Bravyi},
  \citenamefont {Gosset},\ and\ \citenamefont {K{\"o}nig}}]{ShallowCircuits}%
  \BibitemOpen
  \bibfield  {author} {\bibinfo {author} {\bibnamefont {Bravyi}, \bibfnamefont
  {Sergey}}, \bibinfo {author} {\bibfnamefont {David}\ \bibnamefont {Gosset}},
  and\ \bibinfo {author} {\bibfnamefont {Robert}\ \bibnamefont {K{\"o}nig}}}
  (\bibinfo {year} {2018}),\ \bibfield  {title} {\enquote {\bibinfo {title}
  {{Quantum advantage with shallow circuits}},}\ }\href
  {https://doi.org/10.1126/science.aar3106} {\bibfield  {journal} {\bibinfo
  {journal} {Science}\ }\textbf {\bibinfo {volume} {362}}~(\bibinfo {number}
  {6412}),\ \bibinfo {pages} {308--311}}\BibitemShut {NoStop}%
\bibitem [{\citenamefont {Bravyi}\ \emph {et~al.}(2020)\citenamefont {Bravyi},
  \citenamefont {Gosset}, \citenamefont {K{\"o}nig},\ and\ \citenamefont
  {Tomamichel}}]{NoisyShallowCircuits}%
  \BibitemOpen
  \bibfield  {author} {\bibinfo {author} {\bibnamefont {Bravyi}, \bibfnamefont
  {Sergey}}, \bibinfo {author} {\bibfnamefont {David}\ \bibnamefont {Gosset}},
  \bibinfo {author} {\bibfnamefont {Robert}\ \bibnamefont {K{\"o}nig}}, and\
  \bibinfo {author} {\bibfnamefont {Marco}\ \bibnamefont {Tomamichel}}}
  (\bibinfo {year} {2020}),\ \bibfield  {title} {\enquote {\bibinfo {title}
  {{Quantum advantage with noisy shallow circuits}},}\ }\href
  {https://doi.org/10.1038/s41567-020-0948-z} {\bibfield  {journal} {\bibinfo
  {journal} {Nature Physics}\ }\textbf {\bibinfo {volume} {16}}~(\bibinfo
  {number} {10}),\ \bibinfo {pages} {1040--1045}}\BibitemShut {NoStop}%
\bibitem [{\citenamefont {Brunner}\ \emph {et~al.}(2014)\citenamefont
  {Brunner}, \citenamefont {Cavalcanti}, \citenamefont {Pironio}, \citenamefont
  {Scarani},\ and\ \citenamefont {Wehner}}]{RevModPhys.86.419}%
  \BibitemOpen
  \bibfield  {author} {\bibinfo {author} {\bibnamefont {Brunner}, \bibfnamefont
  {Nicolas}}, \bibinfo {author} {\bibfnamefont {Daniel}\ \bibnamefont
  {Cavalcanti}}, \bibinfo {author} {\bibfnamefont {Stefano}\ \bibnamefont
  {Pironio}}, \bibinfo {author} {\bibfnamefont {Valerio}\ \bibnamefont
  {Scarani}}, and\ \bibinfo {author} {\bibfnamefont {Stephanie}\ \bibnamefont
  {Wehner}}} (\bibinfo {year} {2014}),\ \bibfield  {title} {\enquote {\bibinfo
  {title} {{Bell nonlocality}},}\ }\href
  {https://doi.org/10.1103/RevModPhys.86.419} {\bibfield  {journal} {\bibinfo
  {journal} {Rev. Mod. Phys.}\ }\textbf {\bibinfo {volume} {86}},\ \bibinfo
  {pages} {419--478}}\BibitemShut {NoStop}%
\bibitem [{\citenamefont {Budroni}\ \emph {et~al.}(2021)\citenamefont
  {Budroni}, \citenamefont {Cabello}, \citenamefont {G{\"u}hne}, \citenamefont
  {Kleinmann},\ and\ \citenamefont {Larsson}}]{Budroni2021ContextualityReview}%
  \BibitemOpen
  \bibfield  {author} {\bibinfo {author} {\bibnamefont {Budroni}, \bibfnamefont
  {Costantino}}, \bibinfo {author} {\bibfnamefont {Ad{\'a}n}\ \bibnamefont
  {Cabello}}, \bibinfo {author} {\bibfnamefont {Otfried}\ \bibnamefont
  {G{\"u}hne}}, \bibinfo {author} {\bibfnamefont {Matthias}\ \bibnamefont
  {Kleinmann}}, and\ \bibinfo {author} {\bibfnamefont {Jan-{\AA}ke}\
  \bibnamefont {Larsson}}} (\bibinfo {year} {2021}),\ \bibfield  {title}
  {\enquote {\bibinfo {title} {{Quantum Contextuality}},}\ }\href
  {https://arxiv.org/abs/2102.13036} {\bibinfo  {journal} {arXiv:2102.13036}\
  }\BibitemShut {NoStop}%
\bibitem [{\citenamefont {Cabello}(2021)}]{Cabello2021KSBell}%
  \BibitemOpen
\bibfield  {journal} {  }\bibfield  {author} {\bibinfo {author} {\bibnamefont
  {Cabello}, \bibfnamefont {Ad{\'a}n}}} (\bibinfo {year} {2021}),\ \bibfield
  {title} {\enquote {\bibinfo {title} {{Bell Non-locality and Kochen--Specker
  Contextuality: How are They Connected?}}}\ }\href
  {https://doi.org/10.1007/s10701-021-00466-5} {\bibfield  {journal} {\bibinfo
  {journal} {Foundations of Physics}\ }\textbf {\bibinfo {volume}
  {51}}~(\bibinfo {number} {3}),\ \bibinfo {pages} {61}}\BibitemShut {NoStop}%
\bibitem [{\citenamefont {Cubitt}\ \emph {et~al.}(2015)\citenamefont {Cubitt},
  \citenamefont {Perez-Garcia},\ and\ \citenamefont
  {Wolf}}]{Cubitt2015SpectralGap}%
  \BibitemOpen
  \bibfield  {author} {\bibinfo {author} {\bibnamefont {Cubitt}, \bibfnamefont
  {Toby~S}}, \bibinfo {author} {\bibfnamefont {David}\ \bibnamefont
  {Perez-Garcia}}, and\ \bibinfo {author} {\bibfnamefont {Michael~M.}\
  \bibnamefont {Wolf}}} (\bibinfo {year} {2015}),\ \bibfield  {title} {\enquote
  {\bibinfo {title} {{Undecidability of the spectral gap}},}\ }\href
  {https://doi.org/10.1038/nature16059} {\bibfield  {journal} {\bibinfo
  {journal} {Nature}\ }\textbf {\bibinfo {volume} {528}}~(\bibinfo {number}
  {7581}),\ \bibinfo {pages} {207--211}}\BibitemShut {NoStop}%
\bibitem [{\citenamefont {Dalzell}\ and\ \citenamefont
  {Brand{\~{a}}o}(2019)}]{Dalzell2019locallyaccuratemps}%
  \BibitemOpen
  \bibfield  {author} {\bibinfo {author} {\bibnamefont {Dalzell}, \bibfnamefont
  {Alexander~M}}, and\ \bibinfo {author} {\bibfnamefont {Fernando G. S.~L.}\
  \bibnamefont {Brand{\~{a}}o}}} (\bibinfo {year} {2019}),\ \bibfield  {title}
  {\enquote {\bibinfo {title} {Locally accurate {MPS} approximations for ground
  states of one-dimensional gapped local {H}amiltonians},}\ }\href
  {https://doi.org/10.22331/q-2019-09-23-187} {\bibfield  {journal} {\bibinfo
  {journal} {{Quantum}}\ }\textbf {\bibinfo {volume} {3}},\ \bibinfo {pages}
  {187}}\BibitemShut {NoStop}%
\bibitem [{\citenamefont {De~Chiara}\ and\ \citenamefont
  {Sanpera}(2018)}]{Sanpera2018Review}%
  \BibitemOpen
  \bibfield  {author} {\bibinfo {author} {\bibnamefont {De~Chiara},
  \bibfnamefont {Gabriele}}, and\ \bibinfo {author} {\bibfnamefont {Anna}\
  \bibnamefont {Sanpera}}} (\bibinfo {year} {2018}),\ \bibfield  {title}
  {\enquote {\bibinfo {title} {{Genuine quantum correlations in quantum
  many-body systems: a review of recent progress}},}\ }\href
  {https://doi.org/10.1088/1361-6633/aabf61} {\bibfield  {journal} {\bibinfo
  {journal} {Rep. Prog. Phys.}\ }\textbf {\bibinfo {volume} {81}}~(\bibinfo
  {number} {7}),\ \bibinfo {pages} {074002}}\BibitemShut {NoStop}%
\bibitem [{\citenamefont {G{\"a}rtner}\ and\ \citenamefont
  {Matou{\v{s}}ek}(2012)}]{SDPBook}%
  \BibitemOpen
  \bibfield  {author} {\bibinfo {author} {\bibnamefont {G{\"a}rtner},
  \bibfnamefont {Bernd}}, and\ \bibinfo {author} {\bibfnamefont
  {Ji{\v{r}}{\'\i}}\ \bibnamefont {Matou{\v{s}}ek}}} (\bibinfo {year} {2012}),\
  \href {https://doi.org/10.1007/978-3-642-22015-9} {\emph {\bibinfo {title}
  {{Approximation Algorithms and Semidefinite Programming}}}},\ Universitext\
  (\bibinfo  {publisher} {Springer},\ \bibinfo {address} {Berlin,
  Heidelberg})\BibitemShut {NoStop}%
\bibitem [{\citenamefont {Giustina}\ \emph {et~al.}(2015)\citenamefont
  {Giustina}, \citenamefont {Versteegh}, \citenamefont {Wengerowsky},
  \citenamefont {Handsteiner}, \citenamefont {Hochrainer}, \citenamefont
  {Phelan}, \citenamefont {Steinlechner}, \citenamefont {Kofler}, \citenamefont
  {Larsson}, \citenamefont {Abell\'an}, \citenamefont {Amaya}, \citenamefont
  {Pruneri}, \citenamefont {Mitchell}, \citenamefont {Beyer}, \citenamefont
  {Gerrits}, \citenamefont {Lita}, \citenamefont {Shalm}, \citenamefont {Nam},
  \citenamefont {Scheidl}, \citenamefont {Ursin}, \citenamefont {Wittmann},\
  and\ \citenamefont {Zeilinger}}]{PhysRevLett.115.250401}%
  \BibitemOpen
  \bibfield  {author} {\bibinfo {author} {\bibnamefont {Giustina},
  \bibfnamefont {Marissa}}, \bibinfo {author} {\bibfnamefont {Marijn A.~M.}\
  \bibnamefont {Versteegh}}, \bibinfo {author} {\bibfnamefont {S\"oren}\
  \bibnamefont {Wengerowsky}}, \bibinfo {author} {\bibfnamefont {Johannes}\
  \bibnamefont {Handsteiner}}, \bibinfo {author} {\bibfnamefont {Armin}\
  \bibnamefont {Hochrainer}}, \bibinfo {author} {\bibfnamefont {Kevin}\
  \bibnamefont {Phelan}}, \bibinfo {author} {\bibfnamefont {Fabian}\
  \bibnamefont {Steinlechner}}, \bibinfo {author} {\bibfnamefont {Johannes}\
  \bibnamefont {Kofler}}, \bibinfo {author} {\bibfnamefont {Jan-\AA{}ke}\
  \bibnamefont {Larsson}}, \bibinfo {author} {\bibfnamefont {Carlos}\
  \bibnamefont {Abell\'an}}, \bibinfo {author} {\bibfnamefont {Waldimar}\
  \bibnamefont {Amaya}}, \bibinfo {author} {\bibfnamefont {Valerio}\
  \bibnamefont {Pruneri}}, \bibinfo {author} {\bibfnamefont {Morgan~W.}\
  \bibnamefont {Mitchell}}, \bibinfo {author} {\bibfnamefont {J\"orn}\
  \bibnamefont {Beyer}}, \bibinfo {author} {\bibfnamefont {Thomas}\
  \bibnamefont {Gerrits}}, \bibinfo {author} {\bibfnamefont {Adriana~E.}\
  \bibnamefont {Lita}}, \bibinfo {author} {\bibfnamefont {Lynden~K.}\
  \bibnamefont {Shalm}}, \bibinfo {author} {\bibfnamefont {Sae~Woo}\
  \bibnamefont {Nam}}, \bibinfo {author} {\bibfnamefont {Thomas}\ \bibnamefont
  {Scheidl}}, \bibinfo {author} {\bibfnamefont {Rupert}\ \bibnamefont {Ursin}},
  \bibinfo {author} {\bibfnamefont {Bernhard}\ \bibnamefont {Wittmann}}, and\
  \bibinfo {author} {\bibfnamefont {Anton}\ \bibnamefont {Zeilinger}}}
  (\bibinfo {year} {2015}),\ \bibfield  {title} {\enquote {\bibinfo {title}
  {{Significant-Loophole-Free Test of Bell's Theorem with Entangled
  Photons}},}\ }\href {https://doi.org/10.1103/PhysRevLett.115.250401}
  {\bibfield  {journal} {\bibinfo  {journal} {Phys. Rev. Lett.}\ }\textbf
  {\bibinfo {volume} {115}},\ \bibinfo {pages} {250401}}\BibitemShut {NoStop}%
\bibitem [{\citenamefont {Goldstein}\ \emph {et~al.}(2017)\citenamefont
  {Goldstein}, \citenamefont {Kuna}, \citenamefont {Lebowitz},\ and\
  \citenamefont {Speer}}]{Goldstein2017}%
  \BibitemOpen
  \bibfield  {author} {\bibinfo {author} {\bibnamefont {Goldstein},
  \bibfnamefont {Sheldon}}, \bibinfo {author} {\bibfnamefont {Tobias}\
  \bibnamefont {Kuna}}, \bibinfo {author} {\bibfnamefont {Joel~Louis}\
  \bibnamefont {Lebowitz}}, and\ \bibinfo {author} {\bibfnamefont
  {Eugene~Richard}\ \bibnamefont {Speer}}} (\bibinfo {year} {2017}),\ \bibfield
   {title} {\enquote {\bibinfo {title} {{Translation Invariant Extensions of
  Finite Volume Measures}},}\ }\href
  {https://doi.org/10.1007/s10955-016-1595-8} {\bibfield  {journal} {\bibinfo
  {journal} {Journal of Statistical Physics}\ }\textbf {\bibinfo {volume}
  {166}}~(\bibinfo {number} {3}),\ \bibinfo {pages} {765--782}}\BibitemShut
  {NoStop}%
\bibitem [{\citenamefont {Gottesman}\ and\ \citenamefont
  {Irani}(2013)}]{GottesmanIrani2013Tiling}%
  \BibitemOpen
  \bibfield  {author} {\bibinfo {author} {\bibnamefont {Gottesman},
  \bibfnamefont {Daniel}}, and\ \bibinfo {author} {\bibfnamefont {Sandy}\
  \bibnamefont {Irani}}} (\bibinfo {year} {2013}),\ \bibfield  {title}
  {\enquote {\bibinfo {title} {{The Quantum and Classical Complexity of
  Translationally Invariant Tiling and Hamiltonian Problems}},}\ }\href
  {https://doi.org/10.4086/toc.2013.v009a002} {\bibfield  {journal} {\bibinfo
  {journal} {Theory of Computing}\ }\textbf {\bibinfo {volume} {9}}~(\bibinfo
  {number} {2}),\ \bibinfo {pages} {31--116}}\BibitemShut {NoStop}%
\bibitem [{\citenamefont {Haegeman}\ \emph {et~al.}(2011)\citenamefont
  {Haegeman}, \citenamefont {Cirac}, \citenamefont {Osborne}, \citenamefont
  {Pi\ifmmode~\check{z}\else \v{z}\fi{}orn}, \citenamefont {Verschelde},\ and\
  \citenamefont {Verstraete}}]{Haegeman2011TDVP}%
  \BibitemOpen
  \bibfield  {author} {\bibinfo {author} {\bibnamefont {Haegeman},
  \bibfnamefont {Jutho}}, \bibinfo {author} {\bibfnamefont {J.~Ignacio}\
  \bibnamefont {Cirac}}, \bibinfo {author} {\bibfnamefont {Tobias~J.}\
  \bibnamefont {Osborne}}, \bibinfo {author} {\bibfnamefont {Iztok}\
  \bibnamefont {Pi\ifmmode~\check{z}\else \v{z}\fi{}orn}}, \bibinfo {author}
  {\bibfnamefont {Henri}\ \bibnamefont {Verschelde}}, and\ \bibinfo {author}
  {\bibfnamefont {Frank}\ \bibnamefont {Verstraete}}} (\bibinfo {year}
  {2011}),\ \bibfield  {title} {\enquote {\bibinfo {title} {{Time-Dependent
  Variational Principle for Quantum Lattices}},}\ }\href
  {https://doi.org/10.1103/PhysRevLett.107.070601} {\bibfield  {journal}
  {\bibinfo  {journal} {Phys. Rev. Lett.}\ }\textbf {\bibinfo {volume} {107}},\
  \bibinfo {pages} {070601}}\BibitemShut {NoStop}%
\bibitem [{\citenamefont {Hensen}\ \emph {et~al.}(2015)\citenamefont {Hensen},
  \citenamefont {Bernien}, \citenamefont {Dr{\'e}au}, \citenamefont {Reiserer},
  \citenamefont {Kalb}, \citenamefont {Blok}, \citenamefont {Ruitenberg},
  \citenamefont {Vermeulen}, \citenamefont {Schouten}, \citenamefont
  {Abell{\'a}n}, \citenamefont {Amaya}, \citenamefont {Pruneri}, \citenamefont
  {Mitchell}, \citenamefont {Markham}, \citenamefont {Twitchen}, \citenamefont
  {Elkouss}, \citenamefont {Wehner}, \citenamefont {Taminiau},\ and\
  \citenamefont {Hanson}}]{loopholefree_Delft}%
  \BibitemOpen
  \bibfield  {author} {\bibinfo {author} {\bibnamefont {Hensen}, \bibfnamefont
  {Bas}}, \bibinfo {author} {\bibfnamefont {Hannes}\ \bibnamefont {Bernien}},
  \bibinfo {author} {\bibfnamefont {Ana{\"\i}s~E.}\ \bibnamefont {Dr{\'e}au}},
  \bibinfo {author} {\bibfnamefont {Andreas}\ \bibnamefont {Reiserer}},
  \bibinfo {author} {\bibfnamefont {Norbert}\ \bibnamefont {Kalb}}, \bibinfo
  {author} {\bibfnamefont {Machiel~Sebastiaan}\ \bibnamefont {Blok}}, \bibinfo
  {author} {\bibfnamefont {Just}\ \bibnamefont {Ruitenberg}}, \bibinfo {author}
  {\bibfnamefont {Raymond F.~L.}\ \bibnamefont {Vermeulen}}, \bibinfo {author}
  {\bibfnamefont {Raymond~N.}\ \bibnamefont {Schouten}}, \bibinfo {author}
  {\bibfnamefont {Carlos}\ \bibnamefont {Abell{\'a}n}}, \bibinfo {author}
  {\bibfnamefont {Waldimar}\ \bibnamefont {Amaya}}, \bibinfo {author}
  {\bibfnamefont {Valerio}\ \bibnamefont {Pruneri}}, \bibinfo {author}
  {\bibfnamefont {Morgan W.~W.}\ \bibnamefont {Mitchell}}, \bibinfo {author}
  {\bibfnamefont {Matthew}\ \bibnamefont {Markham}}, \bibinfo {author}
  {\bibfnamefont {Daniel~J.}\ \bibnamefont {Twitchen}}, \bibinfo {author}
  {\bibfnamefont {David}\ \bibnamefont {Elkouss}}, \bibinfo {author}
  {\bibfnamefont {Stephanie}\ \bibnamefont {Wehner}}, \bibinfo {author}
  {\bibfnamefont {Tim~Hugo}\ \bibnamefont {Taminiau}}, and\ \bibinfo {author}
  {\bibfnamefont {Ronald}\ \bibnamefont {Hanson}}} (\bibinfo {year} {2015}),\
  \bibfield  {title} {\enquote {\bibinfo {title} {{Loophole-free Bell
  inequality violation using electron spins separated by 1.3 kilometres}},}\
  }\href {https://doi.org/10.1038/nature15759} {\bibfield  {journal} {\bibinfo
  {journal} {Nature}\ }\textbf {\bibinfo {volume} {526}},\ \bibinfo {pages}
  {682--686}}\BibitemShut {NoStop}%
\bibitem [{\citenamefont {Howard}\ \emph {et~al.}(2014)\citenamefont {Howard},
  \citenamefont {Wallman}, \citenamefont {Veitch},\ and\ \citenamefont
  {Emerson}}]{ContextualityMagic}%
  \BibitemOpen
  \bibfield  {author} {\bibinfo {author} {\bibnamefont {Howard}, \bibfnamefont
  {Mark}}, \bibinfo {author} {\bibfnamefont {Joel}\ \bibnamefont {Wallman}},
  \bibinfo {author} {\bibfnamefont {Victor}\ \bibnamefont {Veitch}}, and\
  \bibinfo {author} {\bibfnamefont {Joseph}\ \bibnamefont {Emerson}}} (\bibinfo
  {year} {2014}),\ \bibfield  {title} {\enquote {\bibinfo {title}
  {{Contextuality supplies the `magic' for quantum computation}},}\ }\href
  {https://doi.org/10.1038/nature13460} {\bibfield  {journal} {\bibinfo
  {journal} {Nature}\ }\textbf {\bibinfo {volume} {510}}~(\bibinfo {number}
  {7505}),\ \bibinfo {pages} {351--355}}\BibitemShut {NoStop}%
\bibitem [{\citenamefont {Huang}(2015)}]{HuangEnergyDensity}%
  \BibitemOpen
  \bibfield  {author} {\bibinfo {author} {\bibnamefont {Huang}, \bibfnamefont
  {Yichen}}} (\bibinfo {year} {2015}),\ \bibfield  {title} {\enquote {\bibinfo
  {title} {{Computing energy density in one dimension}},}\ }\href
  {https://arxiv.org/abs/1505.00772} {\bibinfo  {journal} {arXiv:1505:00772}\
  }\BibitemShut {NoStop}%
\bibitem [{\citenamefont {Huang}(2019)}]{HuangConstantBondDimension}%
  \BibitemOpen
\bibfield  {journal} {  }\bibfield  {author} {\bibinfo {author} {\bibnamefont
  {Huang}, \bibfnamefont {Yichen}}} (\bibinfo {year} {2019}),\ \bibfield
  {title} {\enquote {\bibinfo {title} {{Approximating local properties by
  tensor network states with constant bond dimension}},}\ }\href
  {https://arxiv.org/abs/1903.10048} {\bibinfo  {journal} {arXiv:1903:10048}\
  }\BibitemShut {NoStop}%
\bibitem [{\citenamefont {Huang}(2021)}]{Huang2DQMA}%
  \BibitemOpen
\bibfield  {journal} {  }\bibfield  {author} {\bibinfo {author} {\bibnamefont
  {Huang}, \bibfnamefont {Yichen}}} (\bibinfo {year} {2021}),\ \bibfield
  {title} {\enquote {\bibinfo {title} {{Two-dimensional local Hamiltonian
  problem with area laws is QMA-complete}},}\ }\href
  {https://doi.org/10.1016/j.jcp.2021.110534} {\bibfield  {journal} {\bibinfo
  {journal} {Journal of Computational Physics}\ }\textbf {\bibinfo {volume}
  {443}},\ \bibinfo {pages} {110534}}\BibitemShut {NoStop}%
\bibitem [{\citenamefont {Ji}\ \emph {et~al.}(2020)\citenamefont {Ji},
  \citenamefont {Natarajan}, \citenamefont {Vidick}, \citenamefont {Wright},\
  and\ \citenamefont {Yuen}}]{ji2020mipre}%
  \BibitemOpen
  \bibfield  {author} {\bibinfo {author} {\bibnamefont {Ji}, \bibfnamefont
  {Zhengfeng}}, \bibinfo {author} {\bibfnamefont {Anand}\ \bibnamefont
  {Natarajan}}, \bibinfo {author} {\bibfnamefont {Thomas}\ \bibnamefont
  {Vidick}}, \bibinfo {author} {\bibfnamefont {John}\ \bibnamefont {Wright}},
  and\ \bibinfo {author} {\bibfnamefont {Henry}\ \bibnamefont {Yuen}}}
  (\bibinfo {year} {2020}),\ \bibfield  {title} {\enquote {\bibinfo {title}
  {{MIP*=RE}},}\ }\href {https://arxiv.org/abs/2001.04383} {\bibinfo  {journal}
  {arXiv:2001.04383}\ }\BibitemShut {NoStop}%
\bibitem [{\citenamefont {Kochen}\ and\ \citenamefont
  {Specker}(1967)}]{kochen1967problem}%
  \BibitemOpen
\bibfield  {journal} {  }\bibfield  {author} {\bibinfo {author} {\bibnamefont
  {Kochen}, \bibfnamefont {Simon~Bernhard}}, and\ \bibinfo {author}
  {\bibfnamefont {Ernst~Paul}\ \bibnamefont {Specker}}} (\bibinfo {year}
  {1967}),\ \bibfield  {title} {\enquote {\bibinfo {title} {{The Problem of
  Hidden Variables in Quantum Mechanics}},}\ }\href
  {https://doi.org/10.1007/978-3-0348-9259-9_21} {\bibfield  {journal}
  {\bibinfo  {journal} {J. Math. Mech.}\ }\textbf {\bibinfo {volume}
  {17}}~(\bibinfo {number} {1}),\ \bibinfo {pages} {59--87}}\BibitemShut
  {NoStop}%
\bibitem [{\citenamefont {Kuzmin}\ \emph {et~al.}(2022)\citenamefont {Kuzmin},
  \citenamefont {Zache}, \citenamefont {Kokail}, \citenamefont {Pastori},
  \citenamefont {Celi}, \citenamefont {Baranov},\ and\ \citenamefont
  {Zoller}}]{Kuzmin2021FiniteSimulation}%
  \BibitemOpen
  \bibfield  {author} {\bibinfo {author} {\bibnamefont {Kuzmin}, \bibfnamefont
  {Viacheslav}}, \bibinfo {author} {\bibfnamefont {Torsten~V.}\ \bibnamefont
  {Zache}}, \bibinfo {author} {\bibfnamefont {Christian}\ \bibnamefont
  {Kokail}}, \bibinfo {author} {\bibfnamefont {Lorenzo}\ \bibnamefont
  {Pastori}}, \bibinfo {author} {\bibfnamefont {Alessio}\ \bibnamefont {Celi}},
  \bibinfo {author} {\bibfnamefont {Mikhail}\ \bibnamefont {Baranov}}, and\
  \bibinfo {author} {\bibfnamefont {Peter}\ \bibnamefont {Zoller}}} (\bibinfo
  {year} {2022}),\ \bibfield  {title} {\enquote {\bibinfo {title} {Probing
  infinite many-body quantum systems with finite-size quantum simulators},}\
  }\href {https://doi.org/10.1103/PRXQuantum.3.020304} {\bibfield  {journal}
  {\bibinfo  {journal} {PRX Quantum}\ }\textbf {\bibinfo {volume} {3}},\
  \bibinfo {pages} {020304}}\BibitemShut {NoStop}%
\bibitem [{\citenamefont {Landau}\ \emph {et~al.}(2015)\citenamefont {Landau},
  \citenamefont {Vazirani},\ and\ \citenamefont {Vidick}}]{Landau2015RRG}%
  \BibitemOpen
  \bibfield  {author} {\bibinfo {author} {\bibnamefont {Landau}, \bibfnamefont
  {Zeph}}, \bibinfo {author} {\bibfnamefont {Umesh}\ \bibnamefont {Vazirani}},
  and\ \bibinfo {author} {\bibfnamefont {Thomas}\ \bibnamefont {Vidick}}}
  (\bibinfo {year} {2015}),\ \bibfield  {title} {\enquote {\bibinfo {title} {{A
  polynomial time algorithm for the ground state of one-dimensional gapped
  local Hamiltonians}},}\ }\href {https://doi.org/10.1038/nphys3345} {\bibfield
   {journal} {\bibinfo  {journal} {Nature Physics}\ }\textbf {\bibinfo {volume}
  {11}}~(\bibinfo {number} {7}),\ \bibinfo {pages} {566--569}}\BibitemShut
  {NoStop}%
\bibitem [{\citenamefont {Matou{\v{s}}ek}\ and\ \citenamefont
  {G{\"a}rtner}(2007)}]{LPBook}%
  \BibitemOpen
  \bibfield  {author} {\bibinfo {author} {\bibnamefont {Matou{\v{s}}ek},
  \bibfnamefont {Ji{\v{r}}{\'\i}}}, and\ \bibinfo {author} {\bibfnamefont
  {Bernd}\ \bibnamefont {G{\"a}rtner}}} (\bibinfo {year} {2007}),\ \href
  {https://doi.org/10.1007/978-3-540-30717-4} {\emph {\bibinfo {title}
  {{Understanding and Using Linear Programming}}}},\ Universitext\ (\bibinfo
  {publisher} {Springer},\ \bibinfo {address} {Berlin, Heidelberg})\BibitemShut
  {NoStop}%
\bibitem [{\citenamefont {Mayers}\ and\ \citenamefont
  {Yao}(2004)}]{MayersYao2004SelfTesting}%
  \BibitemOpen
  \bibfield  {author} {\bibinfo {author} {\bibnamefont {Mayers}, \bibfnamefont
  {Dominic}}, and\ \bibinfo {author} {\bibfnamefont {Andrew}\ \bibnamefont
  {Yao}}} (\bibinfo {year} {2004}),\ \bibfield  {title} {\enquote {\bibinfo
  {title} {{Self testing quantum apparatus}},}\ }\href@noop {} {\bibfield
  {journal} {\bibinfo  {journal} {Quantum Information \& Computation}\ }\textbf
  {\bibinfo {volume} {4}}~(\bibinfo {number} {4}),\ \bibinfo {pages}
  {273--286}}\BibitemShut {NoStop}%
\bibitem [{\citenamefont {Milsted}\ \emph {et~al.}(2013)\citenamefont
  {Milsted}, \citenamefont {Haegeman}, \citenamefont {Osborne},\ and\
  \citenamefont {Verstraete}}]{TDVP2}%
  \BibitemOpen
  \bibfield  {author} {\bibinfo {author} {\bibnamefont {Milsted}, \bibfnamefont
  {Ashley}}, \bibinfo {author} {\bibfnamefont {Jutho}\ \bibnamefont
  {Haegeman}}, \bibinfo {author} {\bibfnamefont {Tobias~J.}\ \bibnamefont
  {Osborne}}, and\ \bibinfo {author} {\bibfnamefont {Frank}\ \bibnamefont
  {Verstraete}}} (\bibinfo {year} {2013}),\ \bibfield  {title} {\enquote
  {\bibinfo {title} {{Variational matrix product ansatz for nonuniform dynamics
  in the thermodynamic limit}},}\ }\href
  {https://doi.org/10.1103/PhysRevB.88.155116} {\bibfield  {journal} {\bibinfo
  {journal} {Phys. Rev. B}\ }\textbf {\bibinfo {volume} {88}},\ \bibinfo
  {pages} {155116}}\BibitemShut {NoStop}%
\bibitem [{\citenamefont {Navascu\'es}\ \emph {et~al.}(2007)\citenamefont
  {Navascu\'es}, \citenamefont {Pironio},\ and\ \citenamefont
  {Ac\'\i{}n}}]{NPA1}%
  \BibitemOpen
  \bibfield  {author} {\bibinfo {author} {\bibnamefont {Navascu\'es},
  \bibfnamefont {Miguel}}, \bibinfo {author} {\bibfnamefont {Stefano}\
  \bibnamefont {Pironio}}, and\ \bibinfo {author} {\bibfnamefont {Antonio}\
  \bibnamefont {Ac\'\i{}n}}} (\bibinfo {year} {2007}),\ \bibfield  {title}
  {\enquote {\bibinfo {title} {{Bounding the Set of Quantum Correlations}},}\
  }\href {https://doi.org/10.1103/PhysRevLett.98.010401} {\bibfield  {journal}
  {\bibinfo  {journal} {Phys. Rev. Lett.}\ }\textbf {\bibinfo {volume} {98}},\
  \bibinfo {pages} {010401}}\BibitemShut {NoStop}%
\bibitem [{\citenamefont {Navascu{\'e}s}\ \emph {et~al.}(2008)\citenamefont
  {Navascu{\'e}s}, \citenamefont {Pironio},\ and\ \citenamefont
  {Ac{\'\i}n}}]{NPA2}%
  \BibitemOpen
  \bibfield  {author} {\bibinfo {author} {\bibnamefont {Navascu{\'e}s},
  \bibfnamefont {Miguel}}, \bibinfo {author} {\bibfnamefont {Stefano}\
  \bibnamefont {Pironio}}, and\ \bibinfo {author} {\bibfnamefont {Antonio}\
  \bibnamefont {Ac{\'\i}n}}} (\bibinfo {year} {2008}),\ \bibfield  {title}
  {\enquote {\bibinfo {title} {{A convergent hierarchy of semidefinite programs
  characterizing the set of quantum correlations}},}\ }\href
  {https://doi.org/10.1088/1367-2630/10/7/073013} {\bibfield  {journal}
  {\bibinfo  {journal} {New Journal of Physics}\ }\textbf {\bibinfo {volume}
  {10}}~(\bibinfo {number} {7}),\ \bibinfo {pages} {073013}}\BibitemShut
  {NoStop}%
\bibitem [{\citenamefont {Peruzzo}\ \emph {et~al.}(2014)\citenamefont
  {Peruzzo}, \citenamefont {McClean}, \citenamefont {Shadbolt}, \citenamefont
  {Yung}, \citenamefont {Zhou}, \citenamefont {Love}, \citenamefont
  {Aspuru-Guzik},\ and\ \citenamefont {O'Brien}}]{VQE}%
  \BibitemOpen
  \bibfield  {author} {\bibinfo {author} {\bibnamefont {Peruzzo}, \bibfnamefont
  {Alberto}}, \bibinfo {author} {\bibfnamefont {Jarrod}\ \bibnamefont
  {McClean}}, \bibinfo {author} {\bibfnamefont {Peter}\ \bibnamefont
  {Shadbolt}}, \bibinfo {author} {\bibfnamefont {Man-Hong}\ \bibnamefont
  {Yung}}, \bibinfo {author} {\bibfnamefont {Xiao-Qi}\ \bibnamefont {Zhou}},
  \bibinfo {author} {\bibfnamefont {Peter~J.}\ \bibnamefont {Love}}, \bibinfo
  {author} {\bibfnamefont {Al{\'a}n}\ \bibnamefont {Aspuru-Guzik}}, and\
  \bibinfo {author} {\bibfnamefont {Jeremy~L.}\ \bibnamefont {O'Brien}}}
  (\bibinfo {year} {2014}),\ \bibfield  {title} {\enquote {\bibinfo {title} {A
  variational eigenvalue solver on a photonic quantum processor},}\ }\href
  {https://doi.org/10.1038/ncomms5213} {\bibfield  {journal} {\bibinfo
  {journal} {Nature Communications}\ }\textbf {\bibinfo {volume} {5}}~(\bibinfo
  {number} {1}),\ \bibinfo {pages} {4213}}\BibitemShut {NoStop}%
\bibitem [{\citenamefont {Popescu}\ and\ \citenamefont
  {Rohrlich}(1994)}]{popescu1994quantum}%
  \BibitemOpen
  \bibfield  {author} {\bibinfo {author} {\bibnamefont {Popescu}, \bibfnamefont
  {Sandu}}, and\ \bibinfo {author} {\bibfnamefont {Daniel}\ \bibnamefont
  {Rohrlich}}} (\bibinfo {year} {1994}),\ \bibfield  {title} {\enquote
  {\bibinfo {title} {{Quantum nonlocality as an axiom}},}\ }\href
  {https://doi.org/10.1007/BF02058098} {\bibfield  {journal} {\bibinfo
  {journal} {Foundations of Physics}\ }\textbf {\bibinfo {volume}
  {24}}~(\bibinfo {number} {3}),\ \bibinfo {pages} {379--385}}\BibitemShut
  {NoStop}%
\bibitem [{\citenamefont {Rosset}(2015)}]{DenisThesis}%
  \BibitemOpen
  \bibfield  {author} {\bibinfo {author} {\bibnamefont {Rosset}, \bibfnamefont
  {Denis}}} (\bibinfo {year} {2015}),\ \emph {\bibinfo {title}
  {{Characterization of correlations in quantum networks}}},\ \href
  {https://doi.org/10.13097/archive-ouverte/unige:77401} {Ph.D. thesis}\
  (\bibinfo  {school} {Universit\'e de Gen\`eve})\BibitemShut {NoStop}%
\bibitem [{\citenamefont {Scarani}(2019)}]{ScaraniBook}%
  \BibitemOpen
  \bibfield  {author} {\bibinfo {author} {\bibnamefont {Scarani}, \bibfnamefont
  {Valerio}}} (\bibinfo {year} {2019}),\ \href@noop {} {\emph {\bibinfo {title}
  {{Bell Nonlocality}}}}\ (\bibinfo  {publisher} {Oxford University
  Press})\BibitemShut {NoStop}%
\bibitem [{\citenamefont {Schmied}\ \emph {et~al.}(2016)\citenamefont
  {Schmied}, \citenamefont {Bancal}, \citenamefont {Allard}, \citenamefont
  {Fadel}, \citenamefont {Scarani}, \citenamefont {Treutlein},\ and\
  \citenamefont {Sangouard}}]{JordiBEC}%
  \BibitemOpen
  \bibfield  {author} {\bibinfo {author} {\bibnamefont {Schmied}, \bibfnamefont
  {Roman}}, \bibinfo {author} {\bibfnamefont {Jean-Daniel}\ \bibnamefont
  {Bancal}}, \bibinfo {author} {\bibfnamefont {Baptiste}\ \bibnamefont
  {Allard}}, \bibinfo {author} {\bibfnamefont {Matteo}\ \bibnamefont {Fadel}},
  \bibinfo {author} {\bibfnamefont {Valerio}\ \bibnamefont {Scarani}}, \bibinfo
  {author} {\bibfnamefont {Philipp}\ \bibnamefont {Treutlein}}, and\ \bibinfo
  {author} {\bibfnamefont {Nicolas}\ \bibnamefont {Sangouard}}} (\bibinfo
  {year} {2016}),\ \bibfield  {title} {\enquote {\bibinfo {title} {{Bell
  correlations in a Bose-Einstein condensate}},}\ }\href
  {https://doi.org/10.1126/science.aad8665} {\bibfield  {journal} {\bibinfo
  {journal} {Science}\ }\textbf {\bibinfo {volume} {352}}~(\bibinfo {number}
  {6284}),\ \bibinfo {pages} {441--444}}\BibitemShut {NoStop}%
\bibitem [{\citenamefont {Shalm}\ \emph {et~al.}(2015)\citenamefont {Shalm},
  \citenamefont {Meyer-Scott}, \citenamefont {Christensen}, \citenamefont
  {Bierhorst}, \citenamefont {Wayne}, \citenamefont {Stevens}, \citenamefont
  {Gerrits}, \citenamefont {Glancy}, \citenamefont {Hamel}, \citenamefont
  {Allman}, \citenamefont {Coakley}, \citenamefont {Dyer}, \citenamefont
  {Hodge}, \citenamefont {Lita}, \citenamefont {Verma}, \citenamefont
  {Lambrocco}, \citenamefont {Tortorici}, \citenamefont {Migdall},
  \citenamefont {Zhang}, \citenamefont {Kumor}, \citenamefont {Farr},
  \citenamefont {Marsili}, \citenamefont {Shaw}, \citenamefont {Stern},
  \citenamefont {Abell\'an}, \citenamefont {Amaya}, \citenamefont {Pruneri},
  \citenamefont {Jennewein}, \citenamefont {Mitchell}, \citenamefont {Kwiat},
  \citenamefont {Bienfang}, \citenamefont {Mirin}, \citenamefont {Knill},\ and\
  \citenamefont {Nam}}]{PhysRevLett.115.250402}%
  \BibitemOpen
  \bibfield  {author} {\bibinfo {author} {\bibnamefont {Shalm}, \bibfnamefont
  {Lynden~K}}, \bibinfo {author} {\bibfnamefont {Evan}\ \bibnamefont
  {Meyer-Scott}}, \bibinfo {author} {\bibfnamefont {Bradley~G.}\ \bibnamefont
  {Christensen}}, \bibinfo {author} {\bibfnamefont {Peter}\ \bibnamefont
  {Bierhorst}}, \bibinfo {author} {\bibfnamefont {Michael~A.}\ \bibnamefont
  {Wayne}}, \bibinfo {author} {\bibfnamefont {Martin~J.}\ \bibnamefont
  {Stevens}}, \bibinfo {author} {\bibfnamefont {Thomas}\ \bibnamefont
  {Gerrits}}, \bibinfo {author} {\bibfnamefont {Scott}\ \bibnamefont {Glancy}},
  \bibinfo {author} {\bibfnamefont {Deny~R.}\ \bibnamefont {Hamel}}, \bibinfo
  {author} {\bibfnamefont {Michael~S.}\ \bibnamefont {Allman}}, \bibinfo
  {author} {\bibfnamefont {Kevin~J.}\ \bibnamefont {Coakley}}, \bibinfo
  {author} {\bibfnamefont {Shellee~D.}\ \bibnamefont {Dyer}}, \bibinfo {author}
  {\bibfnamefont {Carson}\ \bibnamefont {Hodge}}, \bibinfo {author}
  {\bibfnamefont {Adriana~E.}\ \bibnamefont {Lita}}, \bibinfo {author}
  {\bibfnamefont {Varun~B.}\ \bibnamefont {Verma}}, \bibinfo {author}
  {\bibfnamefont {Camilla}\ \bibnamefont {Lambrocco}}, \bibinfo {author}
  {\bibfnamefont {Edward}\ \bibnamefont {Tortorici}}, \bibinfo {author}
  {\bibfnamefont {Alan~L.}\ \bibnamefont {Migdall}}, \bibinfo {author}
  {\bibfnamefont {Yanbao}\ \bibnamefont {Zhang}}, \bibinfo {author}
  {\bibfnamefont {Daniel~R.}\ \bibnamefont {Kumor}}, \bibinfo {author}
  {\bibfnamefont {William~H.}\ \bibnamefont {Farr}}, \bibinfo {author}
  {\bibfnamefont {Francesco}\ \bibnamefont {Marsili}}, \bibinfo {author}
  {\bibfnamefont {Matthew~D.}\ \bibnamefont {Shaw}}, \bibinfo {author}
  {\bibfnamefont {Jeffrey~A.}\ \bibnamefont {Stern}}, \bibinfo {author}
  {\bibfnamefont {Carlos}\ \bibnamefont {Abell\'an}}, \bibinfo {author}
  {\bibfnamefont {Waldimar}\ \bibnamefont {Amaya}}, \bibinfo {author}
  {\bibfnamefont {Valerio}\ \bibnamefont {Pruneri}}, \bibinfo {author}
  {\bibfnamefont {Thomas}\ \bibnamefont {Jennewein}}, \bibinfo {author}
  {\bibfnamefont {Morgan~W.}\ \bibnamefont {Mitchell}}, \bibinfo {author}
  {\bibfnamefont {Paul~G.}\ \bibnamefont {Kwiat}}, \bibinfo {author}
  {\bibfnamefont {Joshua~C.}\ \bibnamefont {Bienfang}}, \bibinfo {author}
  {\bibfnamefont {Richard~P.}\ \bibnamefont {Mirin}}, \bibinfo {author}
  {\bibfnamefont {Emanuel}\ \bibnamefont {Knill}}, and\ \bibinfo {author}
  {\bibfnamefont {Sae~Woo}\ \bibnamefont {Nam}}} (\bibinfo {year} {2015}),\
  \bibfield  {title} {\enquote {\bibinfo {title} {{Strong Loophole-Free Test of
  Local Realism}},}\ }\href {https://doi.org/10.1103/PhysRevLett.115.250402}
  {\bibfield  {journal} {\bibinfo  {journal} {Phys. Rev. Lett.}\ }\textbf
  {\bibinfo {volume} {115}},\ \bibinfo {pages} {250402}}\BibitemShut {NoStop}%
\bibitem [{\citenamefont {{\v{S}}upi{\'{c}}}\ and\ \citenamefont
  {Bowles}(2020)}]{Supic2020SelfTestingReview}%
  \BibitemOpen
  \bibfield  {author} {\bibinfo {author} {\bibnamefont {{\v{S}}upi{\'{c}}},
  \bibfnamefont {Ivan}}, and\ \bibinfo {author} {\bibfnamefont {Joseph}\
  \bibnamefont {Bowles}}} (\bibinfo {year} {2020}),\ \bibfield  {title}
  {\enquote {\bibinfo {title} {{Self-testing of quantum systems: a review}},}\
  }\href {https://doi.org/10.22331/q-2020-09-30-337} {\bibfield  {journal}
  {\bibinfo  {journal} {{Quantum}}\ }\textbf {\bibinfo {volume} {4}},\ \bibinfo
  {pages} {337}}\BibitemShut {NoStop}%
\bibitem [{\citenamefont {{The OEIS Foundation
  Inc.}}(2021{\natexlab{a}})}]{oeisNum}%
  \BibitemOpen
  \bibfield  {author} {\bibinfo {author} {\bibnamefont {{The OEIS Foundation
  Inc.}},}} (\bibinfo {year} {2021}{\natexlab{a}}),\ \href
  {https://oeis.org/A027691} {\enquote {\bibinfo {title} {{The On-Line
  Encyclopedia of Integer Sequences: A027691}},}\ }\BibitemShut {NoStop}%
\bibitem [{\citenamefont {{The OEIS Foundation
  Inc.}}(2021{\natexlab{b}})}]{oeisDenom}%
  \BibitemOpen
  \bibfield  {author} {\bibinfo {author} {\bibnamefont {{The OEIS Foundation
  Inc.}},}} (\bibinfo {year} {2021}{\natexlab{b}}),\ \href
  {https://oeis.org/A152948} {\enquote {\bibinfo {title} {{The On-Line
  Encyclopedia of Integer Sequences: A152948}},}\ }\BibitemShut {NoStop}%
\bibitem [{\citenamefont {Tura}\ \emph {et~al.}(2015)\citenamefont {Tura},
  \citenamefont {Augusiak}, \citenamefont {Sainz}, \citenamefont {L{\"u}cke},
  \citenamefont {Klempt}, \citenamefont {Lewenstein},\ and\ \citenamefont
  {Ac{\'\i}n}}]{Tura2015370}%
  \BibitemOpen
  \bibfield  {author} {\bibinfo {author} {\bibnamefont {Tura}, \bibfnamefont
  {Jordi}}, \bibinfo {author} {\bibfnamefont {Remigiusz}\ \bibnamefont
  {Augusiak}}, \bibinfo {author} {\bibfnamefont {Ana~Bel{\'e}n}\ \bibnamefont
  {Sainz}}, \bibinfo {author} {\bibfnamefont {Bernd}\ \bibnamefont
  {L{\"u}cke}}, \bibinfo {author} {\bibfnamefont {Carsten}\ \bibnamefont
  {Klempt}}, \bibinfo {author} {\bibfnamefont {Maciej}\ \bibnamefont
  {Lewenstein}}, and\ \bibinfo {author} {\bibfnamefont {Antonio}\ \bibnamefont
  {Ac{\'\i}n}}} (\bibinfo {year} {2015}),\ \bibfield  {title} {\enquote
  {\bibinfo {title} {{Nonlocality in many-body quantum systems detected with
  two-body correlators}},}\ }\href {https://doi.org/10.1016/j.aop.2015.07.021}
  {\bibfield  {journal} {\bibinfo  {journal} {Annals of Physics}\ }\textbf
  {\bibinfo {volume} {362}},\ \bibinfo {pages} {370 -- 423}}\BibitemShut
  {NoStop}%
\bibitem [{\citenamefont {Tura}\ \emph
  {et~al.}(2014{\natexlab{a}})\citenamefont {Tura}, \citenamefont {Augusiak},
  \citenamefont {Sainz}, \citenamefont {V{\'e}rtesi}, \citenamefont
  {Lewenstein},\ and\ \citenamefont {Ac{\'\i}n}}]{Tura13062014}%
  \BibitemOpen
  \bibfield  {author} {\bibinfo {author} {\bibnamefont {Tura}, \bibfnamefont
  {Jordi}}, \bibinfo {author} {\bibfnamefont {Remigiusz}\ \bibnamefont
  {Augusiak}}, \bibinfo {author} {\bibfnamefont {Ana~Bel{\'e}n}\ \bibnamefont
  {Sainz}}, \bibinfo {author} {\bibfnamefont {Tamas}\ \bibnamefont
  {V{\'e}rtesi}}, \bibinfo {author} {\bibfnamefont {Maciej}\ \bibnamefont
  {Lewenstein}}, and\ \bibinfo {author} {\bibfnamefont {Antonio}\ \bibnamefont
  {Ac{\'\i}n}}} (\bibinfo {year} {2014}{\natexlab{a}}),\ \bibfield  {title}
  {\enquote {\bibinfo {title} {{Detecting nonlocality in many-body quantum
  states}},}\ }\href {https://doi.org/10.1126/science.1247715} {\bibfield
  {journal} {\bibinfo  {journal} {Science}\ }\textbf {\bibinfo {volume}
  {344}}~(\bibinfo {number} {6189}),\ \bibinfo {pages}
  {1256--1258}}\BibitemShut {NoStop}%
\bibitem [{\citenamefont {Tura}\ \emph {et~al.}(2017)\citenamefont {Tura},
  \citenamefont {De~las Cuevas}, \citenamefont {Augusiak}, \citenamefont
  {Lewenstein}, \citenamefont {Ac\'{\i}n},\ and\ \citenamefont
  {Cirac}}]{PhysRevX.7.021005}%
  \BibitemOpen
  \bibfield  {author} {\bibinfo {author} {\bibnamefont {Tura}, \bibfnamefont
  {Jordi}}, \bibinfo {author} {\bibfnamefont {Gemma}\ \bibnamefont {De~las
  Cuevas}}, \bibinfo {author} {\bibfnamefont {Remigiusz}\ \bibnamefont
  {Augusiak}}, \bibinfo {author} {\bibfnamefont {Maciej}\ \bibnamefont
  {Lewenstein}}, \bibinfo {author} {\bibfnamefont {Antonio}\ \bibnamefont
  {Ac\'{\i}n}}, and\ \bibinfo {author} {\bibfnamefont {J.~Ignacio}\
  \bibnamefont {Cirac}}} (\bibinfo {year} {2017}),\ \bibfield  {title}
  {\enquote {\bibinfo {title} {{Energy as a Detector of Nonlocality of
  Many-Body Spin Systems}},}\ }\href
  {https://doi.org/10.1103/PhysRevX.7.021005} {\bibfield  {journal} {\bibinfo
  {journal} {Phys. Rev. X}\ }\textbf {\bibinfo {volume} {7}},\ \bibinfo {pages}
  {021005}}\BibitemShut {NoStop}%
\bibitem [{\citenamefont {Tura}\ \emph
  {et~al.}(2014{\natexlab{b}})\citenamefont {Tura}, \citenamefont {Sainz},
  \citenamefont {V\'ertesi}, \citenamefont {Ac\'\i{}n}, \citenamefont
  {Lewenstein},\ and\ \citenamefont {Augusiak}}]{jordi_ti}%
  \BibitemOpen
  \bibfield  {author} {\bibinfo {author} {\bibnamefont {Tura}, \bibfnamefont
  {Jordi}}, \bibinfo {author} {\bibfnamefont {Ana~Bel{\'e}n}\ \bibnamefont
  {Sainz}}, \bibinfo {author} {\bibfnamefont {Tam\'as}\ \bibnamefont
  {V\'ertesi}}, \bibinfo {author} {\bibfnamefont {Antonio}\ \bibnamefont
  {Ac\'\i{}n}}, \bibinfo {author} {\bibfnamefont {Maciej}\ \bibnamefont
  {Lewenstein}}, and\ \bibinfo {author} {\bibfnamefont {Remigiusz}\
  \bibnamefont {Augusiak}}} (\bibinfo {year} {2014}{\natexlab{b}}),\ \bibfield
  {title} {\enquote {\bibinfo {title} {{Translationally invariant multipartite
  Bell inequalities involving only two-body correlators}},}\ }\href
  {https://doi.org/10.1088/1751-8113/47/42/424024} {\bibfield  {journal}
  {\bibinfo  {journal} {Journal of Physics A: Mathematical and Theoretical}\
  }\textbf {\bibinfo {volume} {47}}~(\bibinfo {number} {42}),\ \bibinfo {pages}
  {424024}}\BibitemShut {NoStop}%
\bibitem [{\citenamefont {Vanderstraeten}\ \emph {et~al.}(2019)\citenamefont
  {Vanderstraeten}, \citenamefont {Haegeman},\ and\ \citenamefont
  {Verstraete}}]{MPSReview}%
  \BibitemOpen
  \bibfield  {author} {\bibinfo {author} {\bibnamefont {Vanderstraeten},
  \bibfnamefont {Laurens}}, \bibinfo {author} {\bibfnamefont {Jutho}\
  \bibnamefont {Haegeman}}, and\ \bibinfo {author} {\bibfnamefont {Frank}\
  \bibnamefont {Verstraete}}} (\bibinfo {year} {2019}),\ \bibfield  {title}
  {\enquote {\bibinfo {title} {{Tangent-space methods for uniform matrix
  product states}},}\ }\href {https://doi.org/10.21468/SciPostPhysLectNotes.7}
  {\bibinfo  {journal} {SciPost Phys. Lect. Notes}\ ,\ \bibinfo {pages}
  {7}}\BibitemShut {NoStop}%
\bibitem [{\citenamefont {Wang}\ and\ \citenamefont
  {Navascu{\'e}s}(2018)}]{ti2d}%
  \BibitemOpen
\bibfield  {journal} {  }\bibfield  {author} {\bibinfo {author} {\bibnamefont
  {Wang}, \bibfnamefont {Zizhu}}, and\ \bibinfo {author} {\bibfnamefont
  {Miguel}\ \bibnamefont {Navascu{\'e}s}}} (\bibinfo {year} {2018}),\ \bibfield
   {title} {\enquote {\bibinfo {title} {{Two-dimensional translation-invariant
  probability distributions: approximations, characterizations and no-go
  theorems}},}\ }\href {https://doi.org/10.1098/rspa.2017.0822} {\bibfield
  {journal} {\bibinfo  {journal} {Proceedings of the Royal Society A:
  Mathematical, Physical and Engineering Sciences}\ }\textbf {\bibinfo {volume}
  {474}}~(\bibinfo {number} {2217}),\ \bibinfo {pages} {20170822}}\BibitemShut
  {NoStop}%
\bibitem [{\citenamefont {Wang}\ \emph {et~al.}(2017)\citenamefont {Wang},
  \citenamefont {Singh},\ and\ \citenamefont {Navascu\'es}}]{Marginal1D}%
  \BibitemOpen
  \bibfield  {author} {\bibinfo {author} {\bibnamefont {Wang}, \bibfnamefont
  {Zizhu}}, \bibinfo {author} {\bibfnamefont {Sukhwinder}\ \bibnamefont
  {Singh}}, and\ \bibinfo {author} {\bibfnamefont {Miguel}\ \bibnamefont
  {Navascu\'es}}} (\bibinfo {year} {2017}),\ \bibfield  {title} {\enquote
  {\bibinfo {title} {{Entanglement and Nonlocality in Infinite 1D Systems}},}\
  }\href {https://doi.org/10.1103/PhysRevLett.118.230401} {\bibfield  {journal}
  {\bibinfo  {journal} {Phys. Rev. Lett.}\ }\textbf {\bibinfo {volume} {118}},\
  \bibinfo {pages} {230401}}\BibitemShut {NoStop}%
\bibitem [{\citenamefont {Zauner-Stauber}\ \emph {et~al.}(2018)\citenamefont
  {Zauner-Stauber}, \citenamefont {Vanderstraeten}, \citenamefont {Fishman},
  \citenamefont {Verstraete},\ and\ \citenamefont
  {Haegeman}}]{Valentine2018VUMPS}%
  \BibitemOpen
  \bibfield  {author} {\bibinfo {author} {\bibnamefont {Zauner-Stauber},
  \bibfnamefont {Valentin}}, \bibinfo {author} {\bibfnamefont {Laurens}\
  \bibnamefont {Vanderstraeten}}, \bibinfo {author} {\bibfnamefont
  {Matthew~T.}\ \bibnamefont {Fishman}}, \bibinfo {author} {\bibfnamefont
  {Frank}\ \bibnamefont {Verstraete}}, and\ \bibinfo {author} {\bibfnamefont
  {Jutho}\ \bibnamefont {Haegeman}}} (\bibinfo {year} {2018}),\ \bibfield
  {title} {\enquote {\bibinfo {title} {{Variational optimization algorithms for
  uniform matrix product states}},}\ }\href
  {https://doi.org/10.1103/PhysRevB.97.045145} {\bibfield  {journal} {\bibinfo
  {journal} {Phys. Rev. B}\ }\textbf {\bibinfo {volume} {97}},\ \bibinfo
  {pages} {045145}}\BibitemShut {NoStop}%
\bibitem [{\citenamefont {Zurel}\ \emph {et~al.}(2020)\citenamefont {Zurel},
  \citenamefont {Okay},\ and\ \citenamefont
  {Raussendorf}}]{PhysRevLett.125.260404}%
  \BibitemOpen
  \bibfield  {author} {\bibinfo {author} {\bibnamefont {Zurel}, \bibfnamefont
  {Michael}}, \bibinfo {author} {\bibfnamefont {Cihan}\ \bibnamefont {Okay}},
  and\ \bibinfo {author} {\bibfnamefont {Robert}\ \bibnamefont {Raussendorf}}}
  (\bibinfo {year} {2020}),\ \bibfield  {title} {\enquote {\bibinfo {title}
  {{Hidden Variable Model for Universal Quantum Computation with Magic States
  on Qubits}},}\ }\href {https://doi.org/10.1103/PhysRevLett.125.260404}
  {\bibfield  {journal} {\bibinfo  {journal} {Phys. Rev. Lett.}\ }\textbf
  {\bibinfo {volume} {125}},\ \bibinfo {pages} {260404}}\BibitemShut {NoStop}%
\end{thebibliography}
%

\newpage
\begin{appendix}
	
\section{Couplings and quantum values of the 322-type Hamiltonians} 
\label{sec:appendix 322-type violated results}

Out of the 2102 equivalent classes of 322-type Hamiltonians, our algorithm found 63 which can violate the classical bound. Table~\ref{all_322_couplings} lists their couplings and the corresponding classical bounds (${\cal{L}}$). Table~\ref{all_322_parameters} gives lowest the ground state energy density (${\cal{Q}}$) with $d$-dimensional local observables. The optimal parameters to reconstruct these local observables ($w$), the numbers of $-1$ in the eigenvalues of each local observable ($[n_x,n_y]$), the bond dimension of the uMPS describing the ground state ($D$), and the $\ltinpa{5}{1}$ lower bound are also given.
\begin{table*}
	\caption{\label{all_322_couplings} Couplings and classical bounds for 63 322-type Hamiltonians.}
	\scriptsize
	\renewcommand\arraystretch{1.3}
	\setlength{\tabcolsep}{0.8mm}{	
		\begin{ruledtabular}
			\begin{tabular}{cccccccccccc}
				No. & ${\cal{L}}$ & $J_x$ &$J_y$ & $J_{xx}^{AB}$ & $J_{xy}^{AB}$ & $J_{yx}^{AB}$ & $J_{yy}^{AB}$ & $J_{xx}^{AC}$ & $J_{xy}^{AC}$
				& $J_{yx}^{AC}$ & $J_{yy}^{AC}$
				\\ \hline
				1	&	-6	&	-6	&	0	&	2	&	3	&	3	&	-2	&	3	&	-1	&	-1	&	1	\\
				2	&	-6	&	-4	&	2	&	2	&	2	&	2	&	-4	&	1	&	-1	&	-1	&	3	\\
				3	&	-3	&	-3	&	1	&	1	&	1	&	1	&	-1	&	1	&	0	&	-1	&	1	\\
				4	&	-4	&	-2	&	-2	&	-2	&	1	&	-1	&	-2	&	1	&	0	&	2	&	1	\\				
				5	&	-8	&	-11	&	1	&	5	&	2	&	2	&	-1	&	4	&	-1	&	-2	&	1	\\
				6	&	-4	&	-3	&	-3	&	2	&	2	&	-1	&	2	&	1	&	1	&	-1	&	0	\\
				7	&	-5	&	-3	&	-3	&	2	&	2	&	2	&	-3	&	1	&	0	&	-1	&	2	\\
				8	&	-4	&	-2	&	-4	&	-2	&	2	&	2	&	2	&	1	&	0	&	0	&	1	\\
				9	&	-4	&	0	&	-4	&	2	&	2	&	-2	&	2	&	0	&	1	&	-1	&	0	\\
				10	&	-5	&	-2	&	2	&	2	&	-2	&	-2	&	-4	&	1	&	1	&	1	&	2	\\
				11	&	-3	&	-1	&	1	&	2	&	-1	&	-1	&	-2	&	1	&	1	&	0	&	1	\\
				12	&	-6	&	-2	&	-6	&	-2	&	4	&	3	&	3	&	1	&	-1	&	0	&	2	\\
				13	&	-11	&	6	&	-8	&	4	&	2	&	-8	&	5	&	-2	&	1	&	-7	&	0	\\
				14	&	-16	&	8	&	-12	&	5	&	2	&	-12	&	7	&	-4	&	1	&	-11	&	0	\\
				15	&	-6	&	3	&	-5	&	2	&	1	&	-4	&	3	&	-1	&	1	&	-3	&	1	\\
				16	&	-7	&	-6	&	4	&	4	&	2	&	-4	&	3	&	0	&	1	&	-3	&	0	\\
				17	&	-14	&	-12	&	6	&	7	&	2	&	-10	&	4	&	0	&	1	&	-9	&	-3	\\
				18	&	-9	&	-8	&	4	&	5	&	2	&	-6	&	3	&	0	&	1	&	-5	&	-1	\\
				19	&	-11	&	-4	&	-12	&	-4	&	6	&	6	&	6	&	1	&	-1	&	-1	&	4	\\
				20	&	-4	&	-1	&	-5	&	-1	&	2	&	2	&	2	&	0	&	0	&	-1	&	2	\\
				21	&	-4	&	-3	&	1	&	3	&	-2	&	-2	&	-1	&	2	&	1	&	0	&	1	\\
				22	&	-5	&	-6	&	0	&	2	&	2	&	1	&	-2	&	2	&	0	&	-1	&	1	\\
				23	&	-8	&	-8	&	2	&	4	&	-4	&	-4	&	-4	&	3	&	1	&	1	&	1	\\
				24	&	-7	&	-5	&	-5	&	2	&	3	&	2	&	-4	&	1	&	1	&	-1	&	3	\\
				25	&	-8	&	-3	&	-7	&	-2	&	4	&	5	&	4	&	2	&	0	&	-2	&	3	\\
				26	&	-4	&	2	&	-4	&	1	&	1	&	-3	&	2	&	-1	&	0	&	-2	&	0	\\
				27	&	-3	&	-2	&	-2	&	2	&	2	&	-1	&	1	&	0	&	1	&	0	&	0	\\
				28	&	-3	&	-2	&	0	&	1	&	2	&	1	&	-2	&	1	&	-1	&	0	&	1	\\
				29	&	-4	&	-4	&	2	&	2	&	-1	&	-2	&	-2	&	1	&	0	&	-1	&	1	\\
				30	&	-4	&	-1	&	-5	&	1	&	2	&	-1	&	3	&	0	&	1	&	-1	&	1	\\
				31	&	-8	&	-6	&	-6	&	3	&	6	&	-2	&	3	&	-1	&	3	&	1	&	-1	\\
				32	&	-7	&	-3	&	3	&	-5	&	3	&	2	&	2	&	4	&	-1	&	-1	&	1	\\
				33	&	-3	&	-1	&	-3	&	-1	&	2	&	1	&	2	&	0	&	0	&	0	&	1	\\
				34	&	-6	&	-4	&	-4	&	2	&	5	&	-1	&	2	&	-1	&	4	&	0	&	-1	\\
				35	&	-6	&	-4	&	-6	&	-3	&	3	&	2	&	2	&	2	&	1	&	0	&	1	\\
				36	&	-6	&	-4	&	-4	&	2	&	5	&	-1	&	2	&	-1	&	3	&	1	&	-1	\\
				37	&	-6	&	-4	&	4	&	3	&	2	&	-3	&	4	&	0	&	1	&	-2	&	1	\\
				38	&	-4	&	-4	&	0	&	1	&	2	&	2	&	-1	&	2	&	-1	&	-1	&	0	\\
				39	&	-5	&	-3	&	1	&	-4	&	2	&	1	&	1	&	3	&	-1	&	-1	&	0	\\
				40	&	-4	&	-1	&	1	&	-3	&	2	&	2	&	1	&	2	&	0	&	-1	&	1	\\
				41	&	-15	&	5	&	-3	&	4	&	9	&	7	&	-9	&	5	&	-5	&	2	&	6	\\
				42	&	-5	&	-1	&	-5	&	-1	&	3	&	1	&	4	&	0	&	0	&	-1	&	3	\\
				43	&	-6	&	2	&	0	&	-5	&	3	&	3	&	1	&	3	&	-1	&	-1	&	1	\\
				44	&	-11	&	-3	&	-11	&	-3	&	7	&	5	&	6	&	2	&	-2	&	-1	&	5	\\
				45	&	-6	&	-2	&	-6	&	-2	&	3	&	3	&	4	&	1	&	-1	&	-1	&	3	\\
				46	&	-6	&	-4	&	0	&	-5	&	2	&	2	&	1	&	4	&	-1	&	-1	&	0	\\
				47	&	-4	&	-2	&	2	&	1	&	1	&	1	&	-3	&	0	&	-1	&	-1	&	2	\\
				48	&	-10	&	-4	&	4	&	-5	&	3	&	-2	&	-6	&	2	&	-6	&	1	&	3	\\
				49	&	-8	&	-4	&	2	&	-6	&	3	&	3	&	2	&	5	&	-1	&	-1	&	1	\\
				50	&	-7	&	-6	&	2	&	5	&	1	&	-5	&	1	&	1	&	0	&	-4	&	-2	\\
				51	&	-10	&	4	&	-8	&	3	&	2	&	-7	&	6	&	-2	&	1	&	-6	&	1	\\
				52	&	-4	&	-2	&	2	&	1	&	0	&	-1	&	2	&	2	&	1	&	-2	&	1	\\
				53	&	-13	&	12	&	4	&	7	&	8	&	6	&	-3	&	6	&	-3	&	-1	&	3	\\
				54	&	-6	&	-6	&	2	&	5	&	2	&	-3	&	2	&	1	&	1	&	-2	&	0	\\
				55	&	-8	&	5	&	-3	&	1	&	-3	&	-5	&	-1	&	4	&	1	&	-4	&	1	\\
				56	&	-7	&	-7	&	1	&	6	&	1	&	-4	&	1	&	2	&	1	&	-3	&	-1	\\
				57	&	-5	&	-1	&	5	&	0	&	0	&	0	&	4	&	-1	&	-1	&	-2	&	3	\\
				58	&	-7	&	-6	&	2	&	5	&	0	&	-2	&	-1	&	4	&	1	&	-3	&	1	\\
				59	&	-8	&	6	&	-2	&	5	&	-1	&	-1	&	1	&	4	&	-2	&	-4	&	-2	\\
				60	&	-8	&	-6	&	-8	&	-4	&	3	&	3	&	2	&	3	&	1	&	1	&	1	\\
				61	&	-5	&	-3	&	3	&	-3	&	-1	&	-2	&	2	&	2	&	1	&	-1	&	1	\\
				62	&	-7	&	6	&	-2	&	6	&	0	&	-1	&	-1	&	3	&	-1	&	-4	&-1		\\
				63	&	-5	&	-3	&	-3	&	0	&	3	&	1	&	-1	&	1	&	-2	&	1	&	2				
			\end{tabular}
	\end{ruledtabular}}
\end{table*}

\begin{table*}
	\caption{\label{all_322_parameters} Ground state energy density, optimal parameters, type of parameterized observables and bond dimension of 63 322-type Hamiltonians under the different dimensions of local observables.}
	\scriptsize
	\begin{ruledtabular}
		\renewcommand\arraystretch{1.3}
			\begin{tabular}{ccccccccccccccccccccc}
				\multicolumn{1}{c}{\multirow{2}{*}{No.}} & 
				\multicolumn{1}{c}{\multirow{2}{*}{${\cal{L}}$}} &  
				\multicolumn{4}{c}{$d=2$} & \multicolumn{4}{c}{$d=3$} & \multicolumn{5}{c}{$d=4$} & \multicolumn{5}{c}{$d=5$} &
				\multicolumn{1}{c}{\multirow{2}{*}{$\ltinpa{5}{1}$}}
				\\ \cline{3-6} \cline{7-10}\cline{11-15}\cline{16-20}
				\multicolumn{1}{c}{}  & \multicolumn{1}{c}{}
				&\multicolumn{1}{c}{${\cal{Q}}_2$} & $w_1$ & $[n_x, n_y]$ & $D$ 
				&\multicolumn{1}{c}{${\cal{Q}}_3$} & $w_1$ & $[n_x, n_y]$  & $D$ 
				&\multicolumn{1}{c}{${\cal{Q}}_4$} & $w_1$ & $w_2$ & $[n_x, n_y]$ & $D$ & \multicolumn{1}{c}{${\cal{Q}}_5$} & $w_1$ & $w_2$ & $[n_x, n_y]$ & $D$ & \multicolumn{1}{c}{} 
				\\ \hline
                1 & -6  & -6.08108  & 1.2472  & $[1,1]$  &  7  
                        & -6.32747  & 0.7811  & $[1,2]$  & 5  
         				&  &  &  &  &  
         				&  &  &  &  &  & -6.32747  \\
				2 & -6  & -6.10943  & 1.2224  & $[1,1]$ & 7  
				 		& -6.33712  & 0.9117  & $[1,2]$   & 5  
				 		&  &  &  &  &  
				 		&  &  &  &  &  & -6.33712  \\
				3 & -3  & -3.04150  & 1.2508  & $[1,1]$ & 8  
					    & -3.20711  & 0.7852  & $[1,2]$ & 5  
					    &  &  &  &  &  
					    &  &  &  &  &  & -3.20711  \\
				4 & -4  & -- & -- & -- & -- 
				        & -4.14623  & 0.7925  & $[1,1]$ & 5  
				        &  &  &  &  &  
				        &  &  &  &  &  & -4.14623  \\
				5 & -8  & -- & -- & -- & -- 
				        & -8.12123  & 0.6735  & $[1,2]$ & 5  
				        &  &  &  &  &  
				        &  &  &  &  &  & -8.12123  \\
				6 & -4  & -- & -- & -- & -- 
				        & -4.02415 & 0.3641 & $[1,1]$ & 16 
				        & -4.10310  & -0.0000  & 0.7031  & $[2,2]$ & 6  
				        &  &  &  &  &  & -4.10310  \\
				7 & -5  & -- & -- & -- & -- 
				        & -5.09951  & 1.1158  & $[1,1]$  & 12  
				        & -5.09951  & 0.0000  & -1.1156  & $[2,2]$ & 12  
				        & -5.29852  & 1.5704  & 0.8509  & $[2,2]$   & 6  & -5.29852  \\
				8 & -4  & -- & -- & -- &  
				        & -4.18655  & 0.9483  & $[1,1]$  & 14  
				        & -4.18655  & -0.9483  & 0.0000  & $[2,2]$ & 14  
				        & -4.33137  & 1.5699  & 0.7854  & $[2,2]$   & 6  & -4.33137  \\
				9 & -4  & -- & -- & -- & -- 
				        & -4.11581  & 1.1034  & $[1,1]$  & 14  
				        & -4.11581  & 0.4671  & 1.5708  & $[2,2]$  & 15  
				        & -4.41421  & 0.7854  & 1.5702  & $[2,2]$ & 5  & -4.41421  \\
				10 & -5 & -- & -- & -- & -- 
						& -5.07058  & 0.3947  & $[1,2]$ & 12  
						& -5.07058  & 0.3948  & 1.5708  & $[2,2]$ & 12  
						& -5.26969  & 1.5708  & 0.6590  & $[3,3]$   & 6  & -5.26969  \\
				11 & -3 & -- & -- & -- & -- 
						& -3.00628  & 0.3540  & $[1,2]$ & 8  
						& -3.00628  & 0.3533  & 1.5708  & $[2,2]$ & 8  
						& -3.11696  & 0.7857  & 1.5704  & $[3,3]$   & 6  & -3.11698  \\
				12 & -6 & -- & -- & -- & -- 
						& -6.06459  & 0.4885  & $[1,1]$ & 14  
						& -6.20259  & 0.0000  & 0.7000  & $[2,2]$ & 6  &  &  &  &  &  & -6.20261  \\
				13 &-11 & -- & -- & -- & --
						& -- & -- & -- & -- 
						& -11.19018  & -1.5708  & -2.2493  & $[2,2]$ &  6  
						&  &  &  &  &  & -11.19036  \\
				14 &-16 & -- & -- & -- & -- 
						& -- & -- & -- & -- 
						& -16.17635  & 0.8977  & 1.5708  & $[2,2]$ & 6  
						&  &  &  &  &  & -16.17654  \\
				15 & -6 & -- & -- & -- & -- 
						& -- & -- & -- & -- 
						& -6.09323  & 1.5708  & 0.8916  & $[2,2]$ & 6  
						&  &  &  &  &  & -6.09347  \\
				16 & -7 & -- & -- & -- & -- 
						& -- & -- & -- & -- 
						& -7.21677  & 1.5708  & 0.8683  & $[2,2]$ & 6  
						&  &  &  &  &  & -7.21747  \\
				17 &-14 & -- & -- & -- & -- 
						& -- & -- & -- & -- 
						& -14.18216 & -1.5708  & 2.2462  & $[2,2]$ & 6  
						&  &  &  &  &  & -14.18291  \\
				18 & -9 & -- & -- & -- & -- 
						& -- & -- & -- & -- 
						& -9.20253  & 1.5708  & 0.8852  & $[2,2]$ & 6  
						&  &  &  &  &  & -9.20406  \\
				19 & -11 & -- & -- & -- & -- 
						& -11.40483  & 0.6551  & $[1,1]$ & 14  	
						& -11.47642  & 0.0000  & -0.7229 & $[2,2]$ & 6  
						&  &  &  &  &  & -11.47839  \\
				20 & -4 & -4.00682  & 1.2170  & $[1,1]$ &  4  
						& -4.02137  & 0.4173  & $[1,1]$&  16  
						& -4.12887  & 1.5708  & 0.7030  & $[2,2]$ & 5  
						&  &  &  &  &  & -4.13251  \\
				21 & -4 & -- & -- & -- & --  
						& -- & -- & -- & -- 
						& -4.06491  & 1.5708  & 0.9817  & $[2,2]$ & 6  
						&  &  &  &  &  & -4.06890  \\
				22 & -5 & -5.00062  & 1.3959  & $[1,1]$ & 4  
						& -5.12457  & 0.7855  & $[1,2]$& 5  
						& -5.12457  & 1.5708  & 0.7854  & $[2,2]$ & 5  
						&  &  &  &  &  & -5.12887  \\
				23 & -8 & -8.01053  & 0.2887  & $[1,1]$ & 5  
						& -8.13694  & 0.6619  & $[1,1]$  & 5  
						& -8.13694  & 0.0000  & 0.6618  & $[2,2]$ & 5  
						&  &  &  &  &  & -8.14126  \\
				24 & -7 & -- & -- & -- &  
						&-7.11178  & 1.1000  & $[1,1]$  & 12  
						& -7.11178  & 0.0000  & 1.1002  & $[2,2]$ & 12  
						& -7.25367  & 0.8707  & 1.5704  & $[2,2]$  & 6  & -7.25827  \\
				25 & -8 & -- & -- & -- & -- 
						& -8.00096  & 0.1630  & $[1,1]$  & 16  
						& -8.14914  & 0.0000  & 0.6379  & $[2,2]$ & 6  
						&  &  &  &  &  & -8.15428  \\
				26 & -4 & -- & -- & -- & -- 
						& -- & -- & -- & -- 
						& -4.07800  & 1.5708  & 1.0105  & $[2,2]$ & 6  
						& -4.14354  & 0.8877  & 1.5707  & $[2,2]$ & 5  & -4.14877  \\
				27 & -3 & -- & -- & -- & -- 
						& -3.11927  & 0.8353  & $[1,1]$ & 8  
						& -3.11932  & 0.0000  & 0.7027  & $[2,2]$  &  6  
						& -3.14915  & 0.7027  & 1.5694  & $[2,2]$   & 5  & -3.15470  \\
				28 & -3 & -3.04030  & 1.2555  & $[1,1]$ & 7  
						& -3.14915  & 0.8685  & $[1,2]$ & 5  
						& -3.14915  & 0.8682  & 1.5708  & $[2,2]$  & 5  
						&  &  &  &  &  & -3.15470  \\
				29 & -4 & -- & -- & -- & -- 
						& -- & -- & -- & -- 
						& -4.00935 & 1.3742 & 0.3202  & $[2,2]$  & 12  
						& -4.09966  & 1.5708  & 0.7860  & $[3,3]$  & 7  & -4.10571  \\
				30 & -4 & -- & -- & -- & -- 
						& -4.03262  & 0.3966 & $[1,1]$ & 19  
						& -4.11218  & 0.0000  & 0.7054 & $[2,2]$  & 6  
						& -4.14023  & 0.7054  & 1.5699  & $[2,2]$   & 5  & -4.14877  \\
				31 & -8 & -- & -- & -- & -- 
						& -- & -- & -- & -- 
						& -8.22037  & 0.0000  & 0.6827  & $[2,2]$ & 6  
						&  &  &  &  &  & -8.23071  \\
				32 & -7 & -7.08653  & 1.2383  &  & 6  
						& -7.25928  & 0.9558  & $[1,2]$ & 5  
						& -7.25928  & 0.9558  & 1.5708  & $[2,2]$ & 5  
						&  &  &  &  &  & -7.27038  \\
				33 & -3 & -- & -- & -- & -- 
						& -3.11932  & 0.7018  & $[1,1]$ & 6 
						& -3.11932  & 0.0000  & 0.7026  & $[2,2]$ & 6  
						&  &  &  &  &  & -3.13058  \\
				34 & -6 & -- & -- & -- & -- 
						& -- & -- & -- & -- 
						& -6.03951  & 0.0000  & 0.4637  & $[2,2]$ & 6  
						&  &  &  &  &  & -6.05216  \\
				35 & -6 & -- & -- & -- & -- 
						& -6.19072  & 0.9694  & $[1,1]$  & 14  
						& -6.19072  & 0.0000  & 0.9700  & $[2,2]$ & 14
						& -6.26479  & 0.8478  & 1.5702  & $[2,2]$  & 6  & -6.27801  \\
				36 & -6 & -- & -- & -- & -- 
						& -- & -- & -- & -- 
						& -6.03951  & 0.0000  & 0.4637  & $[2,2]$  & 6  
						&  &  &  &  &  & -6.05327  \\
				37 & -6 & -- & -- & -- & -- 
						& -- & -- & -- & -- 
						& -6.17482  & 1.5708  & 0.8827  & $[2,2]$  & 6 
						&  &  &  &  &  & -6.18885  \\
				38 & -4 & -4.03281  & 1.2745  & $[1,1]$ & 8  
						& -4.14623  & 0.7767  & $[1,2]$ & 5  
						& -4.14623  & 1.5708  & 0.7775  & $[2,2]$  & 5 
						&  &  &  &  &  & -4.16445  \\
				39 & -5 & -5.06166  & 1.9574  & $[1,1]$ & 7  
						& -5.12466  & 0.9504  & $[1,2]$ & 7  
						& -5.12466  & 1.5708  & 0.9503  & $[2,2]$ & 7 
						&  &  &  &  &  & -5.14311  \\
				40 & -4 & -4.02193  & 1.3015  & $[1,1]$ & 5  
						& -4.06591  & 1.0457  & $[1,2]$ & 5  
						& -4.06591  & 1.0459  & 1.5708  & $[2,2]$ & 5  
						&  &  &  &  &  & -4.08498  \\
				41 & -15 & -15.01150  & 1.3600  & $[1,1]$ & 5  
				  		& -15.01150  & 1.7814  & $[1,1]$  & 5  
				  		& -15.13717  & 4.1651  & 1.5708  & $[2,2]$ & 5 
				  		&  &  &  &  &  & -15.15803  \\
				42 & -5 & -- & -- & -- & -- 
						& -- & -- & -- & -- 
						& -5.01755  & 0.7989  & 0.0884  & $[2,2]$ & 16  
						& -5.04311  & 0.2159  & 0.7668  & $[2,2]$   & 12  & -5.06400  \\
				43 & -6 & -6.01649  & 1.8170  & $[1,1]$ & 5  
						& -6.01649  & 1.8170  & $[1,1]$  & 6  
						& -6.04536  & 1.1166  & 1.5708  & $[2,2]$ &  8  
						&  &  &  &  &  & -6.06803  \\
				44 & -11& -- & -- & -- & -- 
						& -11.00119  & 0.1638  & $[1,1]$  & 12  
						& -11.15383  & 0.0000  & 0.5575  & $[2,2]$ &  6  &  &  &  &  &  & -11.17826  \\
				45 & -6 & -- & -- & -- & -- 
						& -6.18037  & 0.6128  & $[1,1]$  & 16  
						& -6.22875  & 0.0000  & 0.7013  & $[2,2]$ & 6  
						&  &  &  &  &  & -6.25396  \\
				46 & -6 & -6.06990  & 1.9751  & $[1,1]$ & 6  
						& -6.12318  & 0.9210  & $[1,2]$ & 7  
						& -6.12318  & 1.5708  & 0.9210  & $[2,2]$  & 7  
						&  &  &  &  &  & -6.14907  \\
				47 & -4 & -4.07352  & 1.1798  & $[1,1]$ & 7  
						& -4.15468  & 0.9590  & $[1,2]$ & 5  
						& -4.15469  & 1.5708  & 0.9587  & $[2,2]$ &  5  
						&  &  &  &  &  & -4.18142  \\
				48 & -10& -- & -- & -- & -- 
						& -- & -- & -- & -- 
						& -10.14203  & 0.9241  & -1.5708  & $[2,2]$ & 5  
						&  &  &  &  &  & -10.17428  \\
				49 & -8 & -8.07815  & 1.2431  & $[1,1]$ & 7  
						& -8.18189  & 1.0153  & $[1,2]$ & 5  
						& -8.18189  & 1.5708  & 1.0152  & $[2,2]$ &  5  
						&  &  &  &  &  & -8.21614  \\
				50 & -7 & -- & -- & -- & -- 
						&  -- & -- & -- & -- 
						& -7.00917 & 0.7748  & -1.6163  & $[2,2]$  & 12  
						&  &  &  &  &  & -7.04393  \\
				51 & -10& -- & -- & -- & -- 
						&  -- & -- & -- & -- 
						& -10.10510  & -0.9808  & 1.5708  & $[2,2]$ & 6 
						&  &  &  &  &  & -10.14126  \\
				52 & -4 & -- & -- & -- & --  
						& -- & -- & -- & --
						& -4.08523  & 1.5708  & 0.8681  & $[2,2]$ & 9 
						&  &  &  &  &  & -4.12479  \\
				53 & -13& -- & -- & -- & -- 
						& -- & -- & -- & -- 
						& -13.10714 & 0.0000  & -0.5039  & $[2,2]$ & 7  
						&  &  &  &  &  & -13.14828  \\
				54 & -6 & -- & -- & -- & --  
						& -- & -- & -- & -- 
						& -6.16786  & 0.8899  & 1.5708  & $[2,2]$ & 6  
						&  &  &  &  &  & -6.21564  \\
				55 & -8 & -- & -- & -- & --  
						& -- & -- & -- & -- 
						& -8.00371  & 2.3012 & -1.5949  & $[2,2]$ & 16  
						&  &  &  &  &  & -8.05192  \\
				56 & -7 & -- & -- & -- & -- 
						& -- & -- & -- & -- 
						& -7.01064  & 0.8015  & 1.6238  & $[2,2]$ & 12  
						&  &  &  &  &  & -7.05935  \\
				57 & -5 & -- & -- & -- & -- 
				        & -- & -- & -- & -- 
				        & -5.05572  & 1.5708  & 1.0099  & $[2,2]$  &  9  
				        &  &  &  &  &  & -5.10699  \\
				58 & -7 & -- & -- & -- & -- 
						& -- & -- & -- & -- & -7.00144  & 0.3372  & 1.5603  & $[2,2]$ & 11  
						&  &  &  &  &  & -7.05714  \\
				59 & -8 & -- & -- & -- & --  
						& -- & -- & -- & -- 
						& -8.17045  & -1.5708  & 0.8676  & $[2,2]$ & 9  
						& -8.17045  & 0.8686  & 1.5708  & $[2,2]$ & 9  & -8.24220  \\
				60 & -8 & -- & -- & -- & --
						& -8.22013  & 0.9539  & $[1,1]$  & 5  
						& -8.22013  & 0.0000  & 0.9539  & $[2,2]$ & 5  
						&  &  &  &  &  & -8.29902  \\
				61 & -5 & -- & -- & -- & -- 
						& -5.00822  & 0.3766  & $[1,2]$ & 11  
						& -5.00822  & 0.3774  & 1.5708  & $[2,2]$ & 11  
						&  &  &  &  &  & -5.09824  \\
				62 & -7 & -- & -- & -- & -- 
						& -- & -- & -- & -- 
						& -7.00638  & 2.1879  & 1.5295  & $[2,2]$ & 24  
						&  &  &  &  &  & -7.10456  \\
				63 & -5 & -- & -- & -- & -- 
						& -5.00639  & 1.3547  & $[1,1]$  & 8  
						& -5.00639  & 1.3556  & 0.0000  & $[2,2]$ & 8  
						&  &  &  &  &  & -5.13037  \\
		\end{tabular}
	\end{ruledtabular}
\end{table*}

\section{Couplings and parameters of the 232-type Hamiltonians} 
\label{sec:appendix 232-type violated results}

Table~\ref{232couplings} gives the couplings, classical bounds (${\cal{L}}$) of five 232-type TI Hamiltonians presented in the main text. The parameters defining local observables and the numbers of $-1$ in the eigenvalues of each local observables ($[n_x,n_y,n_z]$) are presented in Table~\ref{232_d2_d3} when $d=3$ and in Table~\ref{232_d4} when $d=4$ .

\begin{table*}
	\caption{\label{232couplings} Couplings for 5 232-type TI Hamiltonians presented in Table~\ref{table:232violations}.}
	\scriptsize
	\begin{ruledtabular}
		\renewcommand\arraystretch{1.3}
		\begin{tabular}{ccccccccccccccccccccc}
		No. & ${\cal{L}}$  &  $J_x$ & $J_y$ & $J_z$ & $J_{xx}$ & $J_{xy}$ & $J_{xz}$ & $J_{yx}$ & $J_{yy}$ & $J_{yz}$ & $J_{zx}$ & $J_{zy}$ & $J_{zz}$
			\\ \hline
			1 & -9 & 3 & -2 & 1 & 2 & -2 & 5 & -1 & 1 & 3 & 1 & -2 & -2 \\
			2 & -4 & 2 & 1 & 1 & 2 & 2 & 1 & 0 & -1 & 0 & -1 & 1 & 0 \\
			3 & -5 & 2 & -1 & 1 & 1 & -1 & 2 & -2 & 0 & 2 & -1 & 0 & 0 \\
			4 & -4 & 1 & -1 & 0 & 1 & 1 & 2 & 0 & 1 & -1 & 0 & 1 & -1 \\
			5 & -2 & 0 & 0 & 0 & 1 & 1 & 0 & 0 & 0 & 0 & -1 & 1 & 0 \\			
		\end{tabular}
	\end{ruledtabular}
\end{table*}

\begin{table*}
	\caption{\label{232_d2_d3} Parameters of 5 232-type TI Hamiltonians under the bond dimension $D=10$ and the dimension of local observables $d=3$.}
	\scriptsize
	\begin{ruledtabular}
		\renewcommand\arraystretch{1.3}
		\begin{tabular}{cccccccccccccccc}
			\multicolumn{1}{c}{\multirow{2}{*}{No.}} & \multicolumn{1}{c}{\multirow{2}{*}{${\cal{L}}$}} & \multicolumn{10}{c}{$d=3$} 
			\\ \cline{3-12}
			\multicolumn{1}{c}{} & \multicolumn{1}{c}{} & \multicolumn{1}{c}{$w_1$} & $w_2$ & $w_3$ & $w_4$ & $w_5$ & $w_6$ & $w_7$ & $w_8$ & $w_9$ &  $[n_x, n_y, n_z]$ \\
			1 & -9 & 0.3392  & 0.1055  & 0.4115  & 0.4133  & 0.3652  & 0.9916  & 0.4810  & -0.0351  & -0.3163  & $[2, 1, 1]$\\
			2 & -4 & 0.5943  & 0.0469  & 0.2011  & 0.2756  & 0.4268  & 0.1822  & 0.0277  & 0.7213  & 0.1413  & $[2, 1, 2]$\\
			3 & -5 & 1.0061  & -0.0081  & 0.2303  & 0.3148  & 0.0616  & 0.5666  & -0.2498  & 0.2271  & 0.0076  & $[2, 1, 2]$\\
			4 & -4 & 0.1303  & -0.0032  & 0.5139  & 0.2574  & 0.4723  & 0.6582  & 0.3285  & -0.1727  & 0.2820  & $[2, 1, 1]$\\
			5 & -2 & 0.4899  & -0.1481  & 0.1905  & 0.4931  & 0.4919  & 0.3638  & -0.1067  & 0.7160  & 0.2088  & $[2, 1, 2]$\\
							
		\end{tabular}
	\end{ruledtabular}
\end{table*}

\begin{table*}
	\caption{\label{232_d4} Parameters of 5 232-type TI Hamiltonians under the bond dimension $D=10$ and the dimension of local observables $d=4$.}
	\scriptsize
	\begin{ruledtabular}
		\renewcommand\arraystretch{1.3}
		\setlength{\tabcolsep}{0.2mm}{
		\begin{tabular}{ccccccccccccccccccccc}
			\multicolumn{1}{c}{\multirow{2}{*}{No.}} & \multicolumn{1}{c}{\multirow{2}{*}{${\cal{L}}$}} & \multicolumn{19}{c}{$d=4$}  
			\\ \cline{3-21}
			\multicolumn{1}{c}{} & \multicolumn{1}{c}{}  & \multicolumn{1}{c}{$w_1$}  & $w_2$ & $w_3$ & $w_4$ & $w_5$ & $w_6$ & $w_7$ & $w_8$ & $w_9$ & $w_{10}$ & $w_{11}$ & $w_{12}$ & $w_{13}$ & $w_{14}$ & $w_{15}$ & $w_{16}$ & $w_{17}$ & $w_{18}$ &  $[n_x, n_y, n_z]$ \\
			1 & -9 & -0.0270  & 0.2185  & 0.0934  & 0.2068  & 0.3829  & 0.3010  & 0.3126  & 0.3287  & 0.2716  & 0.1611  & -0.2841  & -0.4393  & 0.1235  & 0.1030  & 0.1933  & 0.1480  & 0.6689  & 1.0601 & $[3,1,1]$\\
			2 & -4 & 0.4490  & 0.5450  & 0.4828  & 0.0387  & 0.1335  & 0.1225  & 0.2003  & 0.2535  & 0.2878  & 0.0661  & 0.2800  & 0.1061  & 0.0102  & -0.0804  & 0.0821  & 0.0320  & 0.3449  & 0.0535  & $[3,2,3]$\\
			3 & -5 & 0.5565  & 1.2916  & 0.8706  & 0.1371  & 0.1162  & 0.9160  & 0.4551  & 0.5068  & 0.1782  & 0.4305  & 0.8979  & 0.1097  & 0.2928  & -0.4064  & 0.1088  & 0.7717  & 0.6469  & 0.9466  & $[3,2,3]$\\
			4 & -4 & -0.2912  & 0.2321  & 0.0211  & 0.0911  & 0.2954  & 0.0986  & 0.6905  & 0.3049  & 0.3104  & 0.2755  & 0.1093  & 0.1728  & 0.0995  & 0.4044  & -0.0148  & 0.1701  & 0.2056  & 0.4232  & $[3,2,1]$\\
			5 & -2 & 0.4998  & 0.0447  & -0.3366  & -0.0197  & 0.3869  & -0.0971  & 0.2528  & 0.2196  & 0.5018  & 0.0579  & 0.1594  & 0.2374  & 0.0932  & 0.1598  & 0.5637  & 0.5072  & 0.2408  & 0.2743 & $[2,1,2]$\\
		\end{tabular}}
	\end{ruledtabular}
\end{table*}

\end{appendix}

\end{document}